\documentclass[aps,twocolumn,prd,superscriptaddress]{revtex4-2}

\usepackage[utf8]{inputenc}

\usepackage{mathtools}
\usepackage{amsfonts}
\usepackage{mathrsfs}
\usepackage{bbm}
\usepackage{physics}
\usepackage{slashed}
\usepackage{tensor}

\usepackage{graphicx}
\usepackage{color, float}
\usepackage{array}
\usepackage[abs]{overpic}

\usepackage{placeins}

\usepackage{makecell}
\usepackage{subcaption}

\usepackage{xspace}
\usepackage{siunitx}
\usepackage{xfrac}
\usepackage{hyperref}
\usepackage[nameinlink]{cleveref}
\usepackage{appendix}

\usepackage{xifthen}
\usepackage{xcolor}
\hypersetup{
	colorlinks,
	linkcolor={red!75!black},
	citecolor={blue!75!black},
	urlcolor={blue!75!black}
}

\usepackage{booktabs}
\usepackage{multirow}

\newcolumntype{C}{>{$}c<{$}}
\AtBeginDocument{
	\heavyrulewidth=.08em
	\lightrulewidth=.05em
	\cmidrulewidth=.03em
	\belowrulesep=.65ex
	\belowbottomsep=0pt
	\aboverulesep=.4ex
	\abovetopsep=0pt
	\cmidrulesep=\doublerulesep
	\cmidrulekern=.5em
	\defaultaddspace=.5em
}

\graphicspath{{./figures/}}

\newcommand{\gettitle}{Physics-informed operator flows and observables }

\newcommand{\getHeidelbergAffiliation}{\affiliation{Institut f{\"u}r Theoretische Physik, Universit{\"a}t Heidelberg, Philosophenweg 16, 69120 Heidelberg, Germany}}
\newcommand{\getEMMIAffiliation}{\affiliation{ExtreMe Matter Institute EMMI, GSI, Planckstr. 1, 64291 Darmstadt, Germany}}

\hypersetup{
	pdftitle={\gettitle},
	pdfauthor={Ihssen, Pawlowski},
	pdfkeywords= {renomalisation group} ,
	bookmarksopen=true,
	bookmarksopenlevel=2,
	bookmarksnumbered=true
}

\begin{document}
	
	\title{\gettitle}
	
	\author{Friederike Ihssen}
	\getHeidelbergAffiliation
	\author{Jan M. Pawlowski}
	\getHeidelbergAffiliation\getEMMIAffiliation
\begin{abstract}
We discuss physics-informed renormalisation group flows (PIRGs) for general operators. We show that operator PIRGs provide comprehensive access to all correlation functions of the quantum field theory under investigation. The operator PIRGs can be seen as a completion of the PIRG approach, whose qualitative computational simplification and structural insights are now fully accessible for general applications. The potential of this setup is assessed within a simple analytic example of the zero-dimensional $\phi^4$-theory for which the generating functions of the fundamental field are computed within a vertex expansion, using the one- to ten-point functions. 

\end{abstract}
	\maketitle

\section{Introduction} 
\label{sec:Introduction} 

A novel functional renormalisation group approach to quantum field theories (QFTs) has been put forward in \cite{Ihssen:2024ihp}, the physics-informed renormalisation group (PIRG). It is based on the generating functional of the theory, which is coupled to a general composite operator instead of the fundamental field. The setup accommodates general reparametrisations of the theory in combination with integrating out degrees of freedom along some RG-scale formulated for the 1PI effective action $\Gamma_\phi$ in \cite{Pawlowski:2005xe}. 

The novel ingredient of PIRG flows is a change of perspective: instead of accessing the physics of the given theory by computing a generating functional in terms of the fundamental field $\varphi$, in most cases the effective action, we split this task into computing the pair of functionals $(\Gamma_T[\phi]\,,\,\dot\phi)$. Here, $\Gamma_T$ is the \textit{target action} and $\dot\phi$ is the \textit{flowing composite field} that describes the change of the composite field with the RG-scale. This allows us to either choose the target action or the flowing field in a \textit{physics-informed} way for every coupling within the approximation; and leaves us with the task of computing the remaining components of the composite operator or target action from the RG flow. 
Seemingly, this only entails that one moves the difficulties of computing the effective action of the fundamental fields by means of the Wetterich flow \cite{Wetterich:1992yh}, a functional non-linear partial differential equation (PDE) of the diffusion-convection type, to that of computing the flowing composite field. However, it is far more: First of all, the composite field satisfies a functional linear ordinary differential equation (ODE) at every RG-scale, which is qualitatively simpler than solving the PDE which defines the effective action. Consequently, the choice of the pair $(\Gamma_T\,,\,\dot\phi)$ or rather the split can minimise the computational effort. Secondly, the formulation gives access to novel structural insights into the theory at hand. 

In \cite{Ihssen:2024ihp, Ihssen:2025cff} we have put forward the conceptual background of this novel approach, both for general QFTs as well as for gauge theories. We have also illustrated its practical applicability and potential within selected examples. In particular, we have shown at the example of the anharmonic oscillator that the PIRG approach gives access to topological properties of quantum theories \cite{Bonanno:2025mon}, which have so far been elusive within the fRG-approach. 

In the present work we address the pivotal question in the PIRG approach: the computation of correlation functions of the fundamental field $\varphi$, including the standard effective action as well as observables. This has been called the \textit{reconstruction} task in \cite{Ihssen:2024ihp} and we have provided several reconstruction paths with practical examples. This task is fully resolved here with a general reconstruction approach based on the operator flow for general operators derived in \cite{Pawlowski:2005xe}, for respective considerations within the \textit{essential RG} see \cite{Baldazzi:2021ydj}.  
We provide this flow in the present PIRG setup in a concise form and discuss specific physics-informed choices of the operator dependence of the pair $(\Gamma_T\,,\,\dot\phi)$. This allows us to compute general correlation functions of the fundamental field and other observables with simple linear convection-diffusion equations of the heat equation type, and completes the PIRG setup. Among other insights it gives us full access to qualitative computational simplifications, structural insights and direct practical means for computing general correlation functions and observables. 

In \Cref{sec:PIRGs} we briefly describe the PIRG approach, concentrating on the relevant aspects for the generalised flow of general operators. These flows are derived in \Cref{sec:PIRGsforOperators}, where we also discuss constraints for their existence. In \Cref{sec:Applications} we apply the generalised operator flows to the one- and two-point functions of the fundamental field. The mechanisms and potential of these flows are illustrated within the analytically accessible zero-dimensional QFT, where we also discuss different representations of the operator flows. In \Cref{sec:HigherDim} we discuss 
the properties of operator flows in higher dimensions, specifically the distribution of physics amongst the different components of the PIRG approach and the implementation and assessment of systematic expansion schemes. We close with a short conclusion in \Cref{sec:Conclusion}.

\section{PIRGs and their properties} 
\label{sec:PIRGs}

The physics-informed RG (PIRG) approach set up in \cite{Ihssen:2024ihp} accommodates general renormalisation group flows including general reparametrisations of the theory with a classical action $S_\textrm{cl}[\hat\varphi]$ and a fundamental field operator $\hat\varphi_i$, where the index $i$ comprises space-time (or momenta) as well as Lorentz and internal indices and species of fields (DeWitt's condensed notation). While we resort to a real scalar theory for our examples, the following derivations are general for theories with an arbitrary field content. 

In \Cref{sec:Flowingphi+TargetAction} we briefly review the PIRG approach and specifically discuss the change of perspective that is at its root. In \Cref{sec:GenFlow} we review the generalised flow equation that governs the pair $(\Gamma_T\,,\,\dot\phi)$.

\subsection{Flowing fields and target actions} 
\label{sec:Flowingphi+TargetAction}

The PIRG approach is based on the regularised generating functional for correlation functions of a composite operator $\hat\phi_k[\hat\varphi]$, 
\begin{align} 
Z_\phi[J_\phi] \propto \frac{1}{{\cal N}_k} \int d\hat\varphi \, e^{-S_\textrm{cl}[\hat\varphi]- \Delta S[\hat\phi]+\hat\phi_i J^i_\phi }\,, 
	\label{eq:GenFunctZphi}
\end{align}
where the subscript ${}_\phi$ indicates that a composite field $\hat\phi_i[\hat\varphi]$ is coupled to the current. Notably, the infrared cutoff term is quadratic in the composite field $\hat\phi_i$,  
\begin{align} 
	\Delta S[\hat\phi]=\frac12 \hat\phi_i R_\phi^{ij} \hat\phi_j\,.
	\label{eq:Regforhatphi}
\end{align}
\Cref{eq:Regforhatphi} depends on the RG-scale $k$ and suppresses momentum modes in terms of the composite fields. The quadratic form of \labelcref{eq:Regforhatphi} in the composite field, together with the source term for $\hat\phi$, entails that the respective flow equation is one-loop exact in terms of correlation functions of the composite, see \cite{Pawlowski:2005xe}. The composite field operator $\hat\phi_i$ may include the fundamental fields $\hat\varphi$, e.g., 
\begin{align} 
	\hat\phi_1=(\hat\varphi(x), \hat\varphi(x)\hat\varphi(y))\,,\qquad  \hat\phi_2=(\hat\varphi(x) \hat\varphi(x) )\,. 
\label{eq:Examplehatphis}
\end{align}
Note that the index $i$ in $\hat\phi_{1,i}$ and $\hat\phi_{2,i}$ sums over different space-time indices. Moreover, in both cases diverging cutoff terms suppress all degrees of freedom but the setup also accommodates a partial suppression. 

The effective action $\Gamma_\phi[\phi]$ is obtained as the Legendre transform of $\ln Z_\phi[J_\phi]$ with 
\begin{align} 
	\Gamma_\phi[\phi] = \sup_{J_\phi} \left(\phi_i \, J_\phi^i\, - \ln Z[J_\phi] \right) - \Delta S[\phi] \,, 
	\label{eq:Gammaphi}
\end{align}	
where the supremum $J_\phi[\phi]$ is determined such that the expectation value of $\hat\phi$ is the given mean field $\phi$, 
\begin{align} 
		\phi =\langle \hat \phi_k[\hat\varphi]\rangle \,.
\label{eq:Meanphi}
\end{align}	
The subscript $_\phi$ in $\Gamma_\phi$ denotes that the Legendre transformation is taken with respect to the current $J_\phi$.
We emphasise that by definition of the Legendre transform, $\phi$ is an independent variable and does not depend on $t$ or any other parameter in the generating functional. The examples in \labelcref{eq:Examplehatphis} have been chosen such that the effective action $\Gamma_{\phi_1}$ is a variant of the two-particle irreducible effective action. The effective action $\Gamma_{\phi_2}$ is a variant of the density functional.  

Inserting \labelcref{eq:Gammaphi} into \labelcref{eq:GenFunctZphi} leads us to a functional integro-differential equation for the effective action, 
\begin{align} 
	\hspace{-.1cm}e^{-\Gamma_\phi } = \frac{1}{{\cal N}_k} \int\! d\hat\varphi \, e^{-S_\textrm{cl}[\hat\varphi]-\Delta S[\hat\phi-\phi]+ (\hat\phi-\phi)_i\frac{\delta\Gamma_\phi}{\delta\phi_i}} \,,
	\label{eq:GammaIntegral}
\end{align}
which shall be used for the derivation of the generalised flow equation of the effective action as well as for the derivation of the generalised operator flows. \Cref{eq:GammaIntegral} makes it evident, that the effective action $\Gamma_\phi[\phi]$ in general does not agree with that of the fundamental field, $\Gamma_\varphi[\phi=\varphi]$, if not evaluated on the equations of motion. 

\begin{figure*}
	\centering%
	\begin{subfigure}[t]{.45\linewidth}
		\includegraphics[height=0.6\linewidth]{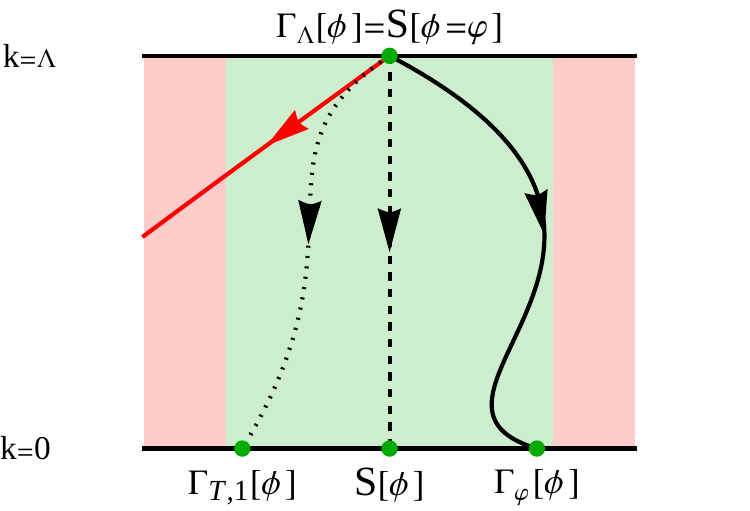}
		\caption{Target action flows: A large set of target actions is allowed at $k=0$, including the classical target action with $\partial_t \Gamma_T\equiv 0$. This extreme choice only illustrates the generality of the setup and is not optimised, for a detailed discussion see \cite{Ihssen:2024ihp}. \hspace*{\fill}}
		\label{fig:TargetAction}
	\end{subfigure}%
	\hspace{0.1\linewidth}%
	\begin{subfigure}[t]{.45\linewidth}
		\includegraphics[height=0.6\linewidth]{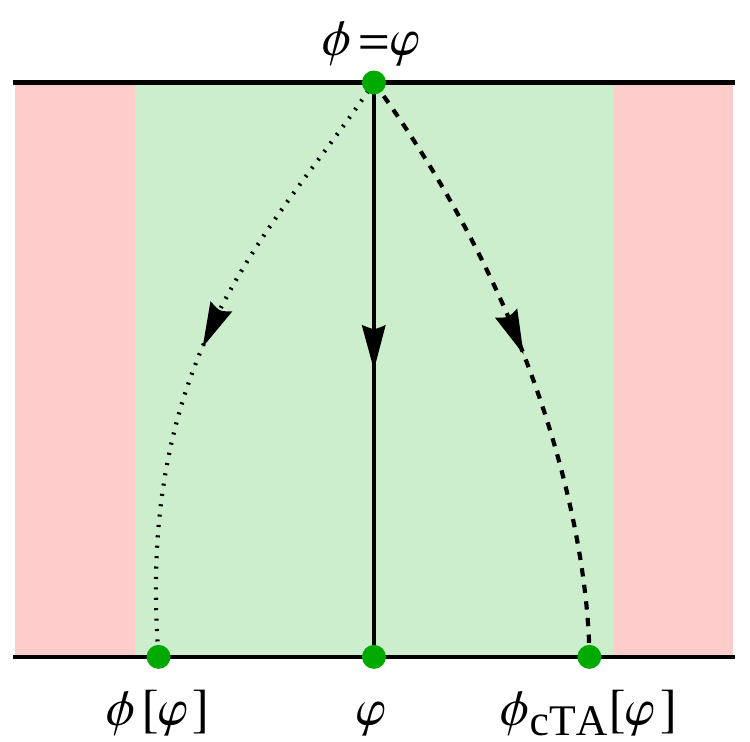}
		\caption{Flowing fields: Cutoff-time integral of the flowing field, $\phi_k=\varphi+\int_\Lambda^k \dot\phi \, dk^\prime/k^\prime$, corresponding to the target action flows depicted in \Cref{fig:TargetAction}. The non-triviality of the flowing field is growing with the increasing triviality of the flow of the target action.   \hspace*{\fill}}
		\label{fig:dotPhi}
	\end{subfigure}
	\caption{Schematic depiction of the flows for the \textit{target action} and \textit{flowing field} pair $(\Gamma_T,\dot\phi)$ defined in \labelcref{eq:PairGamm-phi}. Allowed areas are depicted in \textit{green}, whereas prohibited areas are \textit{red}. Notably, some solutions may exist locally \textit{(red arrow)} but do not correspond to a valid field transformation. We show two explicit examples of physics-informed pairs: The standard Wetterich pair $(\Gamma_\varphi[\phi], \phi = \varphi )$ \textit{(black)} and the classical target action $(S[\phi], \phi_{\mathrm{cTA}})$ \textit{(dashed, black)}, which is discussed later in \Cref{sec:Applications} around \labelcref{eq:Zd0+GTScl} and \Cref{app:cumuFlow}.\hspace*{\fill}}
	\label{fig:SchematicPics}
\end{figure*}

PIRG flows are obtained by formulating the functional renormalisation group for the pair  
\begin{align} 
	\bigl( \Gamma_\phi[\phi]\,,\,\dot\phi[\phi]\bigl)\,,
	\label{eq:PairGamm-phi}
\end{align}
where $\Gamma_\phi[\phi]$ is the \textit{target action}, also called $\Gamma_T[\phi]$ and $\dot\phi[\phi]$ is the expectation value of the flow of $\hat\phi$, 
\begin{align}
	\dot\phi[\phi]= \langle \partial_t \hat \phi_k[\hat\varphi]\rangle \,,
	\label{eq:dotphi}
\end{align}
where the RG-time is given by $t=\log k/k_\textrm{ref}$ with some reference scale $k_\textrm{ref}$. 
The field $\phi$ with the flow \labelcref{eq:dotphi} is the \textit{flowing composite field} or emergent composite \labelcref{eq:Meanphi}. We emphasise that $\partial_t \phi=0$ as $\phi$ is the variable in the Legendre transform. 

The important change of perspective advertised in \cite{Ihssen:2024ihp} is the following: the flowing field $\dot \phi[\phi]$ in the pair \labelcref{eq:PairGamm-phi} is at our disposal and so is the target action $\Gamma_T$. This flexibility can be used to construct a wide variety of target actions with the corresponding flowing fields. Limitations have been discussed judiciously in \cite{Ihssen:2024ihp, Ihssen:2025cff}. We recall the two important constraints, 
\begin{itemize} 
	\item[(i)] \textit{Local} existence of $\dot \phi_k[\phi]$ in \labelcref{eq:dotphi}: One has to show that there is an operator $\partial_t \hat\phi_k[\hat\phi]$ leading to $\dot \phi_k[\phi]$, which is an inverse problem. 
	\item[(ii)] \textit{Global} existence of $\phi_k[\varphi]$: Its flow is integrated from $k=\Lambda$ to $k=0$. The result has to be free of divergences.   
\end{itemize} 
These requirements and their implications are discussed at length in Section II in \cite{Ihssen:2024ihp} and further comments can be found in \cite{Ihssen:2025cff}. A schematic visualisation of the PIRG pair as well as allowed and prohibited choices are provided in \Cref{fig:SchematicPics}. Notably, different choices of PIRG pairs, lead to different results for the effective action. 

In \Cref{sec:Constraints+InitialConditions} we further elaborate the above requirements in the context of operator flows. Finally, \Cref{fig:SchematicPics2} illustrates how different PIRG pairs can be inserted into operator flows to derive the same physical observables. 

Finally, we would like to iterate the remark below \labelcref{eq:GammaIntegral} that in general, $\Gamma_\phi$ is not the effective action of the fundamental field, $\Gamma_\varphi$, rewritten in terms of the flowing field but a different generating functional. Still, it carries the same physics information as $\Gamma_\varphi$.
Notably, we do not need to know the underlying transformation $\partial_t \hat\phi_k[\hat\varphi]$, \labelcref{eq:dotphi}, or rather $\partial_t \hat\phi_k[\hat\phi]$ for this setup to reconstruct physical observables: Correlation functions of the fundamental field can directly be recovered from the pair $(\Gamma_\phi, \dot\phi)$ using the operator flow, which is discussed in \Cref{sec:PIRGsforOperators}.

\subsection{Generalised flows} 
\label{sec:GenFlow}

The pair \labelcref{eq:PairGamm-phi} of a target action and the respective flowing field satisfies the generalised flow equation \cite{Pawlowski:2005xe}, 
\begin{align}
\left( \partial_t + \dot{\phi}_i \frac{\delta}{\delta \phi_i} \right) \Gamma_\phi =\frac{1}{2} G^{ij}_{\phi}\left({\cal D}_t \, R_\phi\right)_{ij} +\partial_t {\cal \ln N}_\phi \,, 
	\label{eq:GenFlow} 
\end{align}
where $\partial_t {\cal \ln N}_\phi $ originates in the normalisation of the path integral and adds to the constant part of the effective action. It plays an important rôle for the definition of cumulants-preserving flows, see \cite{Ihssen:2024ihp} or \Cref{app:Details}. 
The generalised flow equation \labelcref{eq:GenFlow} is governed by the propagator $G_\phi^{ij}$ of the flowing field $\phi$, 
\begin{align} 
	G_{\phi}^{ij}[\phi] =\left[\frac{1}{\Gamma_{\phi}^{(2)}[\phi]+R_\phi}\right]^{ij}\,,
	\label{eq:Gphiphi}
\end{align}
with the $n$th derivative of a functional $F[\phi]$ with respect to the fields (or any other function), 
\begin{align} 
	 F_{\phi,i_1\cdots i_n}^{(n)}[\phi]=\frac{\delta^n F_\phi[\phi]}{\delta\phi^{i_1}\cdots \delta \phi^{i_n}}\,, 
	\label{eq:Gn}
\end{align}

and the covariant derivative 
\begin{align} 
{\cal D}_t[\phi]=\partial_t + 2 \gamma_\phi[\phi]\,,\qquad \gamma^{ij}_{\phi}[\phi]\ = \frac{\delta \dot\phi^j }{\delta \phi_i}\,,  
\label{eq:Dt}
\end{align}
see \cite{Pawlowski:2005xe}. \Cref{eq:Dt} makes it explicit that $\gamma^{ij}_\phi[\phi]$ can be understood as a field-dependent anomalous dimension. \Cref{eq:GenFlow} is based on flowing field operators $\partial_t \hat\phi$ on the \textit{microscopic} level within the notation introduced in \cite{Wetterich:2024uub}. Specifically this allows us to regularise the flowing composite degrees of freedom. In turn, transformations on the level of the \textit{macroscopic} mean fields have been considered in \cite{Wetterich:1996kf, Gies:2001nw, Gies:2002hq, Floerchinger:2009uf}. Both,  microscopic and macroscopic flowing fields can be used in the PIRG setup.

\Cref{eq:GenFlow} will play a pivotal rôle in the derivation and assessment of the generalised operator flow, since the latter can be derived directly from it. Alternatively, the generalised operator flow can also be obtained from \labelcref{eq:GammaIntegral}, similarly to the derivation of the generalised flow \labelcref{eq:GenFlow}. 

For this reason we briefly review the derivation of \labelcref{eq:GenFlow} by taking a $t$-derivative of \labelcref{eq:GammaIntegral}. Then, the left hand side follows from the $t$-derivative of the left hand side of \labelcref{eq:GammaIntegral} as well as hitting the source term leading to $\langle \partial_t\hat\phi^i_k\rangle  \delta\Gamma_\phi/\delta\phi^i$ and using \labelcref{eq:dotphi}. The right hand side of \labelcref{eq:GenFlow} follows from the $t$-derivative of $\Delta S[\hat\phi-\phi]$ and using 
\begin{align}
	\left\langle (\hat\phi-\phi)_i R_\phi^{ij} \partial_t \hat\phi_j\right \rangle = G_{\phi,il}\frac{\delta \dot\phi_j}{\delta\phi_l} R_\phi^{ij}=G_{\phi,il}\gamma_\phi{}^l{}_j R_\phi^{ij}. 
	\label{eq:GenFlowterms}  
\end{align}
The last step in \labelcref{eq:GenFlowterms} follows with the definition of $\gamma_\phi[\phi]$ in \labelcref{eq:Dt}. Together with the standard flow term $1/2 G_\phi^{ij} \partial_t R_{ij}$ and the derivative of the normalisation ${\cal N}_k$ this adds up to the right hand side of \labelcref{eq:GenFlow}. 

This completes the PIRG setup. In short, it allows us to use the pair \labelcref{eq:PairGamm-phi} in combination with the generalised flow \labelcref{eq:GenFlow} to simplify the conceptual and numerical tasks at hand. 
As already mentioned before, for non-trivial flowing fields $\dot\phi$, the target action $\Gamma_T$ is not the generating functional for correlation functions of the fundamental field $\varphi$, but that of the composite field $\phi$. If the set of composite fields also includes the fundamental ones, this also gives us access to the the correlation functions of the fundamental field as well as the effective action $\Gamma_\varphi$, for the respective discussion see \cite{Ihssen:2024ihp}, and for explicit examples in QCD see \cite{Fu:2019hdw, Ihssen:2024miv}.

Direct access to the $\varphi$-quantities is lost if the composite field does not include $\varphi$. This happens if the field basis is changed, e.g.~in an O(N) theory, where a polar basis in the broken phase is physics-informed as the theory is reparametrised in terms of the dynamical degrees of freedom, see \cite{Lamprecht2007, Isaule:2018mxt, Isaule:2019pcm}. This is an example for the ground state expansion, formalised in \cite{Ihssen:2022xjv, Ihssen:2024ihp}. Finally, the latter expansion scheme also encompasses expansions about the two point function \cite{Salmhofer:2006pn, Ihssen:2022xjv}, as well as the minimal essential scheme within the essential RG \cite{Wegner_1974, Baldazzi:2021ydj, Baldazzi:2021orb, Baldazzi:2021fye, Knorr:2022ilz, Knorr:2023usb, Baldazzi:2023pep, Knorr:2024yiu, Ohta:2025xxo}.	

Consequently, for the computation of correlation functions of the fundamental field or even the effective action $\Gamma_\varphi$, the PIRG approach has to be complemented with a computational scheme for these correlation functions. 
In \cite{Ihssen:2024ihp} we have initiated this discussion by constructing \textit{cumulants-preserving} PIRGs. In the present work we add to the PIRG approach by introducing generalised operator flows which can be used for the direct computation of correlation functions of the fundamental fields. This reconstruction idea has already been proposed in \cite{Baldazzi:2021ydj} in the context of the essential RG.  
While at its core these flows are provided by the operator flows set up in \cite{Pawlowski:2005xe}, see also \cite{Pagani:2016pad, Baldazzi:2021ydj}, the formalism hosts some intricacies, related to the implicit definition of the composite operators: In short, for expectation values $\langle \hat \varphi^n\rangle$ of operators $\hat\varphi^n$, the generalised operator flows contain additional terms and the standard form \cite{Pawlowski:2005xe} is only recovered if allowing for additional terms in the operator itself. These intricacies and the general setup are discussed in the following Section.

\section{Operator PIRGs and their properties}
\label{sec:PIRGsforOperators}
 
The physics-informed setup can be used to derive flow equations for general operators. The derivation uses the generalised flow equation \labelcref{eq:GenFlow} and follows the one of general operator flows in \cite{Pawlowski:2005xe} for $\hat{\cal O}$ with 
\begin{subequations}
\label{eq:OperatorO}
\begin{align} 
	{\cal O}[\phi]= \langle \hat {\cal O}_k[\hat\phi,J_\phi]\rangle \,,
	\label{eq:OperatorODef}
\end{align} 
and 
\begin{align} 
 \hspace{-0.05cm} \hat {\cal O}_k\left[\frac{\delta}{\delta J_\phi} ,J_\phi\right] = e^{-\Delta S\left[\frac{\delta}{\delta J_\phi}\right]} \,\hat {\cal O}\left[\frac{\delta}{\delta J_\phi} ,J_\phi\right]\, e^{\Delta S\left[\frac{\delta}{\delta J_\phi}\right]}. 
\label{eq:hatODer}
\end{align} 
\end{subequations}
The regulator term $\Delta S$ is defined in \labelcref{eq:Regforhatphi}, for more details on the general setup see \cite{Pawlowski:2005xe}, Section III. Moreover, we restrict the following derivations to $t$-independent $\hat{\cal O}$ with 
\begin{align}
	\left.\partial_t\right|_{\hat\varphi} \hat{\cal O}[\hat\phi_k\left[\hat\varphi],J_\phi\right] =0\,.
\label{eq:ONotDep}
\end{align}
\Cref{eq:ONotDep} ensures that $\langle \left.\partial_t\right|_{\hat\varphi} \hat{\cal O}\rangle $ is absent in the operator flow $\partial_t {\cal O}[\phi]$. For the general case we refer to \cite{Pawlowski:2005xe}. The restriction \labelcref{eq:ONotDep} can be reformulated as follows: all considered operators $\hat{\cal O}$ are cutoff-independent, if considered as functions of the fundamental field $\hat\varphi$ (seen as a functional of $\hat\phi$) and the current $J_\phi$. Hence, in a slight abuse of notation we write 
\begin{align}
	\hat {\cal O}= \hat{\cal O}\left[\hat\varphi,J_\phi\right]\,,\qquad \hat\varphi=\hat\varphi\left[\hat\phi=\frac{\delta}{\delta J_\phi}\right]\,,
	\label{eq:hatOhatvarphi}
\end{align}
where the equation on the right hand side is used for \labelcref{eq:hatODer}. 

We remark that the construction in \cite{Pawlowski:2005xe} is general and also accommodates operator flows with flowing fields. However, the operator flow with flowing fields for $J_\phi$-dependent operators \labelcref{eq:hatOhatvarphi} is complicated. This general setup will be considered elsewhere and we shall only consider generalised operator flows with $J_\phi$-independent operators \labelcref{eq:hatOhatvarphi} in the following.

In the applications to date, operator flows in terms of the fundamental mean fields $\varphi$ have been computed, see in particular in \cite{Herbst:2015ona,Becker:2018quq, Becker:2019tlf,Houthoff:2020zqy}, but also for machine learning applications of operator flows \cite{Luo:2024wlw}. The general setup with \labelcref{eq:hatOhatvarphi} gives us access to the correlation functions of the fundamental field $\langle \hat\varphi(x_1)\cdots   \hat\varphi(x_n)\rangle$ and to general observables of the fundamental field as well as the effective action $\Gamma_\varphi$ in terms of the 1PI-parts of the correlation functions.

We start with a brief review of this case in \Cref{sec:FlowsforOperators}, which serves as a benchmark and point of reference. As such it is used to illustrate the differences to the operator PIRGs. In \Cref{sec:PIRGsforOperators2} we discuss the general case of operator flows and their physics-informed properties. Finally, we discuss existence criteria for operator PIRGs in \Cref{sec:Constraints+InitialConditions}.

\subsection{Operator flows with fundamental fields}
\label{sec:FlowsforOperators}

We initiate our discussion with a brief review of operator flows for \labelcref{eq:OperatorO} with $\hat\phi=\hat\varphi$. Note that this case readily extends to general composites $\phi$ with $\dot\phi=0$. The respective relations are obtained by simply substituting $\varphi\to\phi$ in the reminder of this Section. For $\hat\phi=\hat\varphi$, the flow of a general operator ${\cal O}[\varphi]$ is given by 
\begin{subequations}
	\label{eq:OpFlow}
\begin{align} 
 \hspace{-0.1cm} 	\partial_t {\cal O}[\varphi] = - \frac12 \left[G_\varphi\left(\partial_t R\right) G_\varphi\right]^{ij}\, {\cal O}_{ji}^{(2)}[\varphi\bigl],
	\label{eq:FlowOp}
\end{align}
with $\partial_t \ln {\cal N}_\varphi=0$. In particular, this includes the choice $ \hat {\cal O}=J_\varphi$, which is explicitly discussed below around \labelcref{eq:O-J}. For the derivation we defer the reader to \cite{Pawlowski:2005xe}. We will also recover it as a special case of the derivation in the PIRG setup, even though only for operators $\hat{\cal O}[\hat\varphi]={\cal O}[\hat\varphi,0]$ without a $J_\phi$-dependence. 

\Cref{eq:FlowOp} entails that the total $t$-derivative of the operator ${\cal O}$ on the left hand side has the representation 
\begin{align} 
\partial_t = 	-\frac12\left[G_\varphi\left(\partial_t R_\varphi\right) G_\varphi\right]^{ij}
	\frac{\delta^2}{\delta\varphi^j \delta\varphi^i }\,.
	\label{eq:dtRepresentation}
\end{align}
\end{subequations}
The operator flow \labelcref{eq:OpFlow} holds true for a large set of composite operators, including the expectation values ${\cal O}_n = \langle \hat \varphi_1 \cdots \hat \varphi_n\rangle$. This set spans the space of operators in the given quantum field theory. We emphasise that the ${\cal O}_n$ are the complete $n$-point functions including their connected and disconnected parts \textit{with respect to $\hat\phi=\hat\varphi$}. The necessity of including the disconnected parts is readily seen at the example of the two-point function 
\begin{align} 
	{\cal O}^{ij}_{2} = G^{ij}_{\varphi} + \varphi^i \varphi^j\,.
\label{eq:O2}
\end{align} 
Inserting \labelcref{eq:O2} into the operator flow \labelcref{eq:OpFlow} leads us to the flow of the propagator which can be also derived from the flow equation of the effective action. 
Note that the disconnected term $ \varphi^i \varphi^j$ drops out on the left hand side of \labelcref{eq:FlowOp} as its flow vanishes. One may be tempted to extend \labelcref{eq:FlowOp} to connected correlation functions, and specifically $G^{ij}_{\varphi}$. 
However, on the right hand side of \labelcref{eq:FlowOp} the disconnected term leads to -$G_\varphi\partial_t R_\varphi G_\varphi$, which is the trivial flow of the regulator insertion. 
Furthermore, the addition of $J_\varphi$-dependences includes the flow of the effective action or rather its derivative with
\begin{subequations} 
	\label{eq:O-J}
\begin{align} 
	{\hat O}^i[\hat \varphi,J_\varphi]=J^i_{\varphi}\,,  
	\label{eq:O-J1}
\end{align}
and
\begin{align} 
{\hat O}_k[\hat\varphi, J_\varphi]=J_\varphi  - R^{ij}_{\varphi} \hat \varphi_j\,,\quad \to \quad {\cal O}[\varphi]=\frac{\delta \Gamma_\varphi[\varphi]}{\delta\varphi}\,.
	\label{eq:O-J2}
\end{align}
\end{subequations} 
Inserting \labelcref{eq:O-J2} into the operator flow \labelcref{eq:FlowOp} leads to the $\varphi$-derivative of the Wetterich equation. 

Note also that the operator flow \labelcref{eq:OpFlow} is at the root of functional optimisation as proposed in \cite{Pawlowski:2005xe, Pawlowski:2015mlf}. It can be understood as an optimisation principle of \labelcref{eq:dtRepresentation}: in a given approximation the total $t$-derivative property of the right hand side is lost and optimal regulators minimise the distance to the total $t$-derivative, in the best case restoring it. 
A first non-trivial computational application to an expectation value of a given composite operator is given by the computation of the Polyakov loop expectation value in Yang-Mills theory in \cite{Herbst:2015ona}. The Polyakov loop is an infinite-order correlation function of the gauge field, or rather a series of correlation functions of the gauge field. The numerical result of its operator flow is in quantitative agreement with the respective data from lattice simulations. This result is an impressive benchmark test for the potential of the operator flows.

We close this brief review with the remark that the derivation of the operator flow \labelcref{eq:FlowOp} can be simplified considerably for operators $\hat{\cal O}[\hat\varphi]$. First of all, this restriction implies $\hat{\cal O}_k = \hat{\cal O}$. Moreover, \labelcref{eq:FlowOp} is readily derived from the flow equation of the effective action, see \cite{Pagani:2016pad} and \cite{Becchi:1996an, Igarashi:2009tj}: We simply couple $\hat {\cal O}[\hat\varphi]$ with a respective current $J^{\ }_{\cal O}$ to the theory and take a $J^{\ }_{\cal O}$-derivative of the flow equation for the effective action. 

\begin{figure*}
	\centering%
	\begin{subfigure}[t]{.45\linewidth}
		\includegraphics[height=0.6\linewidth]{OperatorFlow2}
		\caption{Operator flows: The result at $k=0$ is the desired expectation value $\langle \hat{\cal O}\rangle $ for all $\Gamma_T$ and global non-singular $\dot\phi^{(1)}$. The straight \textit{red} line indicates a flow with a flowing field $\dot\phi^{(1)}$ with $\dot\phi^{(1)}_{k=0}\neq 0$, see \Cref{fig:dotphi1Flow}: its $k$-integral diverges for $k=0$. The dashed \textit{red} line indicates a flow where $\dot\phi^{(1)}$ exhibits a singularity at a local $k$. \hspace*{\fill}}
		\label{fig:TargetOperator}
	\end{subfigure}%
	\hspace{0.1\linewidth}%
	\begin{subfigure}[t]{.45\linewidth}
		\includegraphics[height=0.6\linewidth]{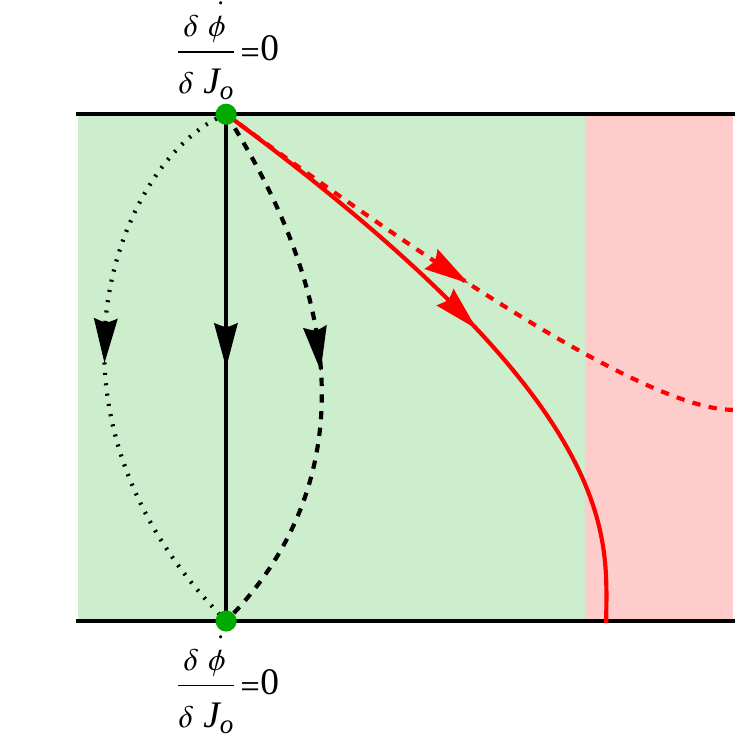}
		\caption{Flowing fields: the $J_{\cal O}$-dependent part $\dot\phi^{(1)}$ of the flowing field defined in \labelcref{eq:dotphiJO}, corresponding to the operator flows depicted in \Cref{fig:TargetOperator}. The flowing field following the straight red line does not vanish at $k=0$, and hence its integral diverges for $k\to 0$. Then, ${\cal O}_{\phi,k=0}\neq \langle \hat{\cal O}\rangle$, see \Cref{fig:TargetOperator}. \hspace*{\fill}}
		\label{fig:dotphi1Flow}
	\end{subfigure}
	\caption{Schematic depiction of operator flows of the derived pairs \textit{operator} and \textit{flowing field} $({\cal O}_\phi, \dot\phi)$ defined in 
		\labelcref{eq:PairO-dotphi1}. Allowed areas are depicted in \textit{green}, whereas prohibited areas are drawn in \textit{red}. The \textit{red} arrow indicates a locally existing solution, whose integrated flow diverges for $k\to 0$, while the \textit{dashed red} arrow indicates a solution which ceases to exist locally at some RG-scale $k$. \hspace*{\fill}}
	\label{fig:SchematicPics2}
\end{figure*}
%

\subsection{Operator flows with flowing fields}
\label{sec:PIRGsforOperators2}

In the following we restrict ourselves to $J_\phi$-independent operators with $ \hat{\cal O}= \hat{\cal O}[\hat\varphi]$, that are also $t$-independent, see \labelcref{eq:ONotDep}. As already mentioned below \labelcref{eq:hatOhatvarphi}, the general case \labelcref{eq:OperatorODef} has been considered in \cite{Pawlowski:2005xe}. Both the $t$- and the $J_\phi$-dependence lead to significantly more involved expressions and potentially higher loop terms. Moreover, the derivations are far more involved. While being interesting in particular for structural investigations as well as functional optimisation, we do not consider it further in the present work and only consider operators $ \hat{\cal O}= \hat{\cal O}[\hat\varphi]$. The general case will be discussed elsewhere.

The restriction leads to the simple relation $\hat{\cal O}_k = \hat{\cal O}$ by inserting $\hat{\cal O}[\hat\varphi[\hat\phi=\delta / \delta J_\phi]]$ in \labelcref{eq:hatODer}. Moreover, we can derive their flows concisely from the generalised flow equation for $\Gamma_\phi[\phi,J_{\cal O}]$ as it is still given by \labelcref{eq:GenFlow}: this is readily seen by augmenting the generating functional \labelcref{eq:GenFunctZphi} with a source term for $\hat{\cal O}$,
\begin{align} 
Z_\phi[J_\phi,J_{\cal O}] \propto \frac{1}{{\cal N}_k} \int d\hat\varphi \, e^{-S_\textrm{cl}[\hat\varphi]-\Delta S[\hat\phi]+\hat\phi_i  J^i_\phi +  J_{\cal O}^j \,\hat{\cal O}_j}.
\label{eq:GenFunctZphiJO}
\end{align}
Note that the (multi-) indices $i$ and $j$ may run over different sets. As discussed around \labelcref{eq:Examplehatphis}, the sum $\hat\phi_i  J^i_\phi$ may include different field species. In the case of the operators, the sum may include one over different operators,  
\begin{align}
J_{\cal O}^j \, \hat{\cal O}_j = \sum_{m=1}^n J_{i_m} {\cal O}^{i_m}\,,
\label{eq:SetofOps}
\end{align} 
with ${\cal O} = ({\cal O}_1\,...,{\cal O}_n)$ and $i=(i_1,...,i_n)$. This will be important in the discussion following \labelcref{eq:GStandardSplit} and in \Cref{sec:O4+6Flows}. For now we restrict ourselves to a single operator species. 
The normalised expectation value \labelcref{eq:OperatorO} of the operator $\hat {\cal O}[\hat\varphi]$ is given by the $J_{\cal O}$-derivative of $\ln Z_\phi$, 
\begin{align} 
 \frac{\delta \ln Z_\phi}{\delta J_{\cal O}}=\langle \hat{\cal O}\rangle \,,
	\label{eq:OExpZ}
\end{align} 
where we have assumed a $J_{\cal O}$-independent $\hat\phi$ and normalisation ${\cal N}_k$. 
The latter corresponds to 
\begin{align} 
	\frac{\delta {\cal N}_k}{\delta J_{\cal O}}=0\,, 
\label{eq:NoJON}
\end{align}
which mirrors the cumulants-preserving constraint in \cite{Ihssen:2024ihp}. 

In our quest for simple or optimised operator flows we want to fully exploit the flexibility of the setup. Hence we shall generalise the setup above and consider $J_{\cal O}$-dependent field operators, 
\begin{align} 
\hat\phi=\hat\phi[\hat\varphi,J_{\cal O}]\,.
\label{eq:hatphivarphiJ} 
\end{align} 
If only considering linear terms in $J_{\cal O}$, this changes the operator that is coupled to the current. Then, the relation \labelcref{eq:OExpZ} turns into 
\begin{align} 
	\frac{\delta \ln Z_\phi}{\delta J_{\cal O}}= \langle \hat{\cal O}\rangle +\left
	\langle \frac{\delta \hat\phi_i }{\delta J_{\cal O}} \left( J_\phi-R_\phi^{ij}\hat\phi_j \right) \right\rangle \,,
	\label{eq:OExpZFull}
\end{align} 
where we have used the $J_{\cal O}$-independence of the normalisation ${\cal N}_k$.

\subsubsection{Generalised operator flow equation}
\label{sec:GenFlowOps}

We proceed by deriving the flow equation for a general operator $\hat {\cal O}$ within this setup. First we remark that the relation
\labelcref{eq:OExpZFull} translates into a similar one for the effective action $\Gamma_\phi[\phi,J_{\cal O}]$. This effective action is obtained as the Legendre transform of $\ln Z_\phi$ with respect to $J_\phi$ while keeping $J_{\cal O}$ as a variable, 
\begin{align} 
	\Gamma_\phi[\phi,J_{\cal O}] = \sup_{J_\phi} \left(\phi_i \, J_\phi^i - \ln Z[J_\phi,J_{\cal O}] \right) - \Delta S[\phi] \,.  
	\label{eq:GammaphiJO}
\end{align}	
Inserting \labelcref{eq:GammaphiJO} into \labelcref{eq:GenFunctZphiJO}, leads to the integro-differential form of the effective action, 
\begin{align} 
\hspace{-.1cm}e^{-\Gamma_\phi } = \frac{1}{{\cal N}_k} \int\! d\hat\varphi \, e^{-S_\textrm{cl}[\hat\varphi]-\Delta S[\hat\phi-\phi]+(\hat\phi-\phi)_i  \frac{\delta\Gamma_\phi}{\delta\phi_i} + J_{\cal O}^j \,\hat{\cal O}_j}, 
	\label{eq:GammaIntegralO}
\end{align}
with $\hat\phi=\hat\phi[\hat\varphi,J_{\cal O}]$ as discussed around \labelcref{eq:hatphivarphiJ}. Then, \labelcref{eq:OExpZFull} turns into 
\begin{align} 
	{\cal O}_\phi\left[\phi,J^{\ }_{\cal O}\right] := - \frac{\delta \Gamma_\phi\left[\phi, J^{\ }_{\cal O}\right]}{\delta J^{\ }_{\cal O}}\,. 
	\label{eq:DefofO}
\end{align}
The subscript $_\phi$ indicates the underlying Legendre transform with respect to $J_\phi$ as well as the additional terms in \labelcref{eq:OExpZFull}, sourced by the possible $J_{\cal O}$-dependence of $\hat\phi$. Importantly, the effective action $\Gamma_\phi[\phi,J_{\cal O}]$ satisfies the generalised flow equation \labelcref{eq:GenFlow}. This follows by taking a $t$-derivative of 	\labelcref{eq:GammaIntegralO} and using the $t$-independence of $\hat{\cal O}$, see \labelcref{eq:ONotDep}. The flow of the correlator ${\cal O}_\phi$ is given by (minus) the $J_{\cal O}$-derivative of the generalised flow as the derivatives with respect to $t$ and $J_{\cal O}$ commute,
\begin{align}
		\partial_t {\cal O}_\phi = - \partial_t \frac{\delta \Gamma_\phi}{\delta J^{\ }_{\cal O}}=  - \frac{\delta }{\delta J^{\ }_{\cal O}}\partial_t \Gamma_\phi\,, 
\label{eq:FlowOpCommute}
\end{align}
and $\partial_t \Gamma_\phi$ is given by \labelcref{eq:GenFlow}. We are led to 
\begin{subequations}
\label{eq:GenFlowOp} 	 
\begin{align}
	\hspace{-.3cm}	\left( \partial_t + \dot{\phi}_i\frac{\delta}{\delta \phi_i} \right) {\cal O}_\phi=- \frac{1}{2} \left[G\left( {\cal D}_t \, R\right) G \right]^{ij} {\cal O}^{(2)}_{\phi,ji} - F^{\ }_{\cal O}, 
		\label{eq:GenFlowOp1} 
\end{align}
with 
\begin{align}
		F^{\ }_{\cal O}=G_{il}\,\frac{\delta^2 \dot\phi_j }{\delta J^{\ }_{\cal O} \delta \phi_l}\, R^{ij}-\frac{\delta \dot{\phi}_i}{\delta J^{\ }_{\cal O}}\, \frac{\delta \Gamma_\phi}{\delta \phi_i}\,. 
		\label{eq:GenFlowOpFphi} 
\end{align}
\end{subequations} 
In \labelcref{eq:GenFlowOp} we dropped the subscript ${}_\phi$ in the propagators and the cutoff function for better readability. We also dropped the arguments $\phi, J_{\cal O}^{\ }$ of the functionals in \labelcref{eq:GenFlowOp}. \Cref{eq:GenFlowOp} constitutes the generalised flow equation for operators, and is a main result of this work. Its schematic visualisation is provided in \Cref{fig:SchematicPics2}, including further explanations. The figure is the analogue to the illustration of the underlying PIRG pair in \Cref{fig:SchematicPics}.

We shall see, that the set of operators that satisfy \labelcref{eq:GenFlowOpFphi} includes all order correlation functions of $\hat\varphi$ (at $k=0)$, and hence spans the complete set of operators in the given QFT. In \Cref{sec:Integrability} we also show, that this set even directly includes the correlation functions ${\cal O}=\langle \hat{\cal O}[\hat\varphi]\rangle$. We close this Section with a few final remarks:

If compared with the operator flow \labelcref{eq:OpFlow} in terms of the fundamental field and effective action $\Gamma_\varphi$, or more generally for composite fields $\phi$ with $\dot\phi=0$, they differ by the term $F_{\cal O}$ in \labelcref{eq:GenFlowOp1}. Additionally, due to the derivation of \labelcref{eq:OpFlow} as a $J_{\cal O}$-derivative of the generalised flow \labelcref{eq:GenFlow}, it also inherits the $\dot \phi$-terms in \labelcref{eq:GenFlow}. We emphasise that $F_{\cal O}\neq 0$ is required for a $J_{\cal O}$-independent field $\hat\phi[\hat\varphi]$. In turn, a $J_{\cal O}$-dependent flowing field, \labelcref{eq:hatphivarphiJ}, is required for $F_{\cal O}= 0$. Consequently, $F_{\cal O}$ can be removed by a suitable choice of $\dot\phi$ or rather its $J_{\cal O}$-derivative. Then, however, ${\cal O}$ cannot be identified with the expectation value of $\hat{\cal O}$ for all cutoff scales $k$ and mean fields $\phi$ or $\varphi$. 

The careful assessment of this choice or rather the implicit one of the respective operator $\hat\phi$ is done in the following two Sections, \Cref{sec:OP-PIRGs} and \Cref{sec:SimpleOP-PIRGs}, in the context of general \textit{operator PIRGs}. 

Finally, we note, that in contradistinction to the standard operator flow discussed in \Cref{sec:FlowsforOperators}, the respective result does not generalise to operators $\hat{\cal O}[\hat \varphi, J_\phi]$. In the latter case there are additional terms, see \cite{Pawlowski:2005xe}. While being specifically interesting for structural insights and for functional optimisation, we defer the discussion of this general case to future work.

\subsubsection{Operator PIRGs}
\label{sec:OP-PIRGs}

With \labelcref{eq:GenFlowOp} we are now fully prepared for extending the PIRG approach to flows of general operators, such as correlation functions of the fundamental field. Specifically we shall use the generality of the flowing field $\dot\phi$, that is that of general choices of the composite operators $\hat\phi$, to unlock the full potential of the generalised operator flows: similarly to the use of PIRGs for the effective action in \cite{Ihssen:2023nqd, Ihssen:2024ihp, Ihssen:2025cff, Bonanno:2025mon} we use them for either 
optimising the computational simplicity or the physics content of the operator flows, as well as gaining structural insights. 

In the spirit of the PIRG approach we consider derived operator pairs  
\begin{subequations} 
\label{eq:PairO-dotphi1}
\begin{align} 
	\Bigl( {\cal O}_\phi[\phi]\,,\, \dot\phi[\phi,J_{\cal O}]\Bigr)_{( \Gamma_\phi,\dot\phi)}\,,
\label{eq:PairO-phi}
\end{align} 
for a given PIRG-pair \labelcref{eq:PairGamm-phi}. Here, $\dot\phi[\phi,J_{\cal O}]$ is defined via 
	\begin{align} 
		\dot\phi[\phi,J_{\cal O}]=\dot\phi[\phi] + J_{\cal O}^n \,\dot\phi^{(1)}_n[\phi,J_{\cal O}]\,,
		\label{eq:dotphiJO}
\end{align} 
\end{subequations} 
where the second term on the right hand side vanishes for $J_{\cal O}=0$. Moreover, for linear perturbations, $\dot\phi^{(1)}$ is nothing but the $J_{\cal O}$-derivative of $\dot\phi$ at vanishing current $J_{\cal O}$. As the full flowing field $\dot\phi$, its $J_{\cal O}$-dependent part \labelcref{eq:dotphiJO} is implicitly defined via the underlying operator $\hat\phi[\hat\varphi,J_{\cal O}]$ with 
\begin{align} 
	\hat\phi_i\left[\hat\varphi, J_{\cal O}\right] = \hat\phi_i[\hat\varphi] + J^j_{\cal O} \,\hat\phi^{(1)}_{i,j}[\hat\varphi,J_{\cal O}] \,.
	\label{eq:hatphiPert} 
\end{align}
In most cases we only consider linear perturbations with $ \hat\phi^{(1)}[\hat\varphi,J_{\cal O}]\to \hat\phi^{(1)}[\hat\varphi,0]$. Then $\hat\phi^{(1)}$ is simply the $J_{\cal O}$-derivative of $\hat\phi$ at vanishing current. Note that for both, linear perturbations \labelcref{eq:hatphiPert} and the general case, the field $\phi$ is $J_{\cal O}$-independent as it enters as the independent variable of the Legendre transform. However, the expectation values of $J_{\cal O}$-derivatives of $\hat\phi$ are in general $J_{\cal O}$-dependent due to operator correlations. 

With the general $\hat\phi$ in \labelcref{eq:hatphiPert}, the effective action satisfies the $J_{\cal O}$-'flow' equation 
\begin{align} 
	\frac{\delta \Gamma_\phi}{\delta J^{n}_{\cal O} } + \phi_{n m}^{\cal O}\,\frac{\delta \Gamma_\phi}{\delta \phi_m } =- \langle \hat {\cal O}_n\rangle 
	+ R_\phi^{ij} G_{il} \frac{\delta \phi^{\cal O}_{n j} }{\delta \phi_l }  \,,
	\label{eq:O-hatOGen}
\end{align}
with 
\begin{align} 
	\phi_{nm}^{\cal O}\left[\phi,J_{\cal O}\right]= \left \langle \frac{\delta \hat\phi_m}{\delta J^{n}_{\cal O}}\right\rangle= \left \langle \hat\phi^{(1)}_{m,n}\right\rangle\,. 
	\label{eq:dJphi} 
\end{align}
The last equality in \labelcref{eq:dJphi} only holds true for linear perturbations \labelcref{eq:hatphiPert}, while in general the right hand side is given by the expectation value of $\hat\phi^{(1)}+ J_{\cal O} \,\delta \hat\phi^{(1)}/\delta J_{\cal O}$. In short, the second terms on both sides of \labelcref{eq:O-hatOGen} are the price to pay for a $J_{\cal O}$-dependence of $\hat\phi$, which is required for the operator PIRG setup. 

We have cast \labelcref{eq:O-hatOGen} already in a form that emphasises its similarity to the generalised flow \labelcref{eq:GenFlow}. Evidently, the implicit definition of $\phi^{\cal O}[\phi,J_{\cal O}]$ mirrors that of $\dot\phi[\phi]$ in \labelcref{eq:dotphi}, and it leads to the form-equivalent terms to those proportional to $\dot\phi[\phi]$ in \labelcref{eq:GenFlow}. Similar to the latter, $\phi^{\cal O}$ can be chosen freely, subject to an underlying operator $\hat\phi^{(1)}$ whose expectation value provides it. For 
$\dot\phi$ this has been discussed comprehensively in \cite{Ihssen:2024ihp, Ihssen:2025cff}, and for $\phi^{\cal O}$ as well as the $J_{\cal O}$-dependent parts of $\dot\phi$ this assessment is done in \Cref{sec:Constraints+InitialConditions}. 

The first term on the right hand side of \labelcref{eq:O-hatOGen} originates in the source term of $\hat{\cal O}$, while the other terms follow similarly to the the $\dot\phi$-terms in the generalised flow \labelcref{eq:GenFlow} derived in \labelcref{eq:GenFlowterms}: the last term in \labelcref{eq:O-hatOGen} originates in the $J_{\cal O}$-derivative of $\hat\phi$ in $\hat \phi \,\delta\Gamma_\phi/\delta\phi$ while the terms proportional to $R_\phi$ originate in 
\begin{align} 
	\left\langle  (\hat \phi-\phi)_i R_\phi^{ij} \frac{\delta \hat\phi_j}{\delta J^n_{\cal O}}\right\rangle = 
	G_{il} \frac{\delta 	\phi_{nj}^{\cal O}}{\delta \phi_l}\,R_\phi^{ij}  \,.
	\label{eq:JOCutoff}
\end{align}
In \labelcref{eq:JOCutoff} we have inserted \labelcref{eq:dJphi}, which leads to the term proportional to $R_\phi$ in \labelcref{eq:O-hatOGen}. 

\Cref{eq:O-hatOGen} and \labelcref{eq:DefofO} entail that the operator ${\cal O}_\phi$ is given by the desired correlation function as well as additional terms, 
\begin{align} 
	{\cal O}_\phi = \langle \hat {\cal O}_n\rangle +  \phi_{n m}^{\cal O}\,\frac{\delta \Gamma_\phi}{\delta \phi_m } 
	- R_\phi^{ij} G_{il} \frac{\delta \phi^{\cal O}_{n j} }{\delta \phi_l }  \,. 
	\label{eq:O-hatOGen+Rest}
\end{align}
\Cref{eq:O-hatOGen+Rest} contains more terms than we bargained for. However, at the physical point the regulator vanishes, $R_\phi=0$. Then, the operator \labelcref{eq:O-hatOGen+Rest} reduces to 
\begin{align}
 	{\cal O}_{\phi,n}\left[\phi,J^{\ }_{\cal O}\right]= \langle \hat {\cal O}_n\rangle\left[\phi,J^{\ }_{\cal O}\right]+\phi_{ni}^{\cal O}\,\frac{\delta \Gamma_\phi\left[\phi,J^{\ }_{\cal O}\right] }{\delta \phi_i } \,. 
	\label{eq:O-hatOGenk0}
\end{align}
The second term on the right hand side of \labelcref{eq:O-hatOGenk0} is proportional to the equation of motion (EoM) of $\phi$ with the solution $\phi_\textrm{\tiny{EoM}}$, 
\begin{align}
	\left. \frac{\delta \Gamma_\phi\left[\phi,J^{\ }_{\cal O}\right] }{\delta\phi}\right|_{\phi=\phi_\textrm{\tiny{EoM}}}=0\,.
	\label{eq:phiEoM}
\end{align}
Hence, we conclude with \labelcref{eq:phiEoM} that 
\begin{align}
	{\cal O}_{\phi,k=0}[\phi_\textrm{\tiny{EoM}},0] =\langle \hat {\cal O}\rangle[\phi_\textrm{\tiny{EoM}},0]\,,
	\label{eq:O-hatOGenk0EoM}
\end{align}
which is the desired expectation value of the composite operator $\hat{\cal O}$. 
We note that \labelcref{eq:O-hatOGenk0EoM} also holds for $J^{\ }_{\cal O}\neq 0$ with the solution $\phi_\textrm{\tiny{EoM}}[J^{\ }_{\cal O}]$. 

This concludes the derivation of the operator PIRGs: we have extended the 
change of perspective, which is at the root of the PIRG approach, to the flow of operators. The generalised operator flow is given by \labelcref{eq:GenFlowOp} for ${\cal O}_\phi$, which is accompanied by the derived PIRG-pair \labelcref{eq:PairO-dotphi1} with the generalised $J_{\cal O}$-flow \labelcref{eq:O-hatOGen} or definition \labelcref{eq:O-hatOGen+Rest} of ${\cal O}_\phi$.

\subsubsection{Simplified operator PIRGs}
\label{sec:SimpleOP-PIRGs}

As a first application of the operator PIRG we consider the choice 
\begin{align} 
	F^{\ }_{\cal O}[\phi,J_{\cal O}] =0\,, 
	\label{eq:FO0}
\end{align}
in the generalised operator flow \labelcref{eq:GenFlowOp} with $F_{\cal O}$ defined in \labelcref{eq:GenFlowOpFphi}. 
This choice was already alluded to at the end of \Cref{sec:GenFlowOps}. 
With the choice \labelcref{eq:FO0} the generalised operator flow reduces to a variant of the operator flow \labelcref{eq:FlowOp} with the effective action $\Gamma_\varphi$ of the fundamental field. 
It is explicitly given below in \labelcref{eq:GenFlowOpSimple} and the only remaining difference to \labelcref{eq:FlowOp} is the occurrence of the $\dot\phi$-terms which mirror the respective ones in the generalised flow of the effective action. 

Before we discuss its explicit form, we want to emphasise some intricacies of the choice \labelcref{eq:FO0}. 
A seemingly trivial way of obtaining \labelcref{eq:FO0} is by simply demanding a $J_{\cal O}$-independent $\dot\phi[\phi]$, 
\begin{align} 
	\frac{\delta \dot\phi}{ \delta J^{\ }_{\cal O} }\equiv 0\,.
	\label{eq:dotphidJO}
\end{align}
Note however, that \labelcref{eq:dotphidJO} necessitates a $J_{\cal O}$-dependent $\hat \phi$. This is readily proven by assuming $\hat\phi$ is $J_{\cal O}$-independent: Then, $\phi^{\cal O}=0$ with \labelcref{eq:dJphi} and we are led to  
\begin{align} 
	\frac{\delta \dot\phi_m }{\delta J^n_{\cal O}} = \langle {\hat O}_n \,\partial_t \hat\phi_m \rangle - {\cal O}_n\, \dot\phi_m-  G_{\phi,ij}\frac{\delta \dot\phi_m}{\delta \phi_j}\frac{\delta {\cal O}_n}{\delta  \phi_i}\,,  
	\label{eq:NonTrivial}
\end{align}   
where ${\cal O}= \langle \hat{\cal O}\rangle $ is ${\cal O}_\phi$ for $J_{\cal O}$-independent $\hat\phi$. In general, the right hand side of \labelcref{eq:NonTrivial} is non-vanishing and we conclude that \labelcref{eq:dotphidJO} requires a $J_{\cal O}$-dependent $\hat\phi$ and the $J_{\cal O}$-derivative of the effective action is governed by \labelcref{eq:O-hatOGen}. This does not invalidate 	\labelcref{eq:O-hatOGenk0EoM} and the simplified operator flow provides us upon integration with $\langle \hat{\cal O} \rangle$. Only the initial condition at a large cutoff scale may change. 

Such an implicit choice of the operator $\hat \phi$ comes with constraints, and not all $\dot\phi[\phi,J_{\cal O}]$ are allowed. These constraints have been comprehensively discussed for the flowing fields $\dot\phi[\phi]$ in \cite{Ihssen:2024ihp, Ihssen:2025cff}, and we shall extend the discussion there to $\dot\phi[\phi,J_{\cal O}]$ in \Cref{sec:Constraints+InitialConditions}. 

For the time being, we simply assume that \labelcref{eq:FO0} is a non-trivial but allowed choice of $\dot\phi[\phi,J_{\cal O}]$. Inserting it into the operator flow \labelcref{eq:GenFlowOp}, we arrive at 
\begin{subequations} 
	\label{eq:PIRGOPSimple} 
\begin{align}
		\left( \partial_t + \dot{\phi}_i\frac{\delta}{\delta \phi_i} \right) {\cal O}_\phi
	=- \frac{1}{2} \left[G\left( {\cal D}_t \, R\right) G \right]^{ij} {\cal O}^{(2)}_{\phi,ji}\,, 
	\label{eq:GenFlowOpSimple} 
\end{align}
analogously to the operator flow \labelcref{eq:FlowOp} for $\dot\phi=0$. As for \labelcref{eq:FlowOp}, the operator flow \labelcref{eq:GenFlowOpSimple} implies an operator identity, 
\begin{align} 
	\left( \partial_t + \dot{\phi}_i\frac{\delta}{\delta \phi_i} \right) = - \frac{1}{2} \left[G\left( {\cal D}_t R\right)\, G \right]_{ij} \frac{\delta^2}{\delta \phi_j\delta\phi_i}\,,
	\label{eq:DtRepresentation}
\end{align}
\end{subequations} 
on the set of operators $\{{\cal O}_\phi[\phi]\}$ with \labelcref{eq:DefofO}. At $k=0$, this set of operators includes all order correlation functions of $\hat\varphi$ and spans the complete set of operators in the given QFT. Note also that the right hand side of \labelcref{eq:DtRepresentation} may be interpreted as a total $t$-derivative in a given coordinate system $\phi$ in configuration space with $(\partial_t  \phi)[\phi]=\dot \phi[\phi]$. This interpretation will be discussed in more detail in \Cref{sec:ReoncstructGvarphi}. In any case, with and without this interpretation, \labelcref{eq:DtRepresentation} is at the root of functional optimisation for PIRGs. It can be used to maximise the physics content of an operator flow within a given approximation, while also providing a bridge to machine learning applications of the PIRG setup. This will be discussed in more detail elsewhere.

We close this Section with a few comments: 
\Cref{eq:GenFlowOpSimple} is implicit in the general operator flows in \cite{Pawlowski:2005xe} and also can be found in \cite{Baldazzi:2021ydj}. In both, the discussion of the key constraint \labelcref{eq:FO0} is missing. In \cite{Pawlowski:2005xe}, the analysis does not go beyond the derivation of the equations themselves. In \cite{Baldazzi:2021ydj}, \labelcref{eq:FO0} or rather \labelcref{eq:dotphidJO} is assumed implicitly, together with ${\cal O} = \langle \hat{\cal O}\rangle$. \Cref{eq:dotphidJO} requires a vanishing right hand side of \labelcref{eq:NonTrivial}, which, however, leads to ${\cal O}\neq \langle \hat{\cal O}\rangle$ for $k\neq 0$ and $\phi\neq \phi_\textrm{\tiny{EoM}}$. While these intricacies have no direct consequences for practical implementation of the flows, they are of key importance for the existence of the respective flowing fields as well as for the initial conditions. Finally, they are relevant for the optimisation of the flows, and specifically for the split of the operators ${\cal O}$ in connected and disconnected parts: an optimal split is chiefly important for the rapid convergence of the results within a given approximation. This discussion will be picked up in \Cref{sec:Applications} and in \cite{Ihssen:2026njd}. 

We also repeat that both operator flows \labelcref{eq:GenFlowOp} and \labelcref{eq:PIRGOPSimple} do not hold true for the general operators $\hat{\cal O}[\hat \varphi,J_\phi]$. This is readily seen by using $\hat{\cal O}[\hat \varphi,J_\phi]=J_\phi$ and comparing the flow of $\Gamma_\phi^{(1)}$ following from \labelcref{eq:GenFlow} and that by simply assuming \labelcref{eq:GenFlowOp} and \labelcref{eq:GenFlowOpSimple} being valid for ${\cal O} = \Gamma_\phi^{(1)}$. The case with operators $\hat{\cal O}[\hat \varphi,J_\phi]$ is mostly interesting for conceptual applications and will also be considered elsewhere. 
 
This concludes the derivation of the PIRG for operators: The general case is given by \labelcref{eq:GenFlowOp} with the reduction \labelcref{eq:PIRGOPSimple} that is form-equivalent to the operator flow with the fundamental field \labelcref{eq:GenFlowOp}. This choice is obtained with \labelcref{eq:FO0}.

\subsection{Constraints, existence and initial conditions}
\label{sec:Constraints+InitialConditions}

We have already mentioned below \labelcref{eq:NonTrivial}, that the choice of $\dot\phi[\phi,J_{\cal O}]$ comes with constraints. In the present Section we analyse the constraints on the set 
\begin{align} 
	\left\{ \dot\phi[\phi,J_{\cal O}]\right\}\,, 
	\label{eq:setdophiJ}
\end{align}
in the derived PIRG-pair \labelcref{eq:PairO-dotphi1}. In the following we restrict our analysis to linear perturbations in \labelcref{eq:dotphiJO}, 
\begin{align} 
	 \dot\phi[\phi,J_{\cal O}]=\dot\phi[\phi] + J_{\cal O}^n \,\dot\phi^{(1)}_n[\phi]\,,
\label{eq:LinearPertdotphi}
\end{align} 
for the sake of simplicity. 

Now we extend the structural analysis of the implicit field $\dot\phi[\phi]$ in \cite{Ihssen:2024miv, Ihssen:2025cff}, see \Cref{sec:Flowingphi+TargetAction}, to $\dot\phi^{(1)}[\phi,J_{\cal O}]$. The analysis is also connected to assessing the nature of the difference between ${\cal O}=\langle \hat{\cal O}\rangle $ and ${\cal O}_\phi$ defined in \labelcref{eq:DefofO}. The respective constraints limit the choices of $F_{\cal O}$ in the generalised operator flow \labelcref{eq:GenFlowOp}. Evidently, this assessment is of crucial importance for the operator PIRG setup, beyond being relevant for the existence of specific choices such as \labelcref{eq:FO0} used for the derivation of the simplified operator flow \labelcref{eq:GenFlowOpSimple} in \Cref{sec:SimpleOP-PIRGs}.

\subsubsection{Local and global existence} 
\label{sec:CorrelationOrder} 

In the following we discuss constraints on the first-order contribution $\dot\phi^{(1)}$ in \labelcref{eq:LinearPertdotphi}. The restrictions mirror those on $\dot\phi$, which were put forward in \cite{Ihssen:2024ihp, Ihssen:2025cff}. We have recalled them below \labelcref{eq:dotphi} in \Cref{sec:Flowingphi+TargetAction}. The additional constraints are labelled accordingly, 
\begin{itemize} 
	\item[(i)] \textit{Local} existence of $\dot \phi^{(1)}_k[\phi]$ in \labelcref{eq:LinearPertdotphi}: One has to show that there is an operator $\partial_t \hat\phi^{(1)}_k[\hat\phi,J_{\cal O}]$ leading to $\dot \phi^{(1)}_k[\phi]$. This inverse problem has to be solved minimally for $J_{\cal O}=0$. 
	\item[(ii)] \textit{Global} existence of $\phi^{(1)}_k[\varphi,J_{\cal O}]$: Its flow is integrated from $k=\Lambda$ to $k=0$. The result has to be free of divergences.  
\end{itemize} 
The analysis of the constraints \textit{(i,ii)} has been done in \cite{Ihssen:2024ihp, Ihssen:2025cff}. Still, due to their importance for the operator flows, we briefly review them: \\[-2ex] 

The first one, \textit{(i)}, constrains flowing fields $\dot\phi$ to those that can be generated by an underlying operator $\hat\phi$ with $\langle \partial_t \hat\phi_k[\hat\varphi]\rangle$ at a given cutoff scale $k$. We consider this constraint as a relatively weak one. It accommodates the set of \textit{local} (in field and momentum space) functionals as discussed in \cite{Ihssen:2024ihp}. 

The second requirement \textit{(ii)} ensures the global existence. Put differently, it ensures the existence of a non-singular operator 
\begin{align} 
\hat\phi_k^{(1)}[\hat \varphi]\,,\qquad k\in[0,\Lambda]\,,
\label{eq:GlobalTrafo} 
\end{align} 
with the initial cutoff scale $\Lambda$. Not every choice $\dot\phi^{(1)}[\phi]$ can be integrated over all cutoff scales without experiencing a singularity, and in the presence of the latter the existence of \labelcref{eq:GlobalTrafo} is at stake. This is depicted in \Cref{fig:SchematicPics2}. \\[-2ex]

We close this discussion by proving the presence of constraints following from \textit{(i,ii)} 
within a simple counter example: Assume that the requirement \textit{(ii)} is always satisfied. Then, $\dot\phi[\phi,J_{\cal O}]$ and hence $F_{\cal O}$ is \textit{globally} unconstrained. Consequently, we could use it to cancel the whole right hand side of \labelcref{eq:GenFlowOp1} for all cutoff scales $k$. This renders the operator flow of ${\cal O}_\phi$ trivial for all $k$ and the operator does not collect any quantum fluctuations. This is obviously incorrect and invalidates the \textit{global} existence of such a transformation.

\subsubsection{Integrability constraint} 
\label{sec:Integrability} 

While the global requirement \textit{(ii)} is far more restrictive than the local one, \textit{(i)}, the latter is still very helpful. A minimal consistency constraint that follows from \textit{(i)} is the integrability of the transformation, 
\begin{align} 
	\left[ \partial_t \,,\, \frac{\delta}{\delta J_{\cal O}}\right] \Gamma_\phi[\phi,J_{\cal O}]=0\,.
	\label{eq:Integrability}	
\end{align}
The use of \labelcref{eq:Integrability}	and its violation is twofold. To begin with, approximations will lead to a violation of \labelcref{eq:Integrability} and restoring integrability or rather the total derivative nature of the flow operator \labelcref{eq:DtRepresentation} for flows of the effective action is one of the cornerstones of functional optimisation. The second use is related to the divergences in the integrated flow: they may also leave their signatures in violations of the integrability condition \labelcref{eq:Integrability}. 

The flow \labelcref{eq:GenFlowOp} provides us with $\frac{\delta}{\delta J_{\cal O}} \partial_t \Gamma_\phi$, and the $t$-derivative of \labelcref{eq:O-hatOGen} provides us with $\partial_t \frac{\delta}{\delta J_{\cal O}} \Gamma_\phi$. The latter representation of the operator flow reads 
\begin{align} 
	\partial_t \frac{\delta \Gamma_\phi}{\delta J^{n}_{\cal O} } + \partial_t\! \left[\phi_{n m}^{\cal O}\,\frac{\delta \Gamma_\phi}{\delta \phi_m }\right] =-\partial_t \langle \hat {\cal O}_n\rangle 
	+ \partial_t\! \left[R_\phi^{ij} G_{il} \frac{\delta \phi^{\cal O}_{n j} }{\delta \phi_l }\right]\!. 
	\label{eq:dtO-hatOGen}
\end{align}
With \labelcref{eq:GenFlowOp} and \labelcref{eq:dtO-hatOGen}, the integrability condition \labelcref{eq:Integrability} reads 
\begin{align} \nonumber 
	& \hspace{-1cm}	\partial_t \langle \hat {\cal O}_n\rangle +\partial_t \left[\phi_{n m}^{\cal O}\,\frac{\delta \Gamma_\phi}{\delta \phi_m }\right] 
	- \partial_t \left[R_\phi^{ij} G_{il} \frac{\delta \phi^{\cal O}_{n j} }{\delta \phi_l }\right]\\[1ex] \nonumber 
	= & - \frac{1}{2} \left[G\left( {\cal D}_t \, R\right) G \right]^{ij} {\cal O}^{(2)}_{\phi,ji}\\[1ex]
	& -G_{il}\,\frac{\delta^2 \dot\phi_j }{\delta J^{\ }_{\cal O} \delta \phi^l}\, R^{ij}+\frac{\delta}{\delta J^{\ }_{\cal O} } \left[ \dot{\phi}_i \frac{\delta \Gamma_\phi}{\delta \phi_i}\right]\,.    
	\label{eq:Integrability2} 
\end{align} 
Before we dissect \labelcref{eq:Integrability2}, we remark that it constitutes yet another flow for the composite operator as $\partial_t \langle \hat{\cal O}\rangle $ survives in the integrability condition. Loosely speaking, the two flows provide us with the different sides of the flow of $\langle \hat{\cal O}\rangle$ in the simplified flow equation \labelcref{eq:GenFlowOpSimple}: 
\Cref{eq:dtO-hatOGen} provides us with $\partial_t \langle \hat{\cal O}\rangle$ on the left hand side and \labelcref{eq:GenFlowOp} provides us with the right hand side.  

For the following detailed analysis we only consider $\phi^{\cal O}$ in 	\labelcref{eq:dJphi} that are perturbations of $\dot\phi[\phi]$. With this constraint we have ensured their \textit{local} existence, if $\dot\phi$ exists locally. Let us first discuss two leading order terms: $\partial_t \langle \hat{\cal O}\rangle $ and the standard operator flow term proportional to ${\cal O}_\phi^{(2)}$. We have already argued that $\dot\phi^{(1)}$ cannot be used to eliminate one of them, see the discussion below \labelcref{eq:GlobalTrafo}. However, we may eliminate their difference. 

We proceed with the two terms proportional to the EoM, the first $\phi$-derivative of $\Gamma_\phi$. We find 
\begin{align}\nonumber  
 &\, \partial_t \left[\phi_{n m}^{\cal O}\,\frac{\delta \Gamma_\phi}{\delta \phi_m }\right] - 
	\frac{\delta}{\delta J^{\ }_{\cal O} } \left[ \dot{\phi}_i \frac{\delta \Gamma_\phi}{\delta \phi_i}\right]\\[1ex] 
	= &\, - \dot{\phi}_i \frac{\delta^2 \Gamma_\phi}{\delta J^{\ }_{\cal O} \delta \phi_i}	+ \left[\partial_t\phi_{n m}^{\cal O}- \frac{\delta \dot{\phi}_i }{\delta J^{\ }_{\cal O} } \right] \frac{\delta \Gamma_\phi}{\delta \phi_m } + \phi_{n m}^{\cal O}\,\frac{\delta \partial_t \Gamma_\phi}{\delta \phi_m }  	\,.
	\label{eq:Reordering1} 
\end{align}
The first term in the second line is potentially a first order correlation order or even tree-level, if basis rotations are considered. In turn, the second and third terms are second order correlation terms: The square bracket vanishes if there are no correlations between $\hat\phi$ and $\hat{\cal O}$ and both terms are already correlation terms. The second term is a product of first order correlations and hence is at least of second order. We conclude that only the first term on the right hand side of the equation is a potentially leading order correlation term. 

Now we apply this correlation order counting to the remaining terms in \labelcref{eq:Integrability2}. We readily conclude that they are at least of second order: they are loop terms and hence of at least first order. They are proportional to $\phi^{\cal O}$ or the $J_{\cal O}$-derivative of $\dot \phi$ and both are first order correlations. There are even additional cancellations between the terms that increase the correlation order of some parts. We do not study these cancellations explicitly as the terms are subleading. 

In summary, only $\partial_t \langle \hat{\cal O}\rangle $, the standard operator flow term and the $\dot\phi\,\frac{ \delta\Gamma_\phi}{\delta\phi} $-term in \labelcref{eq:Reordering1} are of first order. The first consequence is that there is no first correlation order choice $\hat\phi^{(1)}$ that removes either of them from \labelcref{eq:Integrability2}, but of course we can remove the sum of them. More importantly, the other terms are of the same (second) order. This mirrors the similar statement about $F_{\cal O}$ in \Cref{sec:CorrelationOrder}. We may simply choose $\hat\phi^{(1)}$ such that they cancel each other as well as the difference between $\langle \hat {\cal O} \rangle$ and (minus) the $J_{\cal O}$-derivative of $\Gamma_\phi$ in \labelcref{eq:O-hatOGen}. This leads us to yet another simple form of the operator flow, 
\begin{align}
	\left( \partial_t + \dot{\phi}_i\frac{\delta}{\delta \phi_i} \right) \langle \hat {\cal O}\rangle 
	=- \frac{1}{2} \left[G\left( {\cal D}_t \, R\right) G \right]^{ij} \langle \hat {\cal O}\rangle ^{(2)}_{ji}\,. 
	\label{eq:GenFlowOpSimplehatO} 
\end{align}
\Cref{eq:GenFlowOpSimplehatO} is form-equivalent to \labelcref{eq:GenFlowOpSimple}, but applies directly to ${\cal O}$ instead of ${\cal O}_\phi$. While this is seemingly surprising, it simply entails that with appropriate choices of $\hat\phi^{(1)}$, \labelcref{eq:GenFlowOpSimple} holds true for all composite operators that only differ by sub-leading correlation orders. This concludes our analysis.

\subsubsection{Non-triviality of trivialising flowing fields $\dot\phi^{(1)}$} 
\label{sec:TrivialisingFlows} 

We close this Section with an analysis of operator flows with trivialising flowing fields $\dot\phi^{(1)}$: they are defined by a vanishing right hand side of the generalised operator flow \labelcref{eq:OpFlow}, leading to $\partial_t {\cal O}_\phi=0$: the target operator is the one at $k=\Lambda$. This may be seen as the analogue of a classical target action. There, the flowing field accommodates all quantum physics and the flow was preserving all information. It is one of the target actions for which the computational task is significantly reduced: The task of solving a rather complicated PDE is reduced to solving a linear ODE. It is therefore tempting to extend this example to the operator PIRGs, hoping for the same significant simplification. As we are finally interested in the solution on the EoM, we restrict this analysis to $\phi=\phi_{\textrm{\tiny{EoM}}}$. Then, the second term in $F_{\cal O}$ in \labelcref{eq:GenFlowOpFphi} is missing. For linear perturbations \labelcref{eq:LinearPertdotphi} we are led to 
\begin{align} 
	\frac{\delta^2 \dot\phi_j }{\delta J^{\ }_{\cal O} \delta \phi_l}= \frac{\delta \dot\phi^{(1)}_j }{\delta \phi_l}	= {\cal O}_{\phi,lm}^{(2)} \left[ G_\phi\left( {\cal D}_t R_\phi\right) \frac{1}{R_\phi} \right]^m_{\hspace{.3cm}j} \,.
	\label{eq:CancelAll}
\end{align} 
A seeming problem of \labelcref{eq:CancelAll} is the presence of $1/R_\phi$ but the product $\left( {\cal D}_t R_\phi\right) \frac{1}{R_\phi}$ is not problematic for a large set of regulators $R_\phi$. The right hand side is $\dot \phi^{(1)}$-independent and hence $\dot \phi^{(1)}$ is simply given by its $\phi$-integral. This ensures the \textit{local} existence of $\dot \phi^{(1)}$. 

For the present purpose of a structural discussion we simplify this case even further by using $\dot\phi[\phi,0]\equiv 0$ for all $k$, i.e.~we are considering the Wetterich flow for the effective action. Then, $\phi=\varphi$ and with $\dot\phi[\phi,0]\equiv 0$ we fall back to the operator flow \labelcref{eq:FlowOp} in terms of the fundamental field. However, for $\phi=\varphi$ and \labelcref{eq:CancelAll} we are left with
\begin{align} 
	\partial_t {\cal O}_\phi[\varphi_{\textrm{\tiny{EoM}}}] \equiv 0\,, \qquad k\in [0,\Lambda]\,, 
	\label{eq:ZeroFlow}
\end{align} 
with the initial cutoff $\Lambda$ and ${\cal O}_\phi[\varphi_{\textrm{\tiny{EoM}}}]$ follows from \labelcref{eq:O-hatOGen}, 
\begin{align} 
	{\cal O}_\phi[\varphi_{\textrm{\tiny{EoM}}}] =	 \langle \hat {\cal O}_n\rangle 
	- R_\varphi^{ij} G_{\varphi,il} \frac{\delta \phi^{\cal O}_{n j} }{\delta \varphi_l }  \,, 
	\label{eq:Ophi0}
\end{align} 
with $\phi^{\cal O}$ given in \labelcref{eq:dJphi}. With \labelcref{eq:ZeroFlow} we conclude 
\begin{align} 
	R_\varphi^{ij} G_{\varphi,il} \frac{\delta \phi^{\cal O}_{n j} }{\delta \varphi_l } = \langle \hat {\cal O}_n\rangle  - ({\cal O}_{\phi,\Lambda})_n\,.  
	\label{eq:SingChoice}
\end{align} 
For $R_\varphi\to 0$ the left hand side vanishes for regular $\phi^{\cal O}$, thus we conclude that $\phi^{\cal O}$ develops a singularity for $k\to 0$ if we attempt to implement $\partial_t {\cal O}_\phi = 0$. Still, we may use \labelcref{eq:SingChoice} in the limit $k\to 0$ for computing $\langle \hat {\cal O}\rangle$ at $k=0$. This requires $\phi^{\cal O}$ or rather its flow: a $t$-derivative of \labelcref{eq:SingChoice} leads us to 
\begin{align} \nonumber 
\partial_t \langle \hat {\cal O}_n\rangle   = &\,\partial_t \left[R_\varphi^{ij} G_{\varphi,il} \frac{\delta \phi^{\cal O}_{n j} }{\delta \varphi_l }\right]\\[1ex]
= &\,\left[ \partial_t \left(R_\varphi^{ij} G_{\varphi,il}\right) \right]  \, \frac{\delta \phi^{\cal O}_{n j} }{\delta \varphi_l }+ R_\varphi^{ij} G_{\varphi,il} \frac{\delta \partial_t \phi^{\cal O}_{n j} }{\delta \varphi_l }\,.
\label{eq:Trivialisingflow} 
\end{align}
\Cref{eq:Trivialisingflow} is nothing but \labelcref{eq:dtO-hatOGen}, evaluated on-shell. This follows readily with the definition of ${\cal O}_\phi$ in \labelcref{eq:DefofO} and \labelcref{eq:ZeroFlow}. Consequently the trivialising choice of $\dot \phi^{(1)}$ satisfies the integrability condition on-shell and the flow is locally well-defined. 

\begin{figure}
	\includegraphics[width=\linewidth]{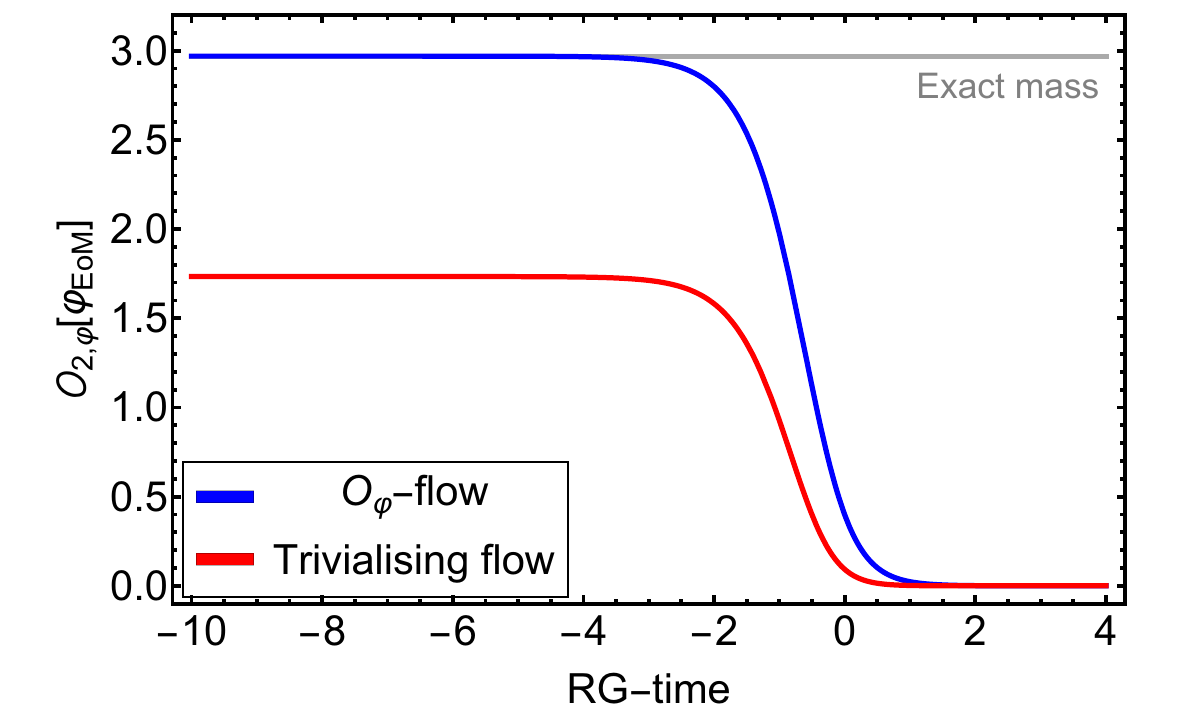}
	\caption{RG-time dependence of the two-point function on the equations of motion: We compare the standard operator flow \labelcref{eq:FlowOp} \textit{(blue)} with the trivialising flow \labelcref{eq:Trivialisingflow} using the assumption $\partial_t \phi^{\cal O} = \dot\phi^{(1)} $ \textit{(red)}. The plot illustrates that the assumption does not hold true and we have \labelcref{eq:dtphiO}. The results are obtained from a zero-dimensional computation of the Wetterich flows with the classical action \labelcref{eq:Scl} with $\Delta{\cal I}\neq 0$.   \hspace*{\fill}}
	\label{fig:MinimalEx}
\end{figure}

The only unknown in \labelcref{eq:Trivialisingflow} is $\phi^{\cal O}[\phi]$. It does not follow directly from $\dot\phi^{(1)}$, which is determined by \labelcref{eq:CancelAll}. It is tempting to use the integrated flow of the latter for computing $\langle \hat {\cal O}\rangle$ as it is (minus) the integrated flow of the standard flow term proportional to ${\cal O}_\phi$. However, for trivialising flows, the latter is proportional to ${\cal O}_{\phi,\Lambda}$ and cannot yield the full result. This is proven by considering a perturbative expansion of the integrated flow: the respective diagrams always carry at least one propagator at $k=\Lambda$. A minimal and illustrative example is given by the two-point function, $\hat{\cal O}=\hat\varphi(x_1)\hat\varphi(x_2)$, which we have also solved numerically for $d=0$, see \Cref{fig:MinimalEx}.   

In short, the flow of $\langle \hat {\cal O}\rangle$ is now partially buried in the standard flow term and partially in $F_{\cal O}$. The above structure begs the question whether we can infer $\partial_t \phi^{\cal O}$ from $\dot\phi^{(1)}$. Both contain $\langle \partial_t \hat\phi^{(1)}\rangle$ and differ by further terms triggered by the $J_{\cal O}-$ and $t$-dependences of the cutoff and source terms. With \labelcref{eq:NonTrivial} and similar relations the difference between $\partial_t \phi^{\cal O}$ from $\dot\phi^{(1)}$ can be provided in a closed form. Structurally it reads 
\begin{align} 
	\partial_t \phi^{\cal O} = \dot\phi^{(1)}+ \Delta {\cal I}\bigl[ \phi^{\cal O} , \dot\phi^{(1)}, {\cal O}_\phi\bigr]\,.  
	\label{eq:dtphiO}
\end{align}
While this relation can be recast in terms of a recursive relation for 
$\partial_t \phi^{\cal O}$, the latter takes a rather complicated form. 
We refrain from discussing it further here, as such an application is safely beyond the scope of the present work. 
Moreover, at present the seeming initial promise of a further structural simplification does not hold. 
In view of this non-triviality we shall stick to the already impressively simple operator flows \labelcref{eq:GenFlowOpSimple} for general applications. 
Note however, that the simplicity of the operator flow is not the only criterion: subject to the theory and operators at hand, other choices of $\dot\phi$ in the generalised operator flow \labelcref{eq:OpFlow} may be better suited.

\section{Applications} 
\label{sec:Applications}

In this Section we illustrate the potential of the operator PIRGs at the example of correlation functions ${\cal O}_{n}$ of the fundamental field with the operator  
\begin{align} 
	\hat {\cal O}_{n,i_1\cdots i_n} = \hat\varphi_{i_1}\cdots \hat\varphi_{i_n}\,,  
\label{eq:On}
\end{align} 
as well as $\dot\phi[\phi]$. While the derivations are general and the results hold true for a general QFT, we will exemplify them with a scalar $\phi^4$-theory in $d$ dimensions with the classical action 
\begin{align}
S_\textrm{cl}[\varphi] = \int \textrm{d}^d x\, \left\{ \frac12 \left(\partial_\mu \varphi\right)^2 +\frac12 m_\varphi^2 \varphi^2 + \frac{\lambda_{\varphi}}{8} \varphi^4\right\}\,.
\label{eq:Scl} 
\end{align}
The respective composites ${\cal O}_{n}$ are given by \labelcref{eq:O-hatOGen} and reduce to the $n$-point functions at $k=0$ and on the EoM, see \labelcref{eq:O-hatOGenk0EoM}. 

An analytic benchmark case is provided by $d=0$, where the theory reduces to a simple one-dimensional integral, 
\begin{subequations} 
\label{eq:Zd0+GTScl}
\begin{align} 
	Z_\phi(J_\phi)= \frac{1}{{\cal N}_\phi} \int\limits_{-\infty}^\infty d\hat\varphi\,e^{-S_\textrm{cl}(\hat\varphi)+ J_\phi \hat\phi(\hat\varphi)}\,, 
\label{eq:Zvarphid0}
\end{align} 
where all functionals reduce to functions, indicated by the parentheses, e.g.~$Z_\phi[J_\phi]\to Z_\phi(J_\phi)$. In this example we use the classical target action, 
\begin{align}
	\Gamma_T(\phi)=S_\textrm{cl}(\phi) + \mathcal{C}_k \,, 
\label{eq:cTA}
\end{align} 
where $\mathcal{C}_k$ is a constant. The PIRG flow for the flowing composite field $\dot\phi$ reads with $R_\phi=k^2$, 
\begin{align}
		\dot{\phi} \left( m_\varphi^2 \phi+ \frac{\lambda_\varphi}{2}\phi^3\right) = \frac{1 +\dot\phi^\prime}{1+\frac{ m_\varphi^2 + \frac{3\lambda_\varphi}{2} \phi^2}{k^2}} +\partial_t {\cal \ln N}_\phi - \partial_t{ \mathcal{C}}_k \,. 
\label{eq:PIRGflowGTScl} 	
\end{align} 
\end{subequations}	
For more details see \Cref{app:cumuFlow}.
\Cref{eq:PIRGflowGTScl} is a simple linear ordinary differential equation for $\dot\phi[\phi]$ and qualitatively simpler than the non-linear, non-algebraic convection-diffusion type equation for the effective action $\Gamma_\varphi$ of the fundamental field. This example was already used in \cite{Ihssen:2024ihp} in the context of the reconstruction of correlation functions of the fundamental field. 
We emphasise that this example is sufficiently complex to illustrate the structure and potential of the approach. Needless to say, a Euclidean quantum field theory is qualitatively different from a one-dimensional integral, but any finite approximation carries the same structural properties. Here, finite approximations encompass lattice formulations on a finite lattice or approximations to functional flows with a finite number of operators for the effective action. In the latter case this includes fully momentum-dependent dressings for these operators. We hasten to add that this statement does not concern the computational complexity, which is qualitatively higher in a quantum field theory, but only the structure. In short, the zero-dimensional example can be used to illustrate the implementation of the approach and also provide a computational proof of the relations. 

We shall explicitly derive the expressions of the flow of ${\cal O}_1, {\cal O}_2$ in general theories in \Cref{sec:O1Flows} and \Cref{sec:O2Flows} respectively. There we also provide solutions in the analytic zero-dimensional example theory with the classical action \labelcref{eq:Scl}. The structure of the flows for ${\cal O}_{1,2}$ also allows us to discuss the general structure and relations for the flows of general ${\cal O}_n$: the flow of an $n$-point function is iterative, in the sense that it only requires the knowledge of the lower order ones with $m<n$. This is exemplified with the four- and six-point functions in \Cref{sec:O4+6Flows} and we show the full reconstruction of generating functionals in terms of $\varphi$ in \Cref{sec:ReoncstructGvarphi}.

\subsection{Flow of the fundamental mean field $\varphi_k[\phi]$} 
\label{sec:O1Flows}

\begin{figure}[t]%
	\centering%
	\includegraphics[width=\linewidth]{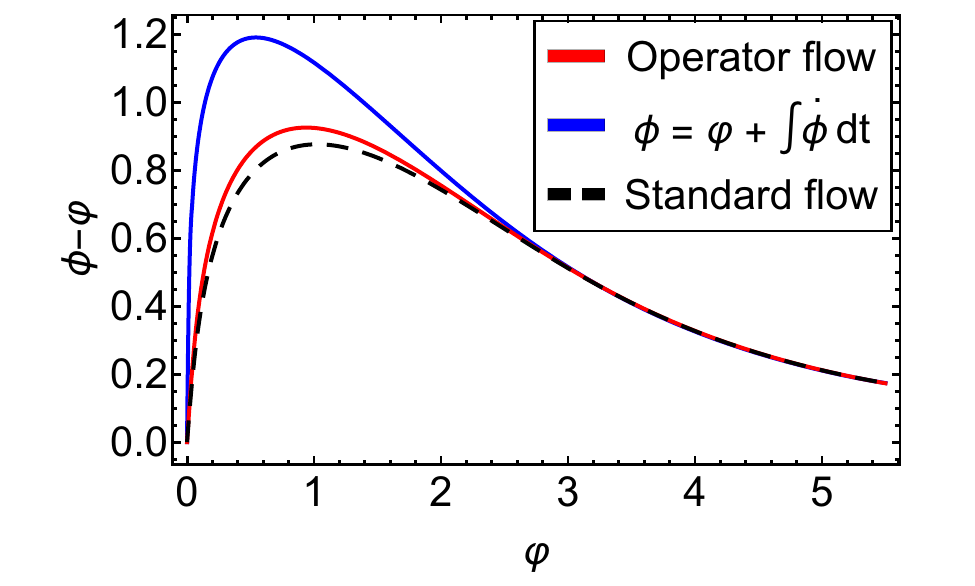}%
	\caption{Different definitions for the map of the field transformation, discussed in \Cref{sec:ReoncstructGvarphi}: Reconstruction by direct integration \labelcref{eq:phidotphi} \textit{(blue)}, as derived from the operator flow 	\labelcref{eq:GenFlowOpvarphi} \textit{(red)} and the aimed-for map which produces the 1PI effective action in terms of $\varphi$ \labelcref{eq:GvarphiGphi} \textit{(black, dashed)}. 
		\hspace*{\fill}}%
	\label{fig:onepoint_function}
\end{figure}

We initiate the discussion with the computation of the expectation value 
\begin{align} 
	{\cal O}_{1,i}[\phi]=\langle \hat\varphi_i\rangle[\phi]\,, \qquad J_{{\cal O}_1} = J_\varphi\,.
\end{align}
The respective current is that of the fundamental field of the theory, and this flow is only non-trivial if $\hat\phi \neq\hat\varphi$. In the latter case we can generate $n$-point functions by $n$-derivatives with respect to $J_{\varphi}$. The total derivative operator flow \labelcref{eq:PIRGOPSimple} reads 
\begin{align}
	\partial_t \varphi_k[\phi]+ \dot{\phi}_i\frac{\delta \varphi_k[\phi]}{\delta \phi_i} 
	=- \frac{1}{2} \left[G_\phi\left( {\cal D}_t \, R_\phi\right) G_\phi \right]^{ij} \varphi^{(2)}_{k,ji}\,, 
	\label{eq:GenFlowOpvarphi} 
\end{align}
with 
\begin{align} 
	\varphi^{(2)}_{k,ji}[\phi] =\frac{\delta^2 \varphi_k[\phi]}{\delta \phi_j\delta \phi_i}\,.
	\end{align} 
We use the trivial initial condition at $J_\varphi=0$, 
\begin{align} 
	\varphi_\Lambda=\phi\,.
	\label{eq:TrivialInitialvarphi}
\end{align}
\Cref{eq:TrivialInitialvarphi} entails that all non-trivial terms in \labelcref{eq:O-hatOGen} are missing at $k=\Lambda$. Moreover, it leads to 
\begin{align}
	\Lambda	\partial_\Lambda \varphi_\Lambda[\phi]= \dot{\phi}_{\Lambda}\,.
	\label{eq:GenFlowOpvarphiLambda} 
\end{align}
It follows from \labelcref{eq:GenFlowOpvarphi}, that the field $\varphi_k[\phi]$ differs from that obtained by inverting the map $\phi_k[\varphi]$, defined by the simple integration of $\dot\phi[\phi]$. This functional underlies the interpretation of the left hand side of \labelcref{eq:DtRepresentation} as a total $t$-derivative. We find, see \cite{Ihssen:2024ihp}, 
\begin{align} 
	\phi_k[\varphi] = \varphi - \int_k^\Lambda \frac{d k^\prime }{k^\prime } \dot \phi[\phi_{k^\prime}]\,. 
	\label{eq:phidotphi}
\end{align}
The underlying map between the two fields is defined by the operator $\hat\phi_k[\hat\varphi]$: if the two agree, the left hand side of 	\labelcref{eq:GenFlowOpvarphi} vanishes and so does the right hand side. This entails a linear relation between $\hat\varphi$ and $\hat\phi$ and reflects the trivial case we have discussed in the introduction of this Section. For non-linear relations the right hand side is non-vanishing and the two functionals differ, in short 
\begin{align} 
	 \varphi_k\bigl[\phi_k[\varphi]\bigr]\neq \varphi\,.
	\label{eq:Map1neqMap2}
\end{align}
We have computed both functions $\varphi_k[\phi]$ and $\phi_k[\varphi]$ within the zero-dimensional example theory with \labelcref{eq:Zd0+GTScl}. The results are displayed in \Cref{fig:onepoint_function}: $\phi[\varphi]=(\varphi_k)^{-1}$ (red) and $\phi_k[\varphi]$ (blue) for $k=0$, together with the solution of \labelcref{eq:GvarphiGphi} (black dashed line), discussed in \Cref{sec:ReoncstructGvarphi}. These maps all differ from each other and their relevance for the reconstruction of the 1PI effective action $\Gamma_\varphi[\varphi]$ is discussed further in \Cref{sec:ReoncstructGvarphi}. 

We close this Section with a discussion of higher correlation functions. The operator flow \labelcref{eq:GenFlowOpvarphi} is readily extended to that of a general $n$-point correlation function of the fundamental field. We use that a $J_{\cal O}$-derivative of ${\cal O}_1$ provides the two-point correlation function 
at $k=0$, if evaluated on the EoM $\phi_\textrm{\tiny{EoM}}$. For $k\neq 0$ and general $\phi$ we find 
\begin{align} 
	{\cal O}_{G_\varphi} := \frac{\delta 	{\cal O}_{1}}{\delta J_\varphi} =\frac{\delta^2 \Gamma_\phi}{\delta J_\varphi^2} =\left\langle \hat\varphi^2\right\rangle +\cdots \,,
\label{eq:OGvarphi} 
\end{align}
where the dots $\cdots$ stand for the other terms in \labelcref{eq:O-hatOGen} that vanish at $k=0$ for $\phi_\textrm{\tiny{EoM}}$. The initial condition is derived from \labelcref{eq:TrivialInitialvarphi} with a further derivative with respect to $J_\varphi$. 
Note however, that the operator ${\cal O}_{G_\varphi}$ and ${\cal O}_{2}$ agree for $k=0$, if evaluated on the EoM: the first term in \labelcref{eq:O-hatOGen} agrees for both operators. For $k\neq 0$ and general $\phi$, the additional terms are different, which leads to different flows and initial conditions. 

%

\subsection{Flow of the propagator $G_\varphi[\phi]$ }
\label{sec:O2Flows}

With the result for the one-point function $\varphi[\phi]$ we can compute the full propagator of the fundamental field. For this purpose we consider the flow of the two-point function of the fundamental field, 
\begin{align} 
		{\cal O}_{2}^{ij}[\phi] = {\cal O}_{2,c_\varphi}^{ij}[\phi]+ \varphi^i[\phi] \varphi^j[\phi]\,.
		\label{eq:GStandardSplit}
	\end{align}
The subscript ${}_{c_\varphi}$ stands for the connected part of the operator in terms of $\hat\varphi$-connectedness at $k=0$ and on the solution $\phi_{\textrm{\tiny{EoM}}}$ of the equations of motion. 
Indeed we find, that the propagator of the fundamental theory is given by 
\begin{align} 
	G^{ij}_\varphi= {\cal O}_{2,c_\varphi}^{ij}[\phi_{\textrm{\tiny{EoM}}}] = \left[{\cal O}_{2}^{ij}[\phi_\textrm{\tiny{EoM}}] - \varphi^i_{\textrm{\tiny{EoM}}}\, \varphi^j_{\textrm{\tiny{EoM}}}\right]_{k=0}\!,
	\label{eq:Gvarphi}
\end{align}
where $\varphi_{\textrm{\tiny{EoM}}}= \varphi[\phi_{\textrm{\tiny{EoM}}}]$. The split \labelcref{eq:GStandardSplit} requires the knowledge of $\varphi^i[\phi]$ as computed in \Cref{sec:O1Flows} with the simplified operator flow \labelcref{eq:PIRGOPSimple}. This is accommodated by considering the simplified operator flow for the operator $\boldsymbol{\cal O}_2 =({\cal O}_1\,,\,{\cal O}_2 )$, as introduced in \labelcref{eq:SetofOps}. 

We also emphasise that \labelcref{eq:GStandardSplit} is not a split into $\hat\varphi$-connected and disconnected parts for $k\neq 0$ which follows already from \labelcref{eq:O-hatOGen+Rest}. Hence, calling ${\cal O}_{2,c_\varphi}^{ij}[\phi]$ a $\hat \varphi$-connected part is a slight abuse of notation for $k\neq 0$. The respective total derivative operator flow \labelcref{eq:PIRGOPSimple} for ${\cal O}_{2,c_\varphi}$ in \labelcref{eq:GStandardSplit} is given by 
\begin{align}\nonumber 
&\partial_t {\cal O}^{nm}_{2,c_\varphi} + \dot{\phi}_i\frac{\delta{\cal O}^{nm}_{2,c_\varphi}}{\delta \phi^i}\\[1ex]\nonumber 
=&\,  -\left[ \left( \partial_t \varphi^n+ \dot{\phi}^i\frac{\delta \varphi^n}{\delta \phi^i} \right) \, \varphi^m+(n\leftrightarrow m) \right] \\[1ex]\nonumber 
&- \frac{1}{2} \left[G_\phi\left( {\cal D}_t R_\phi\right) G_\phi \right]^{ij}\times\left( {\cal O}_{2,ji}^{(2),nm} \right.\\[1ex]
& \hspace{0.5cm}\left.+ \left[ \varphi_j^{(1),n}\varphi_i^{(1),m}+ \varphi_{ji}^{(2),n}\varphi^m +(n\leftrightarrow m)\right]\right)\!, 
	\label{eq:GenFlowOpO2Pre}
\end{align}
\begin{figure}[t]%
	\centering%
	\includegraphics[width=\linewidth]{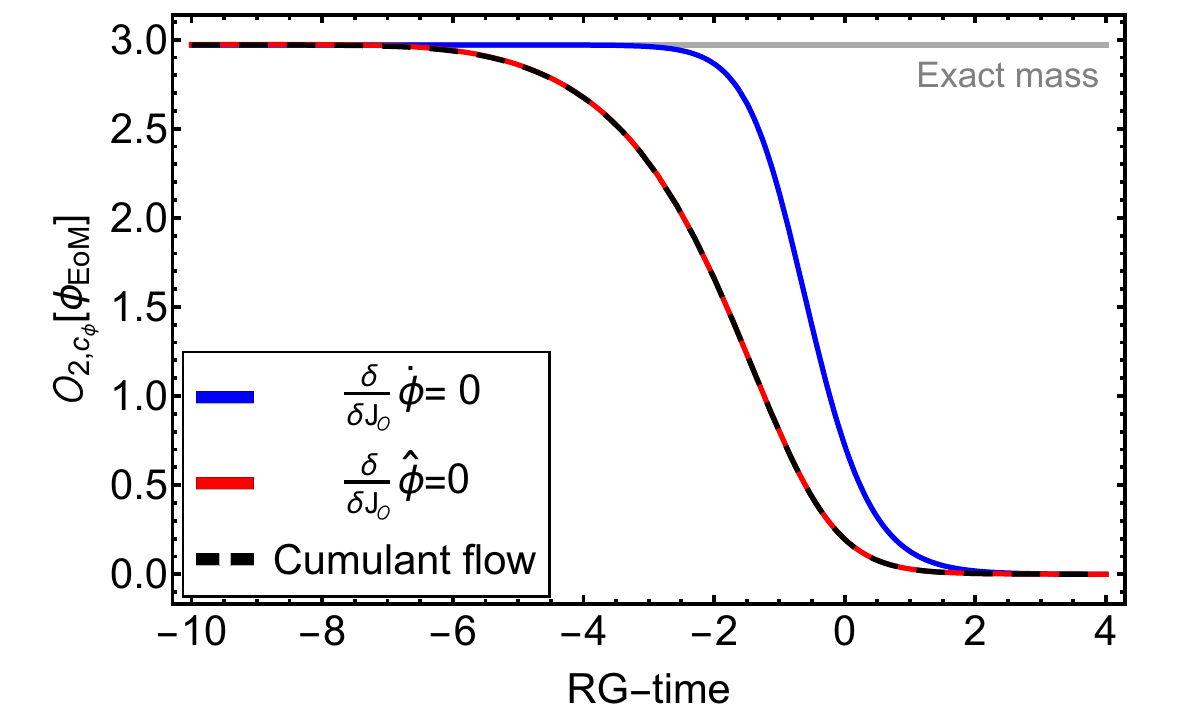}%
	\caption{RG-time dependence of the composite operator flow for different $J_{\mathcal{O}}$-dependences of $\dot \phi$. \hspace*{\fill}}%
	\label{fig:twopoint_function}%
\end{figure}
where $(n\leftrightarrow m)$ denotes the addition of the previous terms with switched indices.
The second line in \labelcref{eq:GenFlowOpO2Pre} is nothing but the right hand side of \labelcref{eq:GenFlowOpvarphi}, multiplied by $\varphi[\phi]$. The second line in \labelcref{eq:GenFlowOpO2Pre} and the last term in the third line proportional to $2 \varphi_{ji}^{(2)}$ cancel out and we arrive at the final expression, 
\begin{align}\nonumber 
\partial_t {\cal O}_{2,c_\varphi} + \dot{\phi}_i\frac{\delta{\cal O}_{2,c_\varphi} }{\delta \phi_i}=& - \frac{1}{2} \left[G_\phi\left( {\cal D}_t \, R_\phi\right) G_\phi \right]^{ij}\\[1ex]
& \hspace{-2.5cm}\times \left( {\cal O}_{2,c_\varphi,ij}^{(2),nm}+  \varphi_j^{(1),n}\varphi_i^{(1),m}+ \varphi_j^{(1),m}\varphi_i^{(1),n}\right), 
	\label{eq:GenFlowOpO2} 
\end{align}
\begin{figure*}
	\centering%
	\begin{subfigure}[t]{.48\linewidth}
		\includegraphics[width=\linewidth]{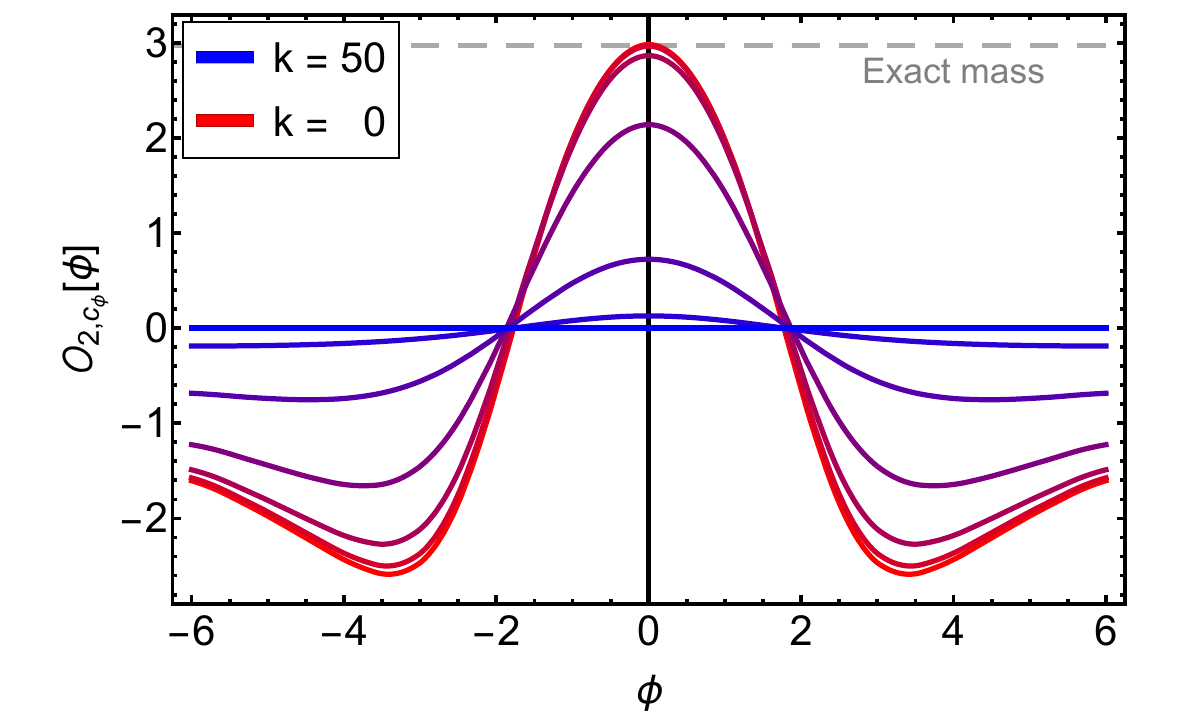}
		\caption{Standard choice with operator flow \labelcref{eq:GenFlowOpO2phi}, fulfilling \labelcref{eq:FO0}. \hspace*{\fill}}
		\label{fig:Dep}
	\end{subfigure}%
	\hspace{0.03\linewidth}%
	\begin{subfigure}[t]{.48\linewidth}
		\includegraphics[width=\linewidth]{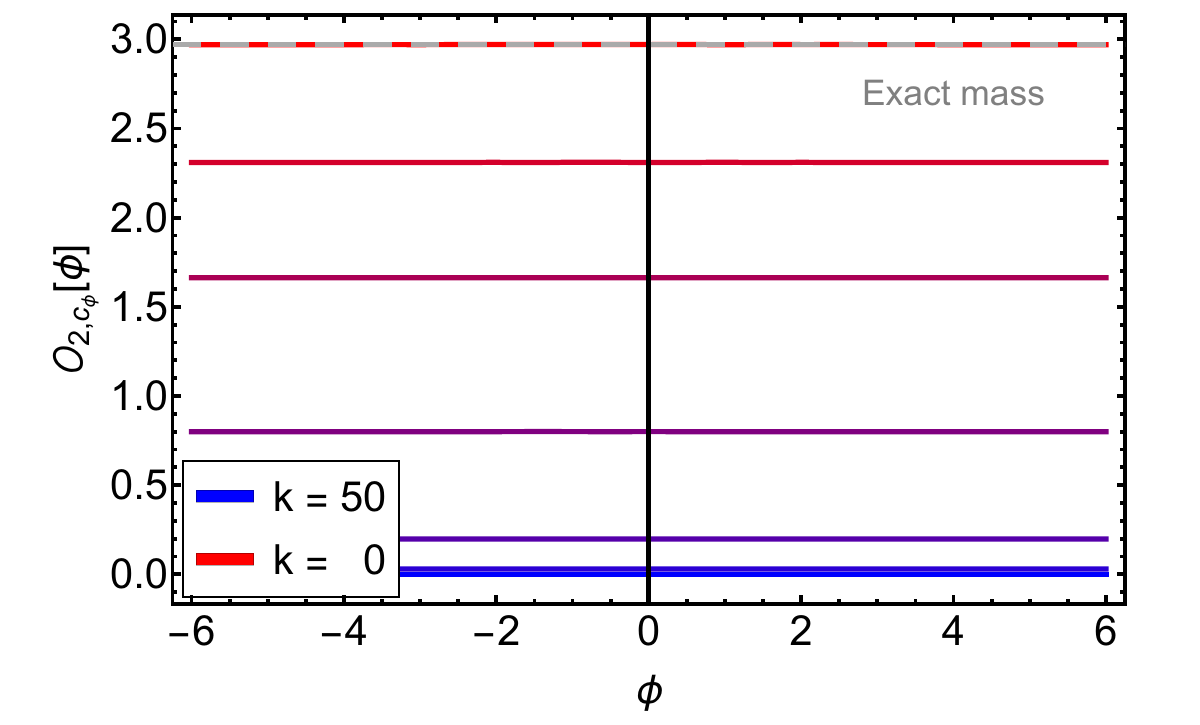}
		\caption{Field independent flow of the composite operator using \labelcref{eq:GenFlowOpO2Jo}. \hspace*{\fill}}
		\label{fig:NoDep}
	\end{subfigure}
	\caption{Field dependence of composite operator flows of the connected part of the two-point function $\mathcal{O}_{2,c_\phi}$ for different $J_{\mathcal{O}}$ dependences of the flowing field. \hspace*{\fill}}
	\label{fig:Fielddependences}
\end{figure*}
which reduces to the standard flow of the propagator $G_\varphi$ for $\phi=\varphi$. In the general case, \labelcref{eq:GenFlowOpO2} requires the knowledge of $\varphi[\phi]$ and hence the solution of \labelcref{eq:GenFlowOpvarphi}. This already illustrates the iterative structure for all correlation functions: the flow of the connected or even 1PI-parts of ${\cal O}_n$ requires the knowledge of the flow of the ${\cal O}_m$ with $m<n$. While this certainly enlarges the computational effort required to compute ${\cal O}_n$, it has the potential of improving a given approximation. We emphasise that while \labelcref{eq:PIRGOPSimple} for ${\cal O}_2$ and \labelcref{eq:GenFlowOpO2} are formally identical, they do not agree if approximations are applied. \Cref{eq:GenFlowOpO2} uses the exact flow of $\varphi={\cal O}_1$, which is then cancelled on the left and right hand sides of \labelcref{eq:GenFlowOpO2Pre}. At $k=0$ this separation is that in connected and disconnected parts with respect to $\varphi$, see \labelcref{eq:Gvarphi}, and we expect that such a flow leads to better results in comparison to, e.g., no split. Indeed, first computations in quantum field theories hint at this scenario, a detailed analysis shall be published elsewhere including a first discussion of functional optimisation of operator PIRGs, \cite{IKP}. In summary, this suggests that the split \labelcref{eq:GStandardSplit} is the natural one.

However, we may also use a split that is guided by computational convenience. For the latter we notice that the flow operator \labelcref{eq:DtRepresentation} is proportional to two derivatives with respect to $\phi$. This reflects the fact that connectedness in the generalised flow equation is defined with respect to correlation functions of the composite field $\hat\phi$. Hence, it is computationally more advantageous to split the correlation function ${\cal O}_2$ in terms of 1PI, connected and disconnected parts in $\hat\phi$. n the present case of the two-point functions this split reads
\begin{align} 
	{\cal O}^{ij}_{2}[\phi] = {\cal O}^{ij}_{2,c_\phi}[\phi] + \phi^i \phi^j\,, 
		\label{eq:GphiSplit}
\end{align}
where the subscript ${}_{c_\phi}$ indicates $\hat\phi$-connectedness. We note in passing that for $k\neq 0$, ${\cal O}_{2,c_\phi}$ is not just the expectation value of $\langle \hat \phi^2\rangle$. Hence, subtracting $\phi^i \phi^j$ from ${\cal O}_{2}$ may in general not give us a $\hat\phi$-connected part of ${\cal O}_{2}$. However, the additional terms in \labelcref{eq:O-hatOGen+Rest} are 1PI, so in contradistinction to the split \labelcref{eq:GStandardSplit} with the fundamental field, the operator ${\cal O}_{2,c_\phi}$ is the $\hat\phi$-connected part of ${\cal O}_{2}$. Finally, we emphasise that this split is purely done for computational convenience and provides the same results as no split in any approximation: in contradistinction to the $\varphi$-split no exact flow is divided out. In summary, we expect that this class of operator flows (no split) works best for physical composite fields such as used in the ground state expansion. In turn, for extreme choices such as the classical target action, \labelcref{eq:GphiSplit} or similar choices may require good approximations in order to work. These arguments are supported by explicit results in non-trivial quantum field theories \cite{IKP}.

With the split \labelcref{eq:GphiSplit} the flow equation for ${\cal O}_{2,c_\phi}$ takes the form 
\begin{align}\nonumber 
&	\partial_t {\cal O}^{nm}_{2,c_\phi} + \dot{\phi}_i\frac{\delta {\cal O}^{nm}_{2,c_\phi} }{\delta \phi_i} + \left( \dot \phi^n \,\phi^m + \phi^n \dot \phi^m\, \right) \\[1ex]
& \hspace{-.12cm}
= - \frac{1}{2} \left[G_\phi\left( {\cal D}_t \, R_\phi\right) G_\phi \right]^{ij}\!\left[ {\cal O}^{(2),nm}_{2,c_\phi,ij}+ \delta^n_i \delta^m_j+\delta^m_i \delta^n_j\right]\!. 
	\label{eq:GenFlowOpO2phi} 
\end{align}
The solution of \labelcref{eq:GenFlowOpO2phi} does not require the knowledge of $\varphi[\phi]$, but the computation of the propagator $G_\varphi$ does for non-vanishing solutions of the EoM, see \labelcref{eq:Gvarphi}. We rewrite \labelcref{eq:Gvarphi} in terms 
of $ {\cal O}^{ij}_{2,c_\phi} $, to wit, 
\begin{align} 
	G^{ij}_\varphi = {\cal O}^{ij}_{2,c_\phi}[\phi_{\textrm{\tiny{EoM}}}] +\left(\phi_{\textrm{\tiny{EoM}}}^i \phi_{\textrm{\tiny{EoM}}}^j - \varphi^i_{\textrm{\tiny{EoM}}}\,\varphi^j_{\textrm{\tiny{EoM}}}\right) \,. 
	\label{eq:GfromtildeGphi}
\end{align} 
For $\phi_{\textrm{\tiny{EoM}}}=0=\varphi_{\textrm{\tiny{EoM}}}$ we do not require the knowledge of $\varphi_{\textrm{\tiny{EoM}}}$, while for non-vanishing mean fields we do. 

We have computed the flow of ${\cal O}_{2,c_\phi}$ in the zero-dimensional example \labelcref{eq:Zd0+GTScl}, the result for $\phi=0$ is depicted in \Cref{fig:twopoint_function} for different implementations of the operator flow as well as the result obtained from the cumulants-preserving flows of the effective action: 
\begin{itemize}
	\item[(i)] Composite fields $\hat \phi$ with $J_{\cal O}$-dependence, chosen such that \labelcref{eq:dotphiJO} holds true. This leads to the simplified operator flow \labelcref{eq:GenFlowOpO2phi} for ${\cal O}_2$: straight blue line in \Cref{fig:twopoint_function}. 
	\item[(ii)] Composite fields $\hat \phi$ without $J_{\cal O}$-dependence, i.e.~\labelcref{eq:hatphinoJ} for all $k$. This is a special case we can derive easily in $d=0$, see \Cref{app:JOdep}. Then, \labelcref{eq:GenFlowOp} holds true: straight red line in \Cref{fig:twopoint_function}. 
	\item[(iii)] Two-point function obtained from an $m_\varphi^2$-derivative of the cumulants-preserving effective action, see \cite{Ihssen:2024ihp} or \Cref{app:cumuFlow}: dashed black line in \Cref{fig:twopoint_function}. 
\end{itemize} 
All results agree at $k=0$ as no approximation has been applied.
Moreover, the flows (ii) and (iii) agree which each other at all scales. The numerical calculation and an explanation of this last finding are given in \Cref{app:Details}. Finally, in \Cref{fig:Fielddependences} we compare the $\phi$-dependence of the two operator flows (i) and (ii): the operator flow for the $J_{\cal O}$-independent composite field (ii) agrees with the $m_\varphi^2$-derivative of the cumulants-preserving effective action and has no field-dependence. In turn, the operator flow (i) provides the full $\phi$-dependent two-point function. However, the operator flow (ii) is hard to access for a quantum theory in $d\geq 1$ and the easy access here is a particularity of $d=0$. In general, we deal with flows of the type (ii).  

We conclude this Section with a remark on the construction of a further effective action that may prove useful: \Cref{eq:GenFlowOpO2} can be seen as the master equation for an effective action $\tilde \Gamma_\varphi$, 
\begin{align} 
	\tilde \Gamma_\varphi^{(2)}\Bigl[\varphi[\phi]\Bigr]:=\left({\cal O} _2- \varphi^2\right)^{-1}[\phi]\,.
\label{eq:Gamma2varphi}
\end{align} 
We refrain from splitting up the regulator term as it is not quadratic in the fundamental field. We also remark that $\tilde \Gamma_\varphi$ does not satisfy the Wetterich equation, but a multi-loop equation. Multiplying \labelcref{eq:GenFlowOpO2} from both sides with $\tilde\Gamma^{(2)}_\varphi$ leads us to 
\begin{align}\nonumber 
		\partial_t \Gamma^{(2)}_{\varphi} + \dot{\phi}_i\frac{\delta \Gamma^{(2)}_{\varphi}}{\delta \phi_i}=&\,  \frac{1}{2} \left[G_\phi\left( {\cal D}_t \, R_\phi\right) G_\phi \right]_{ij}\\[1ex]
		&\hspace{-2.8cm}\times \Biggl( \tilde \Gamma^{(2,2)}_{\varphi,ji} - 2 \tilde \Gamma^{(2,1)}_{\varphi,i}\cdot G_\varphi\cdot\tilde \Gamma^{(2,1)}_{\varphi,j} + 2\, \tilde \Gamma_{\varphi,i}^{(1,1)}\tilde \Gamma_{\varphi,j}^{(1,1)}\Biggr)\,. 
		\label{eq:GenFlowG2varphi} 
\end{align}
The mixed derivatives $\tilde\Gamma_\varphi^{(n,m)}$ originate in the derivatives $\tilde \Gamma_\varphi^{(n+2)}$ with $n\geq 0$ of $\tilde\Gamma^{(2)}$ defined in 
\labelcref{eq:Gamma2varphi}. We define 
\begin{align} 
	\tilde \Gamma_\varphi^{(n,m)} = \frac{ \delta^m \tilde\Gamma^{(n)}_\varphi}{ \delta \phi^m}\,, 
	\qquad 	\tilde \Gamma^{(n,1)}_{\varphi, i_1\cdots i_n,j}
	= \frac{\delta \varphi_{i_{n+1}} }{\delta \phi_j} \tilde \Gamma^{(n+1,0)}_{\varphi,i_1\cdots i_{n+1}}\,, 
\end{align} 
and the mixed derivative $\tilde \Gamma_{\varphi,j}^{(1,1)}$ in \labelcref{eq:Gamma2varphi} is given by 
\begin{align} 
	\tilde \Gamma_{\varphi,i,j}^{(1,1)} = \varphi^{(1)}_{l,j}\tilde \Gamma_{\varphi,il}^{(2,0)}\,.
\end{align} 
It is illustrative to reduce \labelcref{eq:GenFlowG2varphi} to the operator flow \labelcref{eq:OpFlow} with $\phi = \varphi$. For example, the last term simply collapses to $\partial_t R_\varphi $ and originates in hitting the explicit regulator contribution in $1/G_\varphi$ in \labelcref{eq:Gamma2varphi}. In turn, the general form accommodates the transformation of the regulator term. 

Similar relations can be derived for higher correlation functions. Note that the structure of \labelcref{eq:GenFlowOpO2} persists and the explicit flow terms of the lower correlation functions drop out. It is worth emphasising that we cannot use the short cut and simply use further $\varphi$-derivatives of the effective action $\tilde \Gamma_\varphi$ or $G_\varphi$ for computing the higher correlation functions as the mean field $\varphi$ is defined from the current term of $\hat\phi$, this will be discussed in \Cref{sec:ReoncstructGvarphi}.

\subsection{Flow of higher-order correlation functions}
\label{sec:O4+6Flows}

The flows for the higher-order correlation functions follow analogously to those of the propagator. We first consider the generalisation of the $\hat\varphi$-split \labelcref{eq:GStandardSplit} for an $n$-point correlation function ${\hat O}_n$ with \labelcref{eq:On}. Then we have to consider the operator flow of 
\begin{align} 
	\boldsymbol{\cal O}_n = ({\cal O}_1\,,...,\,{\cal O}_n )\,,
\label{eq:O-On}
\end{align}
as the computation within the split requires the knowledge of all lower order correlation functions. As discussed in \Cref{sec:O2Flows} for the case of the two-point function, within this split the flows of the lower correlation functions are divided out, hence guaranteeing exact relations and projecting on the connected part of ${\cal O}_n$. The split is readily implemented as it simply uses the relation of a full $n$-point function and its connected and disconnected parts. We illustrate this explicitly for the four-point function 
\begin{align} 
	{\cal O}_4 =&\, {\cal O}_{4,c_\varphi} -4 \,{\cal O}_{1,c_\varphi}{\cal O}_{3,c_\varphi} \notag\\[1ex]
	&\,-3 \,{\cal O}_{2,c_\varphi}^2 +12\, {\cal O}_{1,c_\varphi}^2{\cal O}_{2,c_\varphi} -6 \, {\cal O}_{1,c_\varphi}^4 \,.
	\label{eq:4Pointcvarphi} 	
\end{align}
\begin{figure*}
	\centering%
	\begin{subfigure}[t]{.49\linewidth}
		\includegraphics[width=\linewidth]{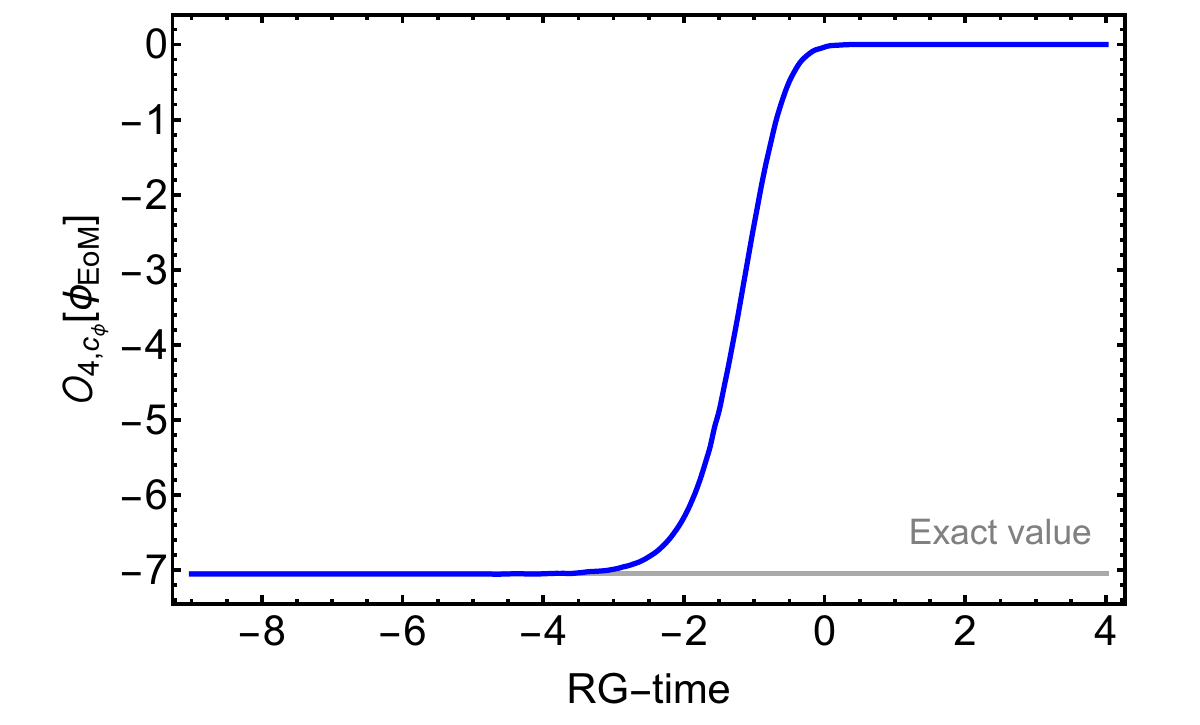}
		\caption{$\hat\phi$-connected part ${\cal O}_{4,c_\phi}$ of the operator ${\cal O}_4$, see \labelcref{eq:O-hatOGen+Rest}. \hspace*{\fill}}
		\label{fig:fourpoint_function}
	\end{subfigure}%
	\hspace{0.01\linewidth}%
	\begin{subfigure}[t]{.49\linewidth}
		\includegraphics[width=\linewidth]{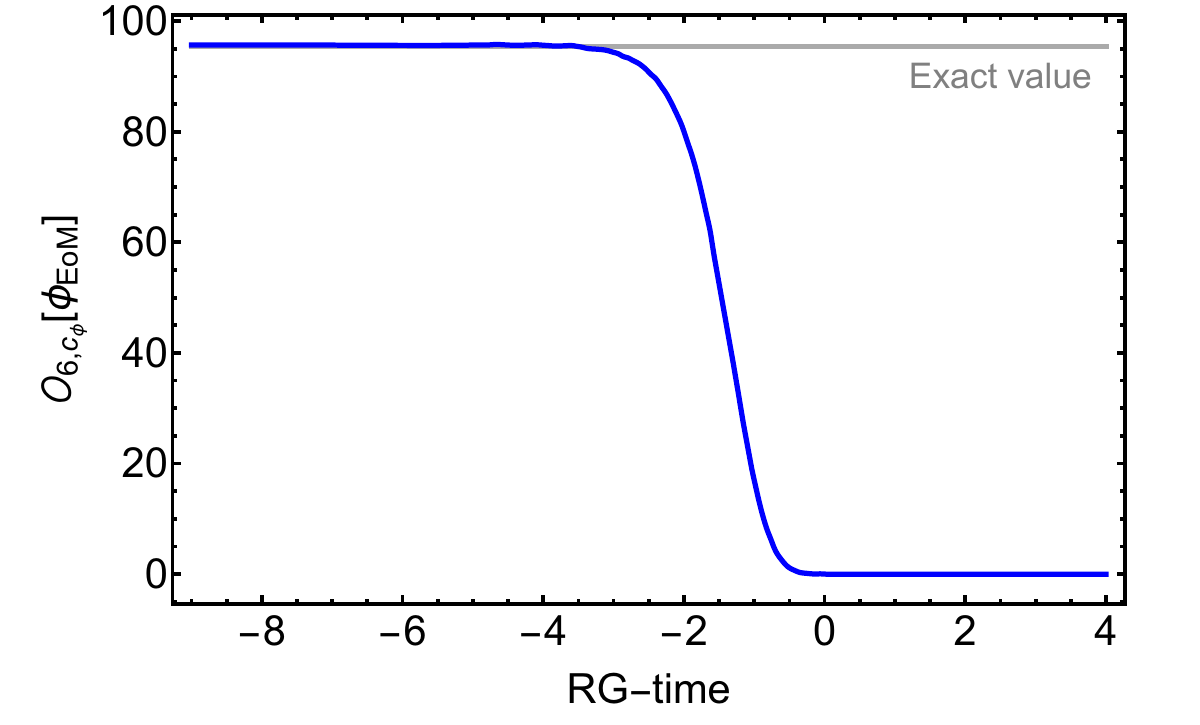}
		\caption{$\hat\phi$-connected part ${\cal O}_{6,c_\phi}$ of the operator ${\cal O}_6$, see \labelcref{eq:O-hatOGen+Rest}.  \hspace*{\fill}}
		\label{fig:sixpoint_function}
	\end{subfigure}
	\caption{RG-time dependence of the operator flows for the $\hat\phi$-connected four- and six-point functions. We show results for the simplified operator flow \Cref{sec:SimpleOP-PIRGs}. This corresponds to the blue curve in \Cref{fig:twopoint_function}. At $k=0$ the $\hat\phi$-connected parts agree with the $\hat\varphi$-connected ones as at $k=0$ we have ${\cal O}_{4,c_\phi}=\langle \hat\varphi^4\rangle_c$ due to 
		$\varphi_\textrm{\tiny{EoM}} =\phi_\textrm{\tiny{EoM}}=0$. \hspace*{\fill}}
	\label{fig:4+6point}
\end{figure*}
This concludes our discussion of the $\hat\varphi$-split and its iterative structure. 
 
In the present work we concentrate on the zero-dimensional case without any approximation. Then, the computation is facilitated by using the split in $\hat \phi$-connected correlation functions: it lowers the redundancies within the computation as the flow is one-particle irreducible in the $\phi$ field. This also improves the numerical convergence. However, as discussed in \Cref{sec:O2Flows}, it should only be considered for physical choice of $\phi$ such as underlying the ground state expansion.

In the present benchmark example both, the numerical challenges as well as those related to approximations, are absent and the computation has been performed without any split in connected and disconnected parts. Still we showcase the split for the four- and six-point functions to provide a glimpse of the systematics. The split of the four-point function is given by \labelcref{eq:4Pointcvarphi} with $c_\varphi\to c_\phi$, and the split of the six-point function is given by 
\begin{align} 
	{\cal O}_6 =& \, {\cal O}_{6,c_\phi} -6 \,{\cal O}_{1,c_\phi}{\cal O}_{5,c_\phi}-\, 15 {\cal O}_{2,c_\phi} {\cal O}_{4,c_\phi} \notag \\[1ex] &
	+ 30 \, \left({\cal O}_{2,c_\phi}^3 + {\cal O}_{1,c_\phi}^2 {\cal O}_{4,c_\phi} \right)-10 \,  {\cal O}_{3,c_\phi}^2 \notag \\[1ex] &+ 120 \,\left( {\cal O}_{1,c_\phi}{\cal O}_{2,c_\phi} {\cal O}_{3,c_\phi}- {\cal O}_{1,c_\phi}^3 {\cal O}_{3,c_\phi}- {\cal O}_{1,c_\phi}^6 \right) \notag \\[1ex] &
	+360 \, {\cal O}_{1,c_\phi}^4 {\cal O}_{2,c_\phi}-270\,  \,{\cal O}_{1,c_\phi}^2{\cal O}_{2,c_\phi}^2
	 \,.
\label{eq:6Point} 
\end{align}
The odd numbered operators vanish on the equations of motion 
due to the $\varphi\to-\varphi$ symmetry. Note that the $c_\phi$-splits also require the input of the lower order correlation functions, that is ${\cal O}_{2,c_\phi}$ for \labelcref{eq:4Pointcvarphi} with $c_\varphi\to c_\phi$ and ${\cal O}_{2,c_\phi}, {\cal O}_{4,c_\phi}$ for \labelcref{eq:6Point}. 
Results for the RG-time dependence of the connected parts ${\cal O}_{4,c_\phi}$ and ${\cal O}_{6,c_\phi}$ on the equations of motion are depicted in \Cref{fig:4+6point}. These showcase results use the simplified operator flow discussed in \Cref{sec:SimpleOP-PIRGs}. We have also computed further correlation functions up to the order ${\cal O}_{10}$, and they also agree with the full numerical results. We shall use them in the next Section for the reconstruction or rather the vertex expansion for the effective action $\Gamma_\varphi$ and the Schwinger functional $W_\varphi$.

\subsection{Reconstruction of the $\varphi$-generating functionals}
\label{sec:ReoncstructGvarphi}

Finally, we discuss the reconstruction or computation of the generating functionals $Z_\varphi$, or rather the Schwinger functional $W_\varphi\simeq \log Z_\varphi$, and $\Gamma_\varphi$. The latter is the key object of interest with the least redundancies and we consider it first: 

It is suggestive that the map $\varphi_k[\phi]$ or rather its inverse $\phi[\varphi]$ can be used for the reconstruction of the full 1PI effective action $\Gamma_\varphi[\varphi]$ of the fundamental field from the effective action $\Gamma_\phi[\phi]$. However we shall see that in general the required map is not $\phi[\varphi]$. What we would need is the functional $\tilde \phi[\varphi]$ at $k=0$ with 
\begin{align} 
	\Gamma_\varphi[\varphi]	= \Gamma_\phi\bigl[\tilde\phi[\varphi]\bigr]\,.	
	\label{eq:GvarphiGphi}
\end{align}
The existence of such a field $\tilde\phi[\varphi]$ has been judiciously discussed in \cite{Falls:2025sxu}. Together with the two maps $\varphi_k[\phi]$ from the operator flow \labelcref{eq:GenFlowOpvarphi} and $\phi_k[\varphi]$ from \labelcref{eq:phidotphi} we are left with three different maps from the fundamental mean field to a composite mean field: 
\begin{enumerate} 
	\item[(i)] $\phi[\varphi]$: this functional is obtained by simply integrating the flowing field $\dot\phi[\phi]$, see \labelcref{eq:phidotphi} and \cite{Ihssen:2024ihp}, see also the discussion below \labelcref{eq:DtRepresentation}. 
	\item[(ii)] $\varphi[\phi]$: this functional is derived from the operator flow 	\labelcref{eq:GenFlowOpvarphi} in \Cref{sec:O1Flows}. 
	\item[(iii)]$\tilde \phi[\varphi]$: this field is simply defined as the aimed-for map in \labelcref{eq:GvarphiGphi}, see \cite{Falls:2025sxu}. 
\end{enumerate} 
We have computed all three functionals within the zero-dimensional example \labelcref{eq:Zd0+GTScl}, where they are one-dimensional functions. They are depicted in \Cref{fig:onepoint_function}, where we show the maps $\phi[\varphi]$ obtained from the definitions (i) -- (iii) below \labelcref{eq:GvarphiGphi}. They all differ from each other and we conclude that 
the effective action $\Gamma_\phi[\phi]$ is \textit{not} the effective action $\Gamma_\varphi$: Evidently, in most practical applications it is possible to find a potentially singular functional $\tilde \phi[\varphi]$ that satisfies \labelcref{eq:GvarphiGphi}. However, $\tilde \phi[\varphi]$ is not directly related to the underlying map $\hat\phi_k[\hat\varphi]$. Note also that \labelcref{eq:GvarphiGphi} does not satisfy the Wetterich equation for any cutoff $R_\varphi$. This is explicitly seen by inserting \labelcref{eq:GvarphiGphi} into the generalised flow equation for the given $\Gamma_\phi$. This leads us to 
\begin{align}\nonumber 
	\partial_t\Gamma_\varphi[\varphi] + \left( \dot{\phi}_i -\partial_t\tilde\phi_i \right) \frac{\delta \Gamma_\phi}{\delta \phi_i} & \\[1ex]  
	& \hspace{-2cm}=\frac{1}{2} G^{ij}_{\phi}\left({\cal D}_t  R_\phi\right)_{ij} +\partial_t {\cal \ln N}_\phi \,. 
	\label{eq:GenFlowGvarphi} 
\end{align}
\begin{figure*}
	\centering%
	\begin{subfigure}[t]{.49\linewidth}
		\includegraphics[width=\linewidth]{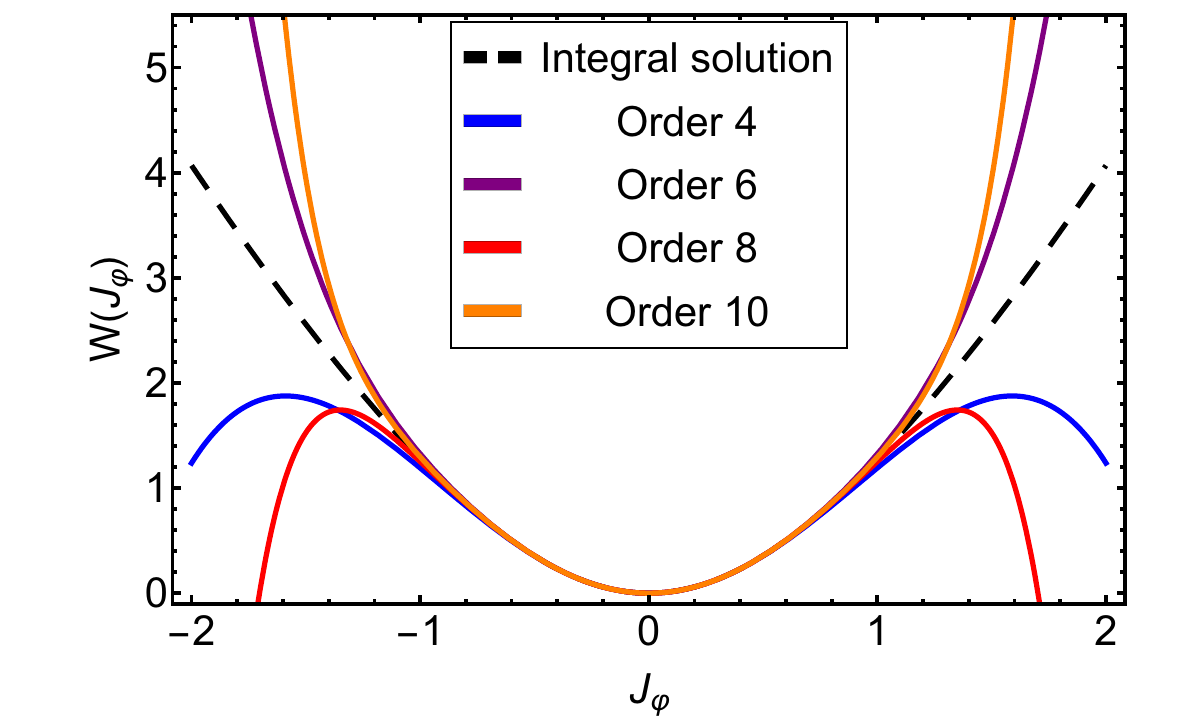}
		\caption{Vertex expansion of the Schwinger functional. \hspace*{\fill}}
		\label{fig:W_plot}
	\end{subfigure}%
	\hspace{0.01\linewidth}%
	\begin{subfigure}[t]{.49\linewidth}
		\includegraphics[width=\linewidth]{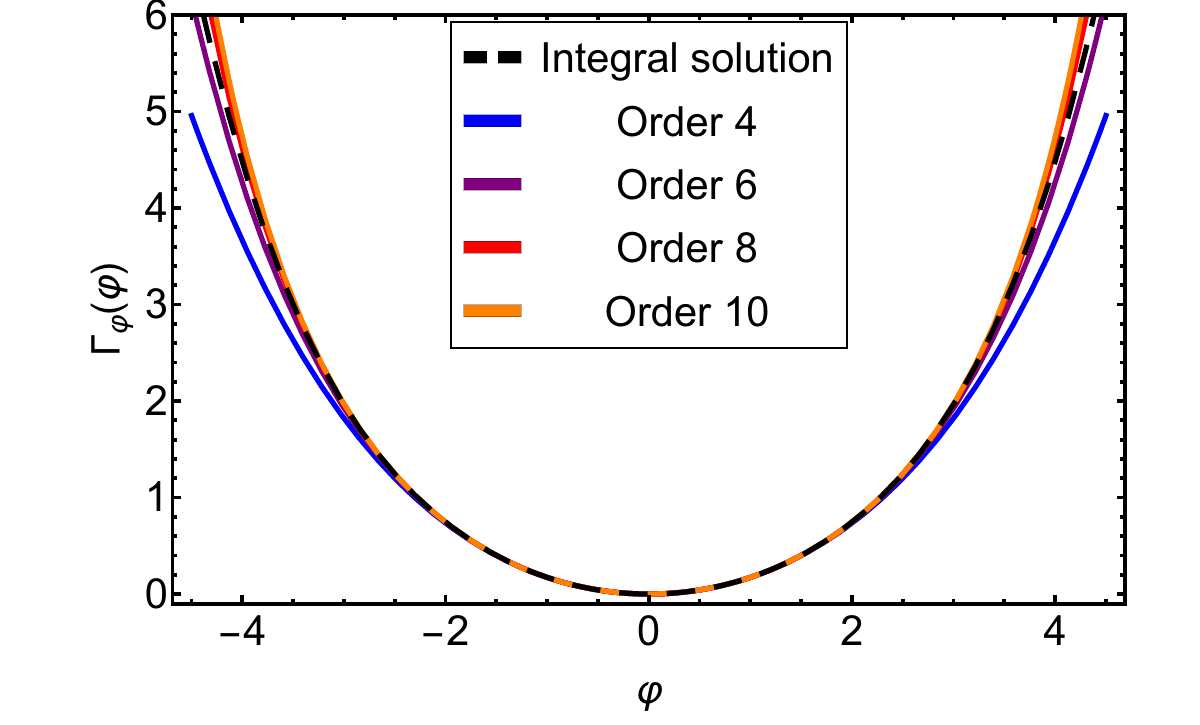}
		\caption{Vertex expansion of the effective action. \hspace*{\fill}}
		\label{fig:G_plot}
	\end{subfigure}
	\caption{Reconstruction of the generating functionals of correlation functions of the fundamental field, the 1PI effective action $\Gamma_\varphi$ and the Schwinger functional $W_\varphi$ from the results for the correlation functions ${\cal O}_n = \langle \hat\varphi^n\rangle$ at $k=0$. The expansion uses the operator flows for $\mathcal{O}_{i,c_\phi}$ with $i=2,4,6$ and \labelcref{eq:GammaExpandOn,eq:WExpandOn}.\hspace*{\fill}}
	\label{fig:functionals}
\end{figure*}
\begin{figure}
	\centering%
	\includegraphics[width=\linewidth]{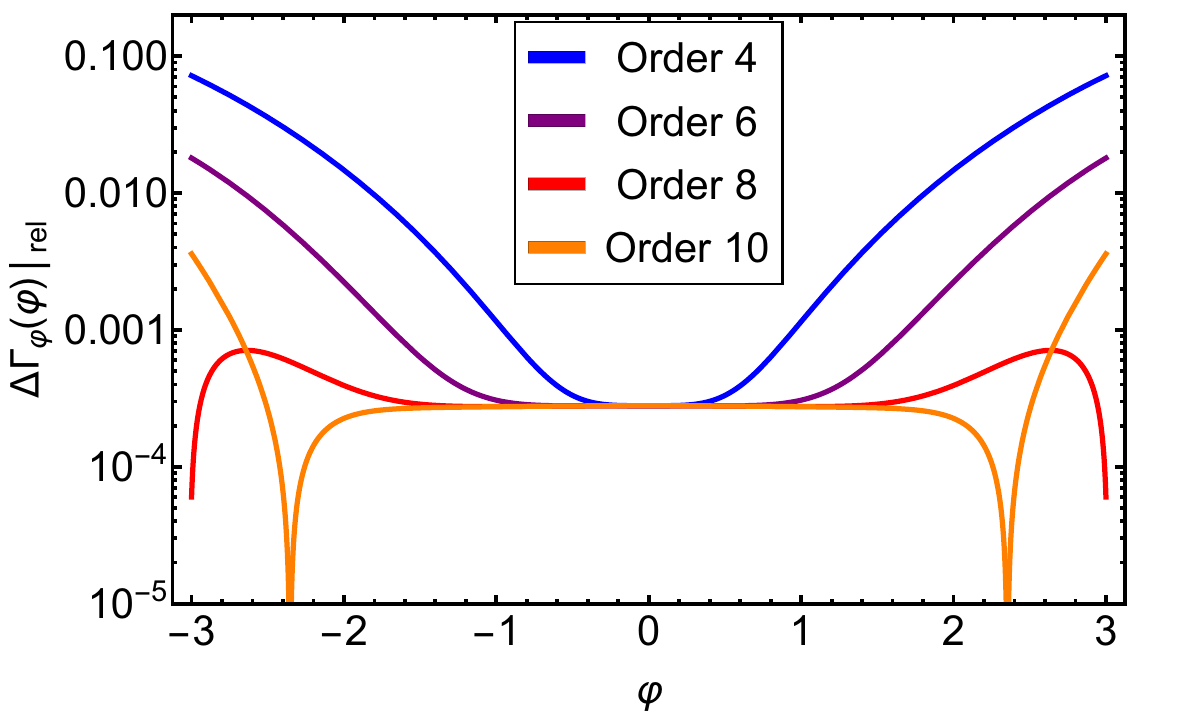}%
	\caption{Relative error $\Delta\Gamma^\textrm{rel}_\varphi$, \labelcref{eq:RelativeError}, of the effective action $\Gamma_\varphi$ of the fundamental field in the $n$th order of the vertex expansion including terms up to the order $\phi^{2n}$. We depict $n=4,6,8,10$. The reference solution is computed numerically from \labelcref{eq:Zd0+GTScl}.
		\hspace*{\fill}}%
	\label{fig:RelativeError}%
\end{figure}
Assume now that $\Gamma_\varphi$ solves the Wetterich equation, 
\begin{align}
	\partial_t\Gamma_\varphi[\varphi] =\frac{1}{2} G^{ij}_{\varphi} \partial_t R_{\varphi, ij}+\partial_t \ln {\cal N}_\varphi\,. 
	\label{eq:WetterichFlow} 
\end{align}
Then, the combination \labelcref{eq:GenFlowGvarphi,eq:WetterichFlow} can be solved for $\partial_t\tilde\phi_i$, to wit, 
\begin{align}\nonumber 
	\partial_t\tilde\phi_i \frac{\delta \Gamma_\phi}{\delta \phi_i} =&\, \dot{\phi}_i \frac{\delta \Gamma_\phi}{\delta \phi_i} \\[1ex]
	&\hspace{-1cm}+ \frac{1}{2} G^{ij}_{\varphi} \partial_t R_{\varphi, ij}- \frac{1}{2} G^{ij}_{\phi}\left({\cal D}_t \, R_\phi\right)_{ij} + \partial_t \ln\frac{ {\cal N}_\varphi}{{\cal N}_\phi}\,. 
	\label{eq:Flowtildephi} 
\end{align}
On the equation of motion of $\varphi$, the left hand side vanishes and so has the right hand side. This induces a specific choice of the flow of the normalisations. As for \labelcref{eq:GvarphiGphi}, we can solve \labelcref{eq:Flowtildephi}, but it is not related to the underlying operator relation $\hat\phi[\hat\varphi]$. This concludes our discussion of the direct reconstruction of $\Gamma_\varphi$ from $\Gamma_\phi$. It will be furthered elsewhere. 

For the reminder of this Section we consider the reconstruction of the 1PI effective action $\Gamma_\varphi[\varphi]$ from 1PI-correlation functions. The respective expansion of the effective action in the $n$-point functions reads  
\begin{align} 
	\Gamma_\varphi[\varphi] = \sum_{n\geq 0} \frac{1}{n!}\,{\cal O}^{(\textrm{1PI})}_{n,i_1\cdots i_n} \,\varphi^{i_1}\cdots \varphi^{i_n}\,, 
	\label{eq:GammaExpandOn}
\end{align}
where ${\cal O}^{(\textrm{1PI})}_n$ are the 1PI-parts of the correlation functions ${\cal O}_n$ and we have used 
\begin{align}
\Gamma_{\varphi}^{(2)}= {\cal O}_{2,c_\varphi}^{-1}\qquad \textrm{and} \quad  \Gamma_{\varphi}^{(n)}= {\cal O}_n^{\textrm{(1PI)}}\,,\quad n\geq 3\,. 
\label{eq:On1PIGamman}	
\end{align} 
\Cref{eq:GammaExpandOn} is the vertex expansion of the effective action and is the standard expansion scheme for quantitative applications in higher-dimensional quantum field theories. If augmented with full effective potentials for emergent composites, it is the expansion scheme used for quantitative applications in QCD, see e.g.~the recent works \cite{Ihssen:2024miv, Fu:2025hcm}.  

The relation between the complete correlation functions, the connected ones and their 1PI parts is provided by the functional identity, see e.g.~\cite{Pawlowski:2005xe}, 
\begin{align}
	\langle \hat\varphi_{i_1} \cdots \hat\varphi_{i_n} \rangle =\prod_{j=1}^n \left[G_{i_j m}\frac{\delta}{\delta \varphi_m}+\varphi_{i_j}\right]\,, 
\label{eq:Correlations-Connected-1PI}
\end{align}
and 
\begin{align} 
\frac{\delta}{\delta \varphi_l}	G_{n m} = - G_{n i} \Gamma_{\varphi,ilj}^{(3)}  G_{j m}\,, 
\end{align} 
with the relation \labelcref{eq:On1PIGamman}. Hence, the ${\cal O}^{(\textrm{1PI})}_n$ can be computed from the set $\{{\cal O}_{m,c_\varphi} \}$ with $m\leq n$. In the explicit computation we have considered ${\cal O}^{(\textrm{1PI})}_{2,...,10}$. They are constructed with \labelcref{eq:Correlations-Connected-1PI} from the connected correlation functions ${\cal O}_{m,c_\varphi}$ with $m=2n$ and $n=1,...,5$, computed in \Cref{sec:O1Flows,sec:O2Flows,sec:O4+6Flows}. 

Furthermore, the ${\cal O}_{m,c_\varphi}$ can be used directly for the expansion of the Schwinger functional $W_\varphi \simeq  \log Z_\varphi$, 
\begin{align} 
	W_\varphi[J_\varphi] = \sum_{n\geq 0} \frac{1}{n!}\, {\cal O}_{n,c_\varphi}^{ i_1\cdots i_n} \,J_{\varphi,i_1}\cdots J_{\varphi, i_n}\,, 
	\label{eq:WExpandOn}
\end{align}
and as for the effective action we use the correlation functions ${\cal O}_{2n,c_\varphi}$ with $n=1,...,5$. The results are displayed in \Cref{fig:functionals}. In \Cref{fig:RelativeError} we depict the relative error $\Delta \Gamma^\textrm{rel}_\varphi$ of the vertex expansion for the effective action, 
\begin{align} 
	\Delta \Gamma^\textrm{rel}_\varphi[\varphi] = \frac{\bigl|\,\Gamma^\textrm{vert}_\varphi[\varphi]-\Gamma_\varphi [\varphi]\,\bigr|}{\Gamma_\varphi[\varphi]}\,,
\label{eq:RelativeError}
\end{align} 
where $\Gamma^\textrm{vert}_\varphi$ is the effective action within a given order of the vertex expansion. We see a rapid convergence of the vertex expansion for fields $|\varphi|\lesssim 2$. For these fields the relative error with is already in the permille regime and below for $2 n=10$.
The exact reference solution is obtained by a numerical evaluation of the $Z_\phi$-integral \labelcref{eq:Zd0+GTScl} and the subsequent numerical Legendre transform. 

For larger field values the effective action converges to the classical one, which is not well-described by the vertex expansion. This regime does not carry any 'physical' significance and the convergence for $|\varphi|\lesssim 2$ indicates a lower estimate for the respective convergence radius.

\section{PIRGs and Operator PIRGs in higher dimensional field theories}
\label{sec:HigherDim}

This Section discusses the use of operator PIRGs in quantum mechanics and quantum field theories. The zero-dimensional examples so far have illustrated the structural properties of operator PIRGs. 
In particular, we have discussed the need and computational validity  of the operator dependence of the field transformation for the concise form \labelcref{eq:GenFlowOpSimple}  as well as the potential use of the general operator flow \labelcref{eq:GenFlowOp}. 
In \Cref{sec:PIRG-SESs} we discuss systematic expansion schemes (SES) and their implementation within the PIRG setup. This continues and extends the respective detailed discussion in \cite{Ihssen:2024ihp, Ihssen:2025cff}. In \Cref{sec:PIRG-Physics+Resonances} we discuss the distribution of physics properties between the components of the two PIRG-pairs $(\Gamma[\phi]\,,\,\dot\phi[\phi])$ and $({\cal O}[\phi]\,,\dot \phi[\phi,J_{\cal O}])$ in \labelcref{eq:PairGamm-phi,eq:PairO-dotphi1} respectively. 

Before analysing these different aspects we would like to issue a statement of caution. The generality of the PIRG approach is both a blessing and a curse: Its inherent maximal flexibility allows for fRG-applications that qualitatively go beyond those considered so far. This generality is important both on the conceptual level as well as on the computational one. On the flip-side, using this maximal flexibility for the first time implies that many properties of the resulting fRG-flows and in particular the systematics and convergence of systematic expansion schemes have to be re-assessed. However, this simply means that applications come with an increased necessity of a systematic error analysis in comparison to the standard fRG-approach. The two following Sections should be read whilst keeping this in mind.

\subsection{Systematic approximations schemes in the PIRG setup} 
\label{sec:PIRG-SESs} 

In all zero-dimensional examples above flows were computed without approximations as both the flow of the effective action and the operator flows are exact within LPA. In higher dimensions, $d>0$, the flows are generally subject to approximations. Importantly, standard approximation schemes such as the derivative expansion or the vertex expansion have to be re-assessed for both parts of the PIRG approach: 

\subsubsection{Systematic expansion schemes for the PIRG-pair}
\label{sec:SES-PIRG}

Firstly, within a given level of a specific SES the computation using an underlying PIRG-pair $(\Gamma[\phi]\,,\,\dot\phi[\phi])$ in \labelcref{eq:PairGamm-phi} leads to different results than the same approximation level used in the standard approach with $\dot\phi\equiv 0$. Moreover, the generality of the PIRG approach gives access to novel expansion schemes like the \textit{ground state expansion} \cite{Ihssen:2023nqd} or the \textit{feed down flows} \cite{Ihssen:2024ihp} that have no counter part in the standard approach. These properties and further ones have been discussed at length in \cite{Ihssen:2024ihp} and we defer the reader to this paper for more details. Roughly speaking, PIRG flows outperform standard flows in a given approximation level of a SES if one can show that the information contained in the standard flow is completely accommodated within the PIRG-pair.
If this is the case, PIRG-flows typically also accommodate further terms in the standard effective action beyond the approximation level used there. Moreover, SESs such as the ground state expansion aim at optimising the convergence for general SES. While this optimisation idea does not guarantee better results within a given order, non-trivial evidence has been collected that the convergence is already qualitatively better in low levels of a given SES: the PIRG approach has been used for the analysis of the instanton-dominated regime of the anharmonic oscillator for small anharmonicities within the ground state expansion. In \cite{Bonanno:2025mon}, the exponential scaling of this regime has been confirmed within the fRG approach for the first time. We emphasise that, while the standard flow approach also offers results in this regime, it is fair to say that the numerical instabilities carried with the standard flow for small anharmonicities do not allow for even a qualitative analysis. Moreover, the results for small anharmonicities even indicate a polynomial scaling of the energy difference between the ground state and the first excited state $\Delta E$. In turn, the PIRG computation in \cite{Bonanno:2025mon} provides the exponential decay constant of $\Delta E$ (by means of a new observable) and agrees with the exact one on the percent level.

\subsubsection{Systematic expansion schemes for the operator PIRG-pair}
\label{sec:SES-OPPIRG}

We now proceed with the discussion of the operator-PIRG pair $({\cal O}[\phi]\,,\dot \phi[\phi,J_{\cal O}])$. Firstly, the operator flows \labelcref{eq:OpFlow,eq:PIRGOPSimple} are closed, i.e.~they do not feed back into the PIRG-pair. This facilitates the discussion of the level of a given SES within these flows. In particular, if the pair  \labelcref{eq:PairO-dotphi1} is resolved such that \labelcref{eq:OpFlow,eq:PIRGOPSimple} contain all terms used in the PIRG-pair $(\Gamma[\phi]\,,\,\dot\phi[\phi])$, the operator is consistent with this expansion-level. A further beneficial property of the operator flows, and specifically their simplified form \labelcref{eq:PIRGOPSimple}, is their form of a (generalised) heat or Fokker-Planck equation. This form makes it more amiable towards the numerical implementation of more sophisticated approximations in comparison to the standard flow equation. This complements the advantages of computations with the underlying PIRG-pair, where specific target actions can leave us with a simple first order partial differential equation for $\dot\phi$, see \cite{Ihssen:2024ihp}. In addition to these qualitative computational advantages, the two-step procedure of the setup allows to go beyond the level of the SES used for the PIRG-pair. For example, one may resolve the PIRG-pair in a given order of the derivative expansion, say the first order. Then, the operator flow of the two-point function of the fundamental field $\langle \varphi(p_1)\,\varphi(p_2)\rangle$ can be solved with its full momentum-dependence (as well as the dependence on constant fields $\varphi_c$). In particular, this provides us with the propagator $G_\varphi(p,\varphi_c)$ of the fundamental field. For a scalar field theory this leads to 
\begin{align} 
	G_\varphi(\varphi_c,p) = \frac{1}{Z_\varphi(\varphi_c,p)}\frac{1}{p^2+m^2_\varphi(\varphi_c)}\,, 
\label{eq:Gvarphip}
\end{align}
with 
\begin{align} 
	G_\varphi(\varphi_c;p,q)= G_\varphi(\varphi_c,p) \,(2 \pi)^d \delta (p-q)\,.
	\label{eq:Gvarphipq}
\end{align}
Similar flows have been used in the standard fRG approach, where full Euclidean and real-time momentum-dependences of propagators and vertices can be read-out from the flows of vertices computed with less momentum-dependences. For an early analysis of the convergence of this procedure see \cite{Helmboldt:2014iya}, for further examples and further analyses see \cite{Tripolt:2014fpa, Pawlowski:2015mia, Ihssen:2024miv}.

\subsubsection{Applications}
\label{sec:Applicationsd>0}

Both aspects of SESs in operator PIRGs are currently used in a computation of correlation functions of the fundamental field for the anharmonic oscillator (aiming at the incorporations of topological effects in the PIRG setup) and higher dimensional scalar field theories (aiming at the study of optimised convergence). These computations go far beyond the scope of the present, mainly conceptual, work and will be presented elsewhere.

\subsection{Physics distribution in the PIRG approach} 
\label{sec:PIRG-Physics+Resonances} 

We close this Section with a discussion of the distribution of physics amongst the different components of the PIRG-pair. This concerns for example physics properties such as the occurrence and location of phase transitions in the physical parameter space spanned by relevant couplings and external parameters such as temperature and density. Furthermore, it concerns the emergence and properties of bound states and resonances, important examples being QCD with its very rich hadron spectrum, but also condensed matter theories with their intricate and dynamical Fermi surfaces, e.g.~in the Hubbard model. These important properties have already been discussed in the literature with an emphasis on different aspects: \\[-1ex]

These properties have been discussed in the context of generalised flow equations and the PIRG approach in \cite{Ihssen:2022xjv, Ihssen:2023nqd, Ihssen:2024ihp, Ihssen:2025cff} with an emphasis on the generality of field transformations that also allow for mapping different theories onto each other. In particular this possibility also underlies the recent application of PIRGs to sampling tasks for lattice field theories and beyond with \textit{physics informed kernels}, see \cite{Ihssen:2025ybn, Ihssen:2026njd}. Similarly to standard Machine Learning architectures such as normalising flows and diffusion models, the efficiency of this generative RG-architecture is based on the map from the original sampling task to that with a simple distribution (or theory) that allows for efficient sampling. 

The properties have also been discussed in the context of the essential RG, \cite{Baldazzi:2021ydj, Baldazzi:2021orb, Falls:2025sxu, Falls:2025tid}. By definition, the essential RG restricts the mappings to those $\dot\phi$ that leave the mass-shell or the equations of motion unchanged. Accordingly, essential RG-transformations have the physics property that they leave the resonance spectrum of the theory unchanged. 

Finally, a comprehensive discussion has been put forward in \cite{Wetterich:2024uub} with a focus on \textit{microscopic} transformations $\hat\phi[\hat\varphi]$ underlying the generalised RG \cite{Pawlowski:2005xe} and hence the PIRG approach, as compared to \textit{macroscopic} transformations that are only defined on the level of the mean field variables $\varphi$ and their flows.

The latter two investigations have in common that they concentrate on specific aspects of transformations while the PIRG approach emphasises and uses the generality of transformations. Roughly speaking, the reliability assessment of the PIRG approach and its implementation in a specific SES certainly benefits from these analyses. However, we would like to emphasise that it is the generality of the PIRG approach which underlies its successful applications beyond the restrictions of the essential RG and the caveats voiced in \cite{Wetterich:2024uub}. In our opinion this is very transparent in its use with PIKs in \cite{Ihssen:2025ybn, Ihssen:2026njd} and their relation to standard Machine Learning architectures. 

In the following we discuss two specific properties of general quantum field theories: (emergent) resonances and (dynamical) Fermi surfaces. These are but two important examples, and the general structure extends to further important properties. 

\subsubsection{Physical resonances and their flow}
\label{sec:PIRGResonances}

We proceed with a discussion of the evolution and storage of resonance properties in the PIRG approach. Respective comments can also be found in \cite{Ihssen:2024miv} in the context of an investigation of the systematic error estimate of expansion schemes in the fRG approach with emergent composites, the latter being a specific implementation of PIRGs. To begin with, \textit{any} momentum space RG with fixed physics parameters (couplings and external parameters), leads to a change of the resonance spectrum of the theory and the Fermi surface of all fields involved. Indeed, it is one of the redeeming properties of momentum space RGs, that they lead from simple (in most cases classical) theories to full quantum field theories. The PIRG approach only redistributes this physics change between the generating functional at hand and the flowing fields. While this typically does not involve the flow or emergence of singularities in the Euclidean or imaginary time branch of the theory, except that of massless composites such as Goldstone modes, it definitely comes with the flow and emergence of singularities in the complex frequency plane of the theory at hand. It cannot be emphasised enough that this is a feature of RG-flows. Still, these flows may be difficult to incorporate in practice, for respective discussions see in particular \cite{Ihssen:2022xjv}. However, as is also discussed there, the PIRG approach does not add to the problem but it rather facilitates the implementation of the flow of singularities and that of emergent resonances. Indeed, the fRG approach with emergent composites, \cite{Gies:2001nw, Gies:2002hq, Pawlowski:2005xe, Floerchinger:2009uf, Fu:2019hdw, Fukushima:2021ctq}, was developed precisely to accommodate emergent composites (resonances) and their flow. Its most general form, the generalised flow equation \cite{Pawlowski:2005xe} underlies the PIRG approach for the effective action, while the Wegner equation \cite{Wegner_1974} underlies the PIRG approach for the flow of probability distributions as used for PIKs, \cite{Ihssen:2025ybn, Ihssen:2026njd}.

\subsubsection{Fermi surfaces and their flow}
\label{sec:FermiSurfacesPIRG}

Finally, we discuss the fate of (dynamical) Fermi surfaces. Respective theories such as the Hubbard model are currently under investigation within the PIRG approach, both within fRG studies but also within the PIK-approach. Naturally, fRG-studies use the ground state expansion as set-up in \cite{Ihssen:2023nqd} and used in \cite{Bonanno:2025mon}. In its simplest form in a scalar theory it uses a trivial wave function $Z_\phi[\phi]\equiv 1$ as suggested in \cite{Baldazzi:2021ydj}. If naïvely implemented in fermionic theories, it suggests the use of a classical dispersion for the fermionic flowing field. This overlaps with the definition of renormalisation-group invariant fermionic fields $\psi_\textrm{ri}$, sometimes also the misnomer renormalised field is used. For relativistic fields they are defined with 
 \begin{align} 
 	\bar \psi_\textrm{ri}(p ) \bar \Gamma^{(2)}_k(p) \psi_\textrm{ri} =    \bar \psi(p ) \Gamma^{(2)}_k(p) \psi\,,
 	\label{eq:psiri}
\end{align}
where $\bar \Gamma^{(2)}_k(p)\simeq S^{(2)}(p)$. Such a rescaling of the fields is a trivial flowing field. The advantage of \labelcref{eq:psiri} and respective transformations for bosonic fields lies in the RG-invariance of fields and $n$-point functions $\bar\Gamma^{(n)}$. In case of a smooth and mildly varying momentum dependence of the two-point functions $\Gamma^{(2)}$ this facilitates and optimises SESs with only a partial momentum-dependence such as the derivative expansion or expansions, where the full momentum dependence of vertices is approximated by that of momentum channels or the symmetric point, for a comprehensive discussion of this scheme see \cite{Ihssen:2024miv}. 

This smoothness property already hints at the problem with Fermi surfaces: they have an intricate momentum dependence and may even include cuts and singularities. Moving these structures from the propagators to the vertices is possible but not advisable, see also the discussion in \cite{Wetterich:2024uub}. 
Note however, that this is a feasibility argument and not a conceptual one. Still, the proper implementation of the ground state expansion forbids such a rescaling as the Fermi surface of the ground state should not be trivialised. The last argument emphasises the systematic use of optimised expansion schemes which in our opinion is qualitatively facilitated by the PIRG approach, as different aspects of such an optimised expansion can be and are naturally spread between the different components $\Gamma_\phi,\dot\phi,{\cal O}$ of the PIRG approach. The price to pay for this advantage is the necessity to solve not only one but two or more differential equations. However, since these differential equations are qualitatively simpler, this comes as a benefit and not a price.

 \subsubsection{Upshot on physics with PIRGs}
 \label{sec:Wrap-up}

We have comprehensively argued that the PIRG approach allows for more general and optimised expansion schemes by trading the solution of the flow equation for the effective action for solving a larger, but far simpler, set of differential equations. This is akin to recasting a second order differential equation into two first order ones, which is commonly used within the solution of differential equations. These clear and structural advantages come with the necessity of re-assessing standard expansion schemes such as the derivative expansion and vertex expansion schemes in the context of the PIRG-approach. 

\section{Conclusion} 
\label{sec:Conclusion}

In the present work we have completed the physics-informed RG (PIRG) setup to quantum field theories, \cite{Ihssen:2024ihp, Ihssen:2025cff}, with a systematic and computationally accessible approach to correlation functions of general composite operators defined by \labelcref{eq:DefofO}. In particular this gives us access to all correlation functions of the fundamental field, which span the space of correlation functions of local operators. The respective operator PIRG flow for correlation functions is given by \labelcref{eq:GenFlowOp} and \labelcref{eq:PIRGOPSimple}. In terms of a vertex expansion this gives us access to the standard effective action of the fundamental field. In short, it offers a direct access to all observables of the quantum field theory at hand in the PIRG setup with its qualitatively simpler numerics and structural insights. 

We have illustrated the potential of the setup within the analytical example of the zero-dimensional field theory. This example was chosen as it allows us to benchmark the PIRG approach with exact results. Moreover, while not being a physical example, it is a showcase for any truncation with a finite number of operators, including a full effective potential and vertices with general momentum dependences.
This paves the way for novel insights and qualitatively improved numerics in strongly correlated systems ranging from condensed matter systems over QCD to quantum gravity. We illustrate this potential by briefly discussing two of the many applications:\\[-2ex]

The first one concerns the embedding of approximate flows in exact PIRG ones with a flowing field that accommodates the missing fluctuations. This embedding has been discussed in \cite{Ihssen:2025cff}, following the classification scheme in complete and consistent flows suggested in \cite{Litim:2002xm} and extended in \cite{Ihssen:2025cff} to PIRG flows. Specifically, this allows us to elevate the background field approximation in Wetterich flows or more generally proper time flows to target actions in an exact PIRG approach. This completes the evaluations started in \cite{Litim:2001ky, Litim:2002xm, Litim:2002hj}. 

These flows are of soothing simplicity, specifically in non-Abelian gauge theories and quantum gravity, where they are seemingly even gauge or diffeomorphism invariant. Consequently, despite their approximate nature they are still in common use, For respective works see the review \cite{Dupuis:2020fhh} and, e.g.,~\cite{Reuter:1993kw, Reuter:1996cp, Schaefer:1999em, Zappala:2002nx, Schaefer:2006ds}, for their potential embedding in or interpretation as exact flows see e.g.~\cite{Dietz:2015owa, Labus:2016lkh, Wetterich:2017aoy, Lippoldt:2018wvi, Bonanno:2019ukb, Falls:2020tmj, Mandric:2022dte, Wetterich:2024ivi, Pagani:2024lcn}. Hence, the present embedding within PIRGs has a great potential, see \cite{Ihssen:2025cff}. Still, it left us with the reconstruction task for observables as they are formulated in terms of the fundamental field. The operator PIRGs discussed here allow us to not only compute these observables but also to minimise the computational costs by chosen optimal operator PIRGs \labelcref{eq:GenFlowOp}. Note that this may or may not lead us to \labelcref{eq:PIRGOPSimple}. Such an evaluation goes beyond the scope of the present work and will be considered elsewhere. 
 
The second application concerns the computation of correlation functions of observables in a given expansion scheme within the PIRG approach. Here we consider the ground state expansion as advertised in \cite{Ihssen:2023nqd, Ihssen:2024ihp, Ihssen:2025cff}, as well as the minimal essential scheme \cite{Baldazzi:2021ydj}. In the ground state expansion the composite field $\phi$ may include transformation of the field basis to one anchored in the (dynamical) ground state of the theory, a simple example being the rotation from the Cartesian field basis in a scalar theory to a polar one in the broken phase as discussed in \cite{Lamprecht2007, Isaule:2018mxt, Isaule:2019pcm}. In both schemes, the ground state expansion and the minimal essential scheme, the field-dependent wave function can be absorbed in the composite field. This leaves us with the zeroth order of the derivative expansion, or local potential approximation (LPA), in the given field as the target action, if formulated with feed-down flows \cite{Ihssen:2024ihp}, even without feedback from the higher order terms. The respective flow for the effective potential is of the diffusion-convection type, and its solution has been studied extensively in the past decades, for a review see e.g.~\cite{Dupuis:2020fhh}. A treatment with modern solvers based on Galerkin methods as used for hydrodynamical applications has been suggested in \cite{Grossi:2019urj}, for further developments and applications see \cite{Grossi:2019urj, Grossi:2021ksl, Koenigstein:2021rxj,Steil:2021cbu, Koenigstein:2021syz, Ihssen:2022xkr, Ihssen:2023qaq, Sattler:2024ozv, Zorbach:2024rre}. Then, the embedding of the LPA approximation in the PIRG approach provides us with far more sophisticated approximation at a numerical cost that is close to that of the LPA computations. This has been demonstrated in \cite{Bonanno:2025mon} at the example of topological scaling of the energy gap $\Delta E$ in the anharmonic oscillator for asymptotically small anharmonicities. However, the topological scaling was extracted from a novel observable based on the flowing fields instead of directly accessing $\Delta E$. In combination with the present operator PIRGs also the latter observable, $\Delta E$, with direct physical relevance is accessible as well as others. This is but one of the many applications of the operator PIRG. 
	
In conclusion, the present work complements the PIRG approach to quantum field theories with a practical computational tool giving access to all correlation functions and hence the generating functional in a comprehensive way. It comes with a qualitatively reduced numerical cost as well as offering many structural insights relevant for optimal formulations of theories. This also is a natural link to novel machine learning approaches to quantum field theories including generating models for lattice simulations. We hope to report on all these exciting applications in the near future.

\subsection*{Acknowledgements}

We thank Kevin Falls, Renzo Kapust, Daniel Krojer and Christof Wetterich for discussions and work on related subjects. 
This work is funded by the Deutsche Forschungsgemeinschaft (DFG, German Research Foundation) under Germany’s Excellence Strategy EXC 2181/1 - 390900948 (the Heidelberg STRUCTURES Excellence Cluster) and the Collaborative Research Centre SFB 1225 (ISOQUANT). It is also supported by EMMI.

\appendix

 \section{Benchmark computations}
 \label{app:Details}

In this Appendix we provide the details on the numerical benchmark computations of the operator flow for the two-point function $\langle \hat\varphi^2\rangle $ in the zero-dimensional scalar theory with the classical action $S_\textrm{cl}[\varphi]$ defined in \labelcref{eq:Scl} and with $\lambda_{\varphi} = 0.1$ and $m_\varphi = 0.01$. The computations are done with a \textit{classical target action} (cTA) $\Gamma_T[\phi]=S_\textrm{cl}[\phi]+{\cal C}_k$ defined in \labelcref{eq:cTA}. In \cite{Ihssen:2024ihp} the cTA was used as an extreme case for the PIRG flow in which the whole dynamics of the theory is stored in the flowing field $\dot\phi(\phi)$. 
In the zero-dimensional case ($d=0$), the zeroth order derivative expansion (local potential approximation) is exact. This allows us to confront the results with those obtained by simply integrating the path integral, which reduces to a one-dimensional integral for $d=0$. In summary this example is a perfect non-trivial benchmark case: it illustrates the potential of the PIRG approach and its impressive flexibility. Note however, that in most practical applications the selection of the target action is guided by a maximisation of its physics content while still keeping numerical simplicity. For example, in the quantum mechanics computation in \cite{Bonanno:2025mon} (anharmonic oscillator) the target action was chosen such that the zeroth order derivative expansion is computed from the Wetterich flow, as this action is readily computed, and the first order of the expansion scheme is absorbed into the field.

Our results from the operator flows are also compared with those obtained from the cumulants-preserving flows discussed in \cite{Ihssen:2024ihp}. In particular we discuss different choices for the $J_{\mathcal{O}}$ dependence of $\dot \phi$ at the example of the operator flow of the two-point function. Note also that the cTA reduces the resolution of the PIRG flow from solving a PDE of convection-diffusion type for $\Gamma_k(\varphi)$ to that of solving a linear ODE for $ \dot \phi$, see also the discussion in \Cref{app:cumuFlow}. There, we recapitulate the computation of cumulants of the fundamental fields via \textit{cumulants-preserving} flows and solve the ODE for $ \dot \phi$. 

In \Cref{app:TotalDerivative} and \Cref{app:JOdep} we use the generalised operator flow \labelcref{eq:GenFlowOp} for the computation of the two-point function. We use two different versions of the generalised operator flow: in \Cref{app:TotalDerivative} we use the simple version \labelcref{eq:PIRGOPSimple}. In this representation, the total $t$-derivative structure of the flow is most apparent, see \labelcref{eq:DtRepresentation}. This representation is achieved with a $J_{\cal O}$-dependent $\hat\phi$, whose 
$J_{\cal O}$-dependence is used to eliminate the terms induced by correlations of the operator $\hat{\cal O}$ and the composite field operator $\hat\phi$. In \Cref{app:JOdep} we use a $J_{\cal O}$-independent $\hat\phi$. We show that this case reduces to the cumulants-preserving flow.

\subsection{Cumulants-preserving computation of $\langle \hat\varphi^2\rangle$}
\label{app:cumuFlow}
 
For the classical target action \labelcref{eq:cTA} the flow of the effective action reduces to a simple shift, 
\begin{align}
 	\partial_t \Gamma_T(\phi) = \partial_t { \mathcal{C}}_k \,.
\end{align}
Then, the generalised flow equation provides a linear ODE for $\dot\phi(\phi)$, see \labelcref{eq:PIRGflowGTScl}. This leaves us with two constant degrees of freedom that need to be fixed: the flow of the constant ${\mathcal{C}}_k$ and the integration constant of the linear ODE. They are fixed by the constraints 
\begin{align}
 	\lim_{\phi \to \infty} \dot \phi(\phi) \to 0\,, \quad \mathrm{and} \quad \dot \phi (0) = 0\,.
\end{align}
The first constraint ensures that the flowing field is well defined for large values of $\phi$ and the latter implements the \textit{cumulants preserving} property: this ensures that $n$th order derivatives of the effective action with respect to the classical couplings $m_\varphi^2$ and $\lambda_\varphi$ provide the (subtracted) expectation values of powers of $\hat\varphi^2$ and $\hat\varphi^4$ respectively, the cumulants. For further details see \cite{Ihssen:2024ihp}. 

Using this setup, the cumulants can be reconstructed from any target action by simply taking derivatives with respect to the classical couplings in the effective action, i.e.~
\begin{align}
 	\Bigl\langle \left(\hat \varphi^2\right)^n\Bigr\rangle^{(c)} =&\, -(-2)^n \frac{d^n \Gamma^{(c)} _\phi(\phi_\textrm{\tiny{EoM}})}{d (m_\varphi^2)^n} \,. 
 	\label{eq:varphi2Correlations}
\end{align}
The superscript $ ^{(c)}$ denotes the cumulants preserving property, which requires that the normalisation of the path integral is independent of the classical couplings
\begin{align} 
 	\frac{d^n {\cal N}_\phi^{(c)}}{d (m^2_\varphi)^n} =0= \frac{d^n {\cal N}_\phi^{(c)}}{d (\lambda_\varphi)^n}\,,\qquad \forall n\,. 
 	\label{eq:Nk-muIndep}	 
\end{align}
In summary, for a cumulants-preserving classical target action, the computation of the two-point function is done by simply integrating 
\begin{subequations}
\begin{align}
 		\partial_t \mathcal{O}_{2,k}(\phi_{\textrm{\tiny{EoM}}}) =&\, 2 \frac{d \, \partial_t{\mathcal{C}}}{d (m_\varphi^2)} \,,
 		\label{eq:varphi2Correlations2}
\end{align}
with the initial condition
\begin{align}
 		\mathcal{O}_{2,\Lambda}(\phi_{\textrm{\tiny{EoM}}}) = \frac{1}{\Lambda^2+ m^2_\varphi}\,. 
\end{align}
\end{subequations}
This was successfully done in \cite{Ihssen:2024ihp} for $m_\varphi^2>0$. This yields the flow of the two-point function on the equations of motion, which is depicted in \Cref{fig:twopoint_function} as a dashed black line.

\subsection{PIRG operator flows with total derivative flows}
\label{app:TotalDerivative}
 
We compute the flow of the two-point function from the total derivative flow \labelcref{eq:PIRGOPSimple}. As discussed in \Cref{sec:O2Flows} we may use different splits of the operator ${\cal O}_2$, such as \labelcref{eq:GStandardSplit} or \labelcref{eq:GphiSplit}. Here we use 
\begin{align}
{\cal O}_2 = {\cal O}_{2,c_\phi}+\phi^2\,,
\label{eq:GphiSplitd0}
\end{align}
in \labelcref{eq:GphiSplit}, leading to the representation \labelcref{eq:GenFlowOpO2phi} of the total derivative flow. Its solution requires $\dot \phi$ and $\dot{\phi}'$, which are readily computed from \labelcref{eq:PIRGflowGTScl}. With these preparations we integrate the resulting convection-diffusion equation \labelcref{eq:PIRGOPSimple} starting from the initial condition
\begin{align}
 	\mathcal{O}_{2,\Lambda}(\phi) = \frac{1}{\Lambda^2+ m^2_\varphi + \frac{3 \lambda_{\varphi}}{2}\phi^2} \,.
\label{eq:IntialO2}
\end{align}
We emphasise that the numerical effort of this computation is significantly lower as the respective one with the Wetterich flow: the present total derivative operator flow is a linear convection-diffusion equation. Moreover, the non-trivial input functions $\dot \phi$ and $\dot{\phi}'$ are computed from a linear ODE. Indeed, it can be even solved with a low numerical cost using standard, readily available tools provided in e.g.~Mathematica such as NDSolve \cite{Mathematica}.

The explicit computation is started at an initial scale $\Lambda = \exp(4) \approx 50$ and solved until $k_{0} = \exp(-10)$. Throughout the flow $\mathcal{O}_{2}$ develops a non-trivial $\phi$ dependence, depicted in \Cref{fig:Dep}, but its result on the equations of motion corresponds to the physical expectation value, see \Cref{fig:twopoint_function}.
\vspace{4mm}

\subsection{PIRG operator flows with $J_{\cal O}$-independent composite operator $\hat\phi$}
\label{app:JOdep}

In case one shies away from yet a further implicit definition for the operator $\hat\phi$, which was used for the derivation of the total derivative in \Cref{sec:SimpleOP-PIRGs}, we may use a $J_{\cal O}$-independent composite operator $\hat\phi$ with 
\begin{align} 
	\frac{\partial \hat\phi}{\partial J_{\cal O}}=0\,.
	\label{eq:hatphinoJ}
\end{align}
Then we have to resort to the generalised operator flow \labelcref{eq:GenFlowOp} as \labelcref{eq:hatphinoJ} induces a $J_{\mathcal{O}}$-dependence for $\dot \phi(\phi, J_{\mathcal{O}})$. We emphasise that such a choice is intricate in $d>0$ dimensions and may not give access to general correlation functions but only integrated correlation functions, the cumulants. 
In this context we note that this choice indeed reproduces the cumulants-preserving flow in \Cref{app:cumuFlow}. 

We begin with the observation that $J_{\mathcal{O}_2}$ couples to $\phi^2$ and hence is just a shift of the mass, $m_\varphi^2 \to m_\varphi^2 - 2 J_{{\cal O}_2}$. This allows us to simply identify $J_{{\cal O}_2}= -m_\varphi^2/2 $. Hence, taking an $m_\varphi^2$-derivative of the defining equation for the cTA \labelcref{eq:PIRGflowGTScl} corresponds to the composite operator flow for the two-point function, 
\begin{align}
\mathcal{O}_{2,k}(\phi) = 2 \frac{d\, \Gamma^{(c)} _\phi(\phi)}{d \,(m_\varphi^2)} \,.
\label{eq:JO2=m2}
\end{align}
We may also choose 
\begin{align}
 	\frac{\partial \dot \phi(\phi)}{\partial J_{\mathcal{O}}} = -2 \frac{\partial \dot \phi(\phi)}{\partial m_\varphi^2} \,.
\end{align}
Using the $\phi$-connected split \labelcref{eq:GphiSplitd0} already used in \Cref{app:TotalDerivative}. The remaining flow for the $\phi$-connected part $\partial_t {\cal O}_{2,c_\phi}$ is simply the constant $\dot {\mathcal{C}}_k$ and we recover \labelcref{eq:varphi2Correlations2}. This can be also verified numerically by integrating the flow 
\begin{align}\nonumber 
	&	\partial_t {\cal O}_{2,c_\phi} + \dot{\phi}\, {\cal O}_{2,c_\phi}^{(1)} + 2 \dot \phi \,\phi + 2 \frac{\delta \dot{\phi}}{\delta (m^2_\varphi)}\, \Gamma_\phi^{(1)} \\[1ex]
	& \hspace{-.12cm}
	= - \frac{1}{2} \left[G_\phi\left( {\cal D}_t \, R_\phi\right) G_\phi \right]\!\left[ {\cal O}^{(2)}_{2,c_\phi}+ 2\right] + 2 G_\phi\,\frac{\delta \dot\phi' }{\delta (m^2_\varphi)}\, R_\phi \,, 
	\label{eq:GenFlowOpO2Jo} 
\end{align}
which is shown in \Cref{fig:NoDep}.

\begingroup
\allowdisplaybreaks
	
	\bibliographystyle{apsrev4-2}
	\bibliography{references}

\begin{thebibliography}{74}%
\makeatletter
\providecommand \@ifxundefined [1]{%
 \@ifx{#1\undefined}
}%
\providecommand \@ifnum [1]{%
 \ifnum #1\expandafter \@firstoftwo
 \else \expandafter \@secondoftwo
 \fi
}%
\providecommand \@ifx [1]{%
 \ifx #1\expandafter \@firstoftwo
 \else \expandafter \@secondoftwo
 \fi
}%
\providecommand \natexlab [1]{#1}%
\providecommand \enquote  [1]{``#1''}%
\providecommand \bibnamefont  [1]{#1}%
\providecommand \bibfnamefont [1]{#1}%
\providecommand \citenamefont [1]{#1}%
\providecommand \href@noop [0]{\@secondoftwo}%
\providecommand \href [0]{\begingroup \@sanitize@url \@href}%
\providecommand \@href[1]{\@@startlink{#1}\@@href}%
\providecommand \@@href[1]{\endgroup#1\@@endlink}%
\providecommand \@sanitize@url [0]{\catcode `\\12\catcode `\$12\catcode
  `\&12\catcode `\#12\catcode `\^12\catcode `\_12\catcode `\%12\relax}%
\providecommand \@@startlink[1]{}%
\providecommand \@@endlink[0]{}%
\providecommand \url  [0]{\begingroup\@sanitize@url \@url }%
\providecommand \@url [1]{\endgroup\@href {#1}{\urlprefix }}%
\providecommand \urlprefix  [0]{URL }%
\providecommand \Eprint [0]{\href }%
\providecommand \doibase [0]{https://doi.org/}%
\providecommand \selectlanguage [0]{\@gobble}%
\providecommand \bibinfo  [0]{\@secondoftwo}%
\providecommand \bibfield  [0]{\@secondoftwo}%
\providecommand \translation [1]{[#1]}%
\providecommand \BibitemOpen [0]{}%
\providecommand \bibitemStop [0]{}%
\providecommand \bibitemNoStop [0]{.\EOS\space}%
\providecommand \EOS [0]{\spacefactor3000\relax}%
\providecommand \BibitemShut  [1]{\csname bibitem#1\endcsname}%
\let\auto@bib@innerbib\@empty
\bibitem [{\citenamefont {Ihssen}\ and\ \citenamefont
  {Pawlowski}(2025{\natexlab{a}})}]{Ihssen:2024ihp}%
  \BibitemOpen
  \bibfield  {author} {\bibinfo {author} {\bibfnamefont {F.}~\bibnamefont
  {Ihssen}}\ and\ \bibinfo {author} {\bibfnamefont {J.~M.}\ \bibnamefont
  {Pawlowski}},\ }\href {https://doi.org/10.1016/j.aop.2025.170177} {\bibfield
  {journal} {\bibinfo  {journal} {Annals Phys.}\ }\textbf {\bibinfo {volume}
  {481}},\ \bibinfo {pages} {170177} (\bibinfo {year} {2025}{\natexlab{a}})},\
  \Eprint {https://arxiv.org/abs/2409.13679} {arXiv:2409.13679 [hep-th]}
  \BibitemShut {NoStop}%
\bibitem [{\citenamefont {Pawlowski}(2007)}]{Pawlowski:2005xe}%
  \BibitemOpen
  \bibfield  {author} {\bibinfo {author} {\bibfnamefont {J.~M.}\ \bibnamefont
  {Pawlowski}},\ }\href {https://doi.org/10.1016/j.aop.2007.01.007} {\bibfield
  {journal} {\bibinfo  {journal} {Annals Phys.}\ }\textbf {\bibinfo {volume}
  {322}},\ \bibinfo {pages} {2831} (\bibinfo {year} {2007})},\ \Eprint
  {https://arxiv.org/abs/hep-th/0512261} {arXiv:hep-th/0512261} \BibitemShut
  {NoStop}%
\bibitem [{\citenamefont {Wetterich}(1993)}]{Wetterich:1992yh}%
  \BibitemOpen
  \bibfield  {author} {\bibinfo {author} {\bibfnamefont {C.}~\bibnamefont
  {Wetterich}},\ }\href {https://doi.org/10.1016/0370-2693(93)90726-X}
  {\bibfield  {journal} {\bibinfo  {journal} {Phys. Lett. B}\ }\textbf
  {\bibinfo {volume} {301}},\ \bibinfo {pages} {90} (\bibinfo {year} {1993})},\
  \Eprint {https://arxiv.org/abs/1710.05815} {arXiv:1710.05815 [hep-th]}
  \BibitemShut {NoStop}%
\bibitem [{\citenamefont {Ihssen}\ and\ \citenamefont
  {Pawlowski}(2025{\natexlab{b}})}]{Ihssen:2025cff}%
  \BibitemOpen
  \bibfield  {author} {\bibinfo {author} {\bibfnamefont {F.}~\bibnamefont
  {Ihssen}}\ and\ \bibinfo {author} {\bibfnamefont {J.~M.}\ \bibnamefont
  {Pawlowski}},\ }\href {https://doi.org/10.1103/jhqd-cjzk} {\bibfield
  {journal} {\bibinfo  {journal} {Phys. Rev. D}\ }\textbf {\bibinfo {volume}
  {112}},\ \bibinfo {pages} {105005} (\bibinfo {year} {2025}{\natexlab{b}})},\
  \Eprint {https://arxiv.org/abs/2503.22638} {arXiv:2503.22638 [hep-th]}
  \BibitemShut {NoStop}%
\bibitem [{\citenamefont {Bonanno}\ \emph {et~al.}(2026)\citenamefont
  {Bonanno}, \citenamefont {Ihssen},\ and\ \citenamefont
  {Pawlowski}}]{Bonanno:2025mon}%
  \BibitemOpen
  \bibfield  {author} {\bibinfo {author} {\bibfnamefont {A.}~\bibnamefont
  {Bonanno}}, \bibinfo {author} {\bibfnamefont {F.}~\bibnamefont {Ihssen}},\
  and\ \bibinfo {author} {\bibfnamefont {J.~M.}\ \bibnamefont {Pawlowski}},\
  }\href {https://doi.org/10.21468/SciPostPhysCore.9.1.005} {\bibfield
  {journal} {\bibinfo  {journal} {SciPost Phys. Core}\ }\textbf {\bibinfo
  {volume} {9}},\ \bibinfo {pages} {005} (\bibinfo {year} {2026})},\ \Eprint
  {https://arxiv.org/abs/2504.03437} {arXiv:2504.03437 [hep-th]} \BibitemShut
  {NoStop}%
\bibitem [{\citenamefont {Baldazzi}\ \emph
  {et~al.}(2022{\natexlab{a}})\citenamefont {Baldazzi}, \citenamefont
  {Zinati},\ and\ \citenamefont {Falls}}]{Baldazzi:2021ydj}%
  \BibitemOpen
  \bibfield  {author} {\bibinfo {author} {\bibfnamefont {A.}~\bibnamefont
  {Baldazzi}}, \bibinfo {author} {\bibfnamefont {R.~B.~A.}\ \bibnamefont
  {Zinati}},\ and\ \bibinfo {author} {\bibfnamefont {K.}~\bibnamefont
  {Falls}},\ }\href {https://doi.org/10.21468/SciPostPhys.13.4.085} {\bibfield
  {journal} {\bibinfo  {journal} {SciPost Phys.}\ }\textbf {\bibinfo {volume}
  {13}},\ \bibinfo {pages} {085} (\bibinfo {year} {2022}{\natexlab{a}})},\
  \Eprint {https://arxiv.org/abs/2105.11482} {arXiv:2105.11482 [hep-th]}
  \BibitemShut {NoStop}%
\bibitem [{\citenamefont {Wetterich}(2024)}]{Wetterich:2024uub}%
  \BibitemOpen
  \bibfield  {author} {\bibinfo {author} {\bibfnamefont {C.}~\bibnamefont
  {Wetterich}},\ }\href {https://doi.org/10.1016/j.nuclphysb.2024.116707}
  {\bibfield  {journal} {\bibinfo  {journal} {Nucl. Phys. B}\ }\textbf
  {\bibinfo {volume} {1008}},\ \bibinfo {pages} {116707} (\bibinfo {year}
  {2024})},\ \Eprint {https://arxiv.org/abs/2402.04679} {arXiv:2402.04679
  [hep-th]} \BibitemShut {NoStop}%
\bibitem [{\citenamefont {Wetterich}(1996)}]{Wetterich:1996kf}%
  \BibitemOpen
  \bibfield  {author} {\bibinfo {author} {\bibfnamefont {C.}~\bibnamefont
  {Wetterich}},\ }\href {https://doi.org/10.1007/s002880050232} {\bibfield
  {journal} {\bibinfo  {journal} {Z. Phys. C}\ }\textbf {\bibinfo {volume}
  {72}},\ \bibinfo {pages} {139} (\bibinfo {year} {1996})},\ \Eprint
  {https://arxiv.org/abs/hep-ph/9604227} {arXiv:hep-ph/9604227} \BibitemShut
  {NoStop}%
\bibitem [{\citenamefont {Gies}\ and\ \citenamefont
  {Wetterich}(2002)}]{Gies:2001nw}%
  \BibitemOpen
  \bibfield  {author} {\bibinfo {author} {\bibfnamefont {H.}~\bibnamefont
  {Gies}}\ and\ \bibinfo {author} {\bibfnamefont {C.}~\bibnamefont
  {Wetterich}},\ }\href {https://doi.org/10.1103/PhysRevD.65.065001} {\bibfield
   {journal} {\bibinfo  {journal} {Phys. Rev. D}\ }\textbf {\bibinfo {volume}
  {65}},\ \bibinfo {pages} {065001} (\bibinfo {year} {2002})},\ \Eprint
  {https://arxiv.org/abs/hep-th/0107221} {arXiv:hep-th/0107221} \BibitemShut
  {NoStop}%
\bibitem [{\citenamefont {Gies}\ and\ \citenamefont
  {Wetterich}(2004)}]{Gies:2002hq}%
  \BibitemOpen
  \bibfield  {author} {\bibinfo {author} {\bibfnamefont {H.}~\bibnamefont
  {Gies}}\ and\ \bibinfo {author} {\bibfnamefont {C.}~\bibnamefont
  {Wetterich}},\ }\href {https://doi.org/10.1103/PhysRevD.69.025001} {\bibfield
   {journal} {\bibinfo  {journal} {Phys. Rev. D}\ }\textbf {\bibinfo {volume}
  {69}},\ \bibinfo {pages} {025001} (\bibinfo {year} {2004})},\ \Eprint
  {https://arxiv.org/abs/hep-th/0209183} {arXiv:hep-th/0209183} \BibitemShut
  {NoStop}%
\bibitem [{\citenamefont {Floerchinger}\ and\ \citenamefont
  {Wetterich}(2009)}]{Floerchinger:2009uf}%
  \BibitemOpen
  \bibfield  {author} {\bibinfo {author} {\bibfnamefont {S.}~\bibnamefont
  {Floerchinger}}\ and\ \bibinfo {author} {\bibfnamefont {C.}~\bibnamefont
  {Wetterich}},\ }\href {https://doi.org/10.1016/j.physletb.2009.09.014}
  {\bibfield  {journal} {\bibinfo  {journal} {Phys. Lett. B}\ }\textbf
  {\bibinfo {volume} {680}},\ \bibinfo {pages} {371} (\bibinfo {year}
  {2009})},\ \Eprint {https://arxiv.org/abs/0905.0915} {arXiv:0905.0915
  [hep-th]} \BibitemShut {NoStop}%
\bibitem [{\citenamefont {Fu}\ \emph {et~al.}(2020)\citenamefont {Fu},
  \citenamefont {Pawlowski},\ and\ \citenamefont {Rennecke}}]{Fu:2019hdw}%
  \BibitemOpen
  \bibfield  {author} {\bibinfo {author} {\bibfnamefont {W.-j.}\ \bibnamefont
  {Fu}}, \bibinfo {author} {\bibfnamefont {J.~M.}\ \bibnamefont {Pawlowski}},\
  and\ \bibinfo {author} {\bibfnamefont {F.}~\bibnamefont {Rennecke}},\ }\href
  {https://doi.org/10.1103/PhysRevD.101.054032} {\bibfield  {journal} {\bibinfo
   {journal} {Phys. Rev. D}\ }\textbf {\bibinfo {volume} {101}},\ \bibinfo
  {pages} {054032} (\bibinfo {year} {2020})},\ \Eprint
  {https://arxiv.org/abs/1909.02991} {arXiv:1909.02991 [hep-ph]} \BibitemShut
  {NoStop}%
\bibitem [{\citenamefont {Ihssen}\ \emph
  {et~al.}(2024{\natexlab{a}})\citenamefont {Ihssen}, \citenamefont
  {Pawlowski}, \citenamefont {Sattler},\ and\ \citenamefont
  {Wink}}]{Ihssen:2024miv}%
  \BibitemOpen
  \bibfield  {author} {\bibinfo {author} {\bibfnamefont {F.}~\bibnamefont
  {Ihssen}}, \bibinfo {author} {\bibfnamefont {J.~M.}\ \bibnamefont
  {Pawlowski}}, \bibinfo {author} {\bibfnamefont {F.~R.}\ \bibnamefont
  {Sattler}},\ and\ \bibinfo {author} {\bibfnamefont {N.}~\bibnamefont
  {Wink}},\ }\href@noop {} {\  (\bibinfo {year} {2024}{\natexlab{a}})},\
  \Eprint {https://arxiv.org/abs/2408.08413} {arXiv:2408.08413 [hep-ph]}
  \BibitemShut {NoStop}%
\bibitem [{\citenamefont {Lamprecht}(2007)}]{Lamprecht2007}%
  \BibitemOpen
  \bibfield  {author} {\bibinfo {author} {\bibfnamefont {F.}~\bibnamefont
  {Lamprecht}},\ }\href@noop {} {\bibinfo {title} {{Diploma thesis Heidelberg
  University}}} (\bibinfo {year} {2007})\BibitemShut {NoStop}%
\bibitem [{\citenamefont {Isaule}\ \emph {et~al.}(2018)\citenamefont {Isaule},
  \citenamefont {Birse},\ and\ \citenamefont {Walet}}]{Isaule:2018mxt}%
  \BibitemOpen
  \bibfield  {author} {\bibinfo {author} {\bibfnamefont {F.}~\bibnamefont
  {Isaule}}, \bibinfo {author} {\bibfnamefont {M.~C.}\ \bibnamefont {Birse}},\
  and\ \bibinfo {author} {\bibfnamefont {N.~R.}\ \bibnamefont {Walet}},\ }\href
  {https://doi.org/10.1103/PhysRevB.98.144502} {\bibfield  {journal} {\bibinfo
  {journal} {Phys. Rev. B}\ }\textbf {\bibinfo {volume} {98}},\ \bibinfo
  {pages} {144502} (\bibinfo {year} {2018})},\ \Eprint
  {https://arxiv.org/abs/1806.10373} {arXiv:1806.10373 [cond-mat.quant-gas]}
  \BibitemShut {NoStop}%
\bibitem [{\citenamefont {Isaule}\ \emph {et~al.}(2020)\citenamefont {Isaule},
  \citenamefont {Birse},\ and\ \citenamefont {Walet}}]{Isaule:2019pcm}%
  \BibitemOpen
  \bibfield  {author} {\bibinfo {author} {\bibfnamefont {F.}~\bibnamefont
  {Isaule}}, \bibinfo {author} {\bibfnamefont {M.~C.}\ \bibnamefont {Birse}},\
  and\ \bibinfo {author} {\bibfnamefont {N.~R.}\ \bibnamefont {Walet}},\ }\href
  {https://doi.org/10.1016/j.aop.2019.168006} {\bibfield  {journal} {\bibinfo
  {journal} {Annals Phys.}\ }\textbf {\bibinfo {volume} {412}},\ \bibinfo
  {pages} {168006} (\bibinfo {year} {2020})},\ \Eprint
  {https://arxiv.org/abs/1902.07135} {arXiv:1902.07135 [cond-mat.quant-gas]}
  \BibitemShut {NoStop}%
\bibitem [{\citenamefont {Ihssen}\ and\ \citenamefont
  {Pawlowski}(2023{\natexlab{a}})}]{Ihssen:2022xjv}%
  \BibitemOpen
  \bibfield  {author} {\bibinfo {author} {\bibfnamefont {F.}~\bibnamefont
  {Ihssen}}\ and\ \bibinfo {author} {\bibfnamefont {J.~M.}\ \bibnamefont
  {Pawlowski}},\ }\href {https://doi.org/10.21468/SciPostPhys.15.2.074}
  {\bibfield  {journal} {\bibinfo  {journal} {SciPost Phys.}\ }\textbf
  {\bibinfo {volume} {15}},\ \bibinfo {pages} {074} (\bibinfo {year}
  {2023}{\natexlab{a}})},\ \Eprint {https://arxiv.org/abs/2207.10057}
  {arXiv:2207.10057 [hep-th]} \BibitemShut {NoStop}%
\bibitem [{\citenamefont {Salmhofer}(2007)}]{Salmhofer:2006pn}%
  \BibitemOpen
  \bibfield  {author} {\bibinfo {author} {\bibfnamefont {M.}~\bibnamefont
  {Salmhofer}},\ }\href {https://doi.org/10.1002/andp.200610223} {\bibfield
  {journal} {\bibinfo  {journal} {Annalen Phys.}\ }\textbf {\bibinfo {volume}
  {16}},\ \bibinfo {pages} {171} (\bibinfo {year} {2007})},\ \Eprint
  {https://arxiv.org/abs/cond-mat/0607289} {arXiv:cond-mat/0607289}
  \BibitemShut {NoStop}%
\bibitem [{\citenamefont {Wegner}(1974)}]{Wegner_1974}%
  \BibitemOpen
  \bibfield  {author} {\bibinfo {author} {\bibfnamefont {F.~J.}\ \bibnamefont
  {Wegner}},\ }\href {https://doi.org/10.1088/0022-3719/7/12/004} {\bibfield
  {journal} {\bibinfo  {journal} {Journal of Physics C: Solid State Physics}\
  }\textbf {\bibinfo {volume} {7}},\ \bibinfo {pages} {2098} (\bibinfo {year}
  {1974})}\BibitemShut {NoStop}%
\bibitem [{\citenamefont {Baldazzi}\ and\ \citenamefont
  {Falls}(2021)}]{Baldazzi:2021orb}%
  \BibitemOpen
  \bibfield  {author} {\bibinfo {author} {\bibfnamefont {A.}~\bibnamefont
  {Baldazzi}}\ and\ \bibinfo {author} {\bibfnamefont {K.}~\bibnamefont
  {Falls}},\ }\href {https://doi.org/10.3390/universe7080294} {\bibfield
  {journal} {\bibinfo  {journal} {Universe}\ }\textbf {\bibinfo {volume} {7}},\
  \bibinfo {pages} {294} (\bibinfo {year} {2021})},\ \Eprint
  {https://arxiv.org/abs/2107.00671} {arXiv:2107.00671 [hep-th]} \BibitemShut
  {NoStop}%
\bibitem [{\citenamefont {Baldazzi}\ \emph
  {et~al.}(2022{\natexlab{b}})\citenamefont {Baldazzi}, \citenamefont {Falls},\
  and\ \citenamefont {Ferrero}}]{Baldazzi:2021fye}%
  \BibitemOpen
  \bibfield  {author} {\bibinfo {author} {\bibfnamefont {A.}~\bibnamefont
  {Baldazzi}}, \bibinfo {author} {\bibfnamefont {K.}~\bibnamefont {Falls}},\
  and\ \bibinfo {author} {\bibfnamefont {R.}~\bibnamefont {Ferrero}},\ }\href
  {https://doi.org/10.1016/j.aop.2022.168822} {\bibfield  {journal} {\bibinfo
  {journal} {Annals Phys.}\ }\textbf {\bibinfo {volume} {440}},\ \bibinfo
  {pages} {168822} (\bibinfo {year} {2022}{\natexlab{b}})},\ \Eprint
  {https://arxiv.org/abs/2112.02118} {arXiv:2112.02118 [hep-th]} \BibitemShut
  {NoStop}%
\bibitem [{\citenamefont {Knorr}(2022)}]{Knorr:2022ilz}%
  \BibitemOpen
  \bibfield  {author} {\bibinfo {author} {\bibfnamefont {B.}~\bibnamefont
  {Knorr}},\ }\href@noop {} {\  (\bibinfo {year} {2022})},\ \Eprint
  {https://arxiv.org/abs/2204.08564} {arXiv:2204.08564 [hep-th]} \BibitemShut
  {NoStop}%
\bibitem [{\citenamefont {Knorr}(2024)}]{Knorr:2023usb}%
  \BibitemOpen
  \bibfield  {author} {\bibinfo {author} {\bibfnamefont {B.}~\bibnamefont
  {Knorr}},\ }\href {https://doi.org/10.1103/PhysRevD.110.026001} {\bibfield
  {journal} {\bibinfo  {journal} {Phys. Rev. D}\ }\textbf {\bibinfo {volume}
  {110}},\ \bibinfo {pages} {026001} (\bibinfo {year} {2024})},\ \Eprint
  {https://arxiv.org/abs/2311.12097} {arXiv:2311.12097 [hep-th]} \BibitemShut
  {NoStop}%
\bibitem [{\citenamefont {Baldazzi}\ \emph {et~al.}(2026)\citenamefont
  {Baldazzi}, \citenamefont {Falls}, \citenamefont {Kluth},\ and\ \citenamefont
  {Knorr}}]{Baldazzi:2023pep}%
  \BibitemOpen
  \bibfield  {author} {\bibinfo {author} {\bibfnamefont {A.}~\bibnamefont
  {Baldazzi}}, \bibinfo {author} {\bibfnamefont {K.}~\bibnamefont {Falls}},
  \bibinfo {author} {\bibfnamefont {Y.}~\bibnamefont {Kluth}},\ and\ \bibinfo
  {author} {\bibfnamefont {B.}~\bibnamefont {Knorr}},\ }\href
  {https://doi.org/10.1103/hlrm-d4g2} {\bibfield  {journal} {\bibinfo
  {journal} {Phys. Rev. D}\ }\textbf {\bibinfo {volume} {113}},\ \bibinfo
  {pages} {026005} (\bibinfo {year} {2026})},\ \Eprint
  {https://arxiv.org/abs/2312.03831} {arXiv:2312.03831 [hep-th]} \BibitemShut
  {NoStop}%
\bibitem [{\citenamefont {Knorr}\ and\ \citenamefont
  {Platania}(2025)}]{Knorr:2024yiu}%
  \BibitemOpen
  \bibfield  {author} {\bibinfo {author} {\bibfnamefont {B.}~\bibnamefont
  {Knorr}}\ and\ \bibinfo {author} {\bibfnamefont {A.}~\bibnamefont
  {Platania}},\ }\href {https://doi.org/10.1007/JHEP03(2025)003} {\bibfield
  {journal} {\bibinfo  {journal} {JHEP}\ }\textbf {\bibinfo {volume} {03}},\
  \bibinfo {pages} {003}},\ \Eprint {https://arxiv.org/abs/2405.08860}
  {arXiv:2405.08860 [hep-th]} \BibitemShut {NoStop}%
\bibitem [{\citenamefont {Ohta}\ and\ \citenamefont
  {Yamada}(2025)}]{Ohta:2025xxo}%
  \BibitemOpen
  \bibfield  {author} {\bibinfo {author} {\bibfnamefont {N.}~\bibnamefont
  {Ohta}}\ and\ \bibinfo {author} {\bibfnamefont {M.}~\bibnamefont {Yamada}},\
  }\href {https://doi.org/10.1103/16c5-73q2} {\bibfield  {journal} {\bibinfo
  {journal} {Phys. Rev. D}\ }\textbf {\bibinfo {volume} {112}},\ \bibinfo
  {pages} {066013} (\bibinfo {year} {2025})},\ \Eprint
  {https://arxiv.org/abs/2506.03601} {arXiv:2506.03601 [hep-th]} \BibitemShut
  {NoStop}%
\bibitem [{\citenamefont {Pagani}(2016)}]{Pagani:2016pad}%
  \BibitemOpen
  \bibfield  {author} {\bibinfo {author} {\bibfnamefont {C.}~\bibnamefont
  {Pagani}},\ }\href {https://doi.org/10.1103/PhysRevD.94.045001} {\bibfield
  {journal} {\bibinfo  {journal} {Phys. Rev. D}\ }\textbf {\bibinfo {volume}
  {94}},\ \bibinfo {pages} {045001} (\bibinfo {year} {2016})},\ \Eprint
  {https://arxiv.org/abs/1603.07250} {arXiv:1603.07250 [hep-th]} \BibitemShut
  {NoStop}%
\bibitem [{\citenamefont {Herbst}\ \emph {et~al.}(2015)\citenamefont {Herbst},
  \citenamefont {Luecker},\ and\ \citenamefont {Pawlowski}}]{Herbst:2015ona}%
  \BibitemOpen
  \bibfield  {author} {\bibinfo {author} {\bibfnamefont {T.~K.}\ \bibnamefont
  {Herbst}}, \bibinfo {author} {\bibfnamefont {J.}~\bibnamefont {Luecker}},\
  and\ \bibinfo {author} {\bibfnamefont {J.~M.}\ \bibnamefont {Pawlowski}},\
  }\href@noop {} {\  (\bibinfo {year} {2015})},\ \Eprint
  {https://arxiv.org/abs/1510.03830} {arXiv:1510.03830 [hep-ph]} \BibitemShut
  {NoStop}%
\bibitem [{\citenamefont {Becker}\ and\ \citenamefont
  {Pagani}(2019{\natexlab{a}})}]{Becker:2018quq}%
  \BibitemOpen
  \bibfield  {author} {\bibinfo {author} {\bibfnamefont {M.}~\bibnamefont
  {Becker}}\ and\ \bibinfo {author} {\bibfnamefont {C.}~\bibnamefont
  {Pagani}},\ }\href {https://doi.org/10.1103/PhysRevD.99.066002} {\bibfield
  {journal} {\bibinfo  {journal} {Phys. Rev. D}\ }\textbf {\bibinfo {volume}
  {99}},\ \bibinfo {pages} {066002} (\bibinfo {year} {2019}{\natexlab{a}})},\
  \Eprint {https://arxiv.org/abs/1810.11816} {arXiv:1810.11816 [gr-qc]}
  \BibitemShut {NoStop}%
\bibitem [{\citenamefont {Becker}\ and\ \citenamefont
  {Pagani}(2019{\natexlab{b}})}]{Becker:2019tlf}%
  \BibitemOpen
  \bibfield  {author} {\bibinfo {author} {\bibfnamefont {M.}~\bibnamefont
  {Becker}}\ and\ \bibinfo {author} {\bibfnamefont {C.}~\bibnamefont
  {Pagani}},\ }\href {https://doi.org/10.3390/universe5030075} {\bibfield
  {journal} {\bibinfo  {journal} {Universe}\ }\textbf {\bibinfo {volume} {5}},\
  \bibinfo {pages} {75} (\bibinfo {year} {2019}{\natexlab{b}})}\BibitemShut
  {NoStop}%
\bibitem [{\citenamefont {Houthoff}\ \emph {et~al.}(2020)\citenamefont
  {Houthoff}, \citenamefont {Kurov},\ and\ \citenamefont
  {Saueressig}}]{Houthoff:2020zqy}%
  \BibitemOpen
  \bibfield  {author} {\bibinfo {author} {\bibfnamefont {W.}~\bibnamefont
  {Houthoff}}, \bibinfo {author} {\bibfnamefont {A.}~\bibnamefont {Kurov}},\
  and\ \bibinfo {author} {\bibfnamefont {F.}~\bibnamefont {Saueressig}},\
  }\href {https://doi.org/10.1007/JHEP04(2020)099} {\bibfield  {journal}
  {\bibinfo  {journal} {JHEP}\ }\textbf {\bibinfo {volume} {04}},\ \bibinfo
  {pages} {099}},\ \Eprint {https://arxiv.org/abs/2002.00256} {arXiv:2002.00256
  [hep-th]} \BibitemShut {NoStop}%
\bibitem [{\citenamefont {Luo}\ \emph {et~al.}(2024)\citenamefont {Luo},
  \citenamefont {Luo},\ and\ \citenamefont {Melko}}]{Luo:2024wlw}%
  \BibitemOpen
  \bibfield  {author} {\bibinfo {author} {\bibfnamefont {X.-Z.}\ \bibnamefont
  {Luo}}, \bibinfo {author} {\bibfnamefont {D.}~\bibnamefont {Luo}},\ and\
  \bibinfo {author} {\bibfnamefont {R.~G.}\ \bibnamefont {Melko}},\ }\href@noop
  {} {\  (\bibinfo {year} {2024})},\ \Eprint {https://arxiv.org/abs/2403.03199}
  {arXiv:2403.03199 [quant-ph]} \BibitemShut {NoStop}%
\bibitem [{\citenamefont {Pawlowski}\ \emph {et~al.}(2017)\citenamefont
  {Pawlowski}, \citenamefont {Scherer}, \citenamefont {Schmidt},\ and\
  \citenamefont {Wetzel}}]{Pawlowski:2015mlf}%
  \BibitemOpen
  \bibfield  {author} {\bibinfo {author} {\bibfnamefont {J.~M.}\ \bibnamefont
  {Pawlowski}}, \bibinfo {author} {\bibfnamefont {M.~M.}\ \bibnamefont
  {Scherer}}, \bibinfo {author} {\bibfnamefont {R.}~\bibnamefont {Schmidt}},\
  and\ \bibinfo {author} {\bibfnamefont {S.~J.}\ \bibnamefont {Wetzel}},\
  }\href {https://doi.org/10.1016/j.aop.2017.06.017} {\bibfield  {journal}
  {\bibinfo  {journal} {Annals Phys.}\ }\textbf {\bibinfo {volume} {384}},\
  \bibinfo {pages} {165} (\bibinfo {year} {2017})},\ \Eprint
  {https://arxiv.org/abs/1512.03598} {arXiv:1512.03598 [hep-th]} \BibitemShut
  {NoStop}%
\bibitem [{\citenamefont {Becchi}(1996)}]{Becchi:1996an}%
  \BibitemOpen
  \bibfield  {author} {\bibinfo {author} {\bibfnamefont {C.}~\bibnamefont
  {Becchi}},\ }\href@noop {} {\  (\bibinfo {year} {1996})},\ \Eprint
  {https://arxiv.org/abs/hep-th/9607188} {arXiv:hep-th/9607188} \BibitemShut
  {NoStop}%
\bibitem [{\citenamefont {Igarashi}\ \emph {et~al.}(2010)\citenamefont
  {Igarashi}, \citenamefont {Itoh},\ and\ \citenamefont
  {Sonoda}}]{Igarashi:2009tj}%
  \BibitemOpen
  \bibfield  {author} {\bibinfo {author} {\bibfnamefont {Y.}~\bibnamefont
  {Igarashi}}, \bibinfo {author} {\bibfnamefont {K.}~\bibnamefont {Itoh}},\
  and\ \bibinfo {author} {\bibfnamefont {H.}~\bibnamefont {Sonoda}},\ }\href
  {https://doi.org/10.1143/PTPS.181.1} {\bibfield  {journal} {\bibinfo
  {journal} {Prog. Theor. Phys. Suppl.}\ }\textbf {\bibinfo {volume} {181}},\
  \bibinfo {pages} {1} (\bibinfo {year} {2010})},\ \Eprint
  {https://arxiv.org/abs/0909.0327} {arXiv:0909.0327 [hep-th]} \BibitemShut
  {NoStop}%
\bibitem [{\citenamefont {Ihssen}\ and\ \citenamefont
  {Pawlowski}(2023{\natexlab{b}})}]{Ihssen:2023nqd}%
  \BibitemOpen
  \bibfield  {author} {\bibinfo {author} {\bibfnamefont {F.}~\bibnamefont
  {Ihssen}}\ and\ \bibinfo {author} {\bibfnamefont {J.~M.}\ \bibnamefont
  {Pawlowski}},\ }\href@noop {} {\  (\bibinfo {year} {2023}{\natexlab{b}})},\
  \Eprint {https://arxiv.org/abs/2305.00816} {arXiv:2305.00816 [hep-th]}
  \BibitemShut {NoStop}%
\bibitem [{\citenamefont {Ihssen}\ \emph {et~al.}(2026)\citenamefont {Ihssen},
  \citenamefont {Kapust},\ and\ \citenamefont {Pawlowski}}]{Ihssen:2026njd}%
  \BibitemOpen
  \bibfield  {author} {\bibinfo {author} {\bibfnamefont {F.}~\bibnamefont
  {Ihssen}}, \bibinfo {author} {\bibfnamefont {R.}~\bibnamefont {Kapust}},\
  and\ \bibinfo {author} {\bibfnamefont {J.~M.}\ \bibnamefont {Pawlowski}},\
  }\href@noop {} {\  (\bibinfo {year} {2026})},\ \Eprint
  {https://arxiv.org/abs/2603.03159} {arXiv:2603.03159 [hep-lat]} \BibitemShut
  {NoStop}%
\bibitem [{\citenamefont {Ihssen}\ \emph
  {et~al.}(2025{\natexlab{a}})\citenamefont {Ihssen}, \citenamefont {Krojer},\
  and\ \citenamefont {Pawlowski}}]{IKP}%
  \BibitemOpen
  \bibfield  {author} {\bibinfo {author} {\bibfnamefont {F.}~\bibnamefont
  {Ihssen}}, \bibinfo {author} {\bibfnamefont {D.}~\bibnamefont {Krojer}},\
  and\ \bibinfo {author} {\bibfnamefont {J.~M.}\ \bibnamefont {Pawlowski}},\
  }\href@noop {} {\bibfield  {journal} {\bibinfo  {journal} {in preparation}\ }
  (\bibinfo {year} {2025}{\natexlab{a}})}\BibitemShut {NoStop}%
\bibitem [{\citenamefont {Falls}(2025{\natexlab{a}})}]{Falls:2025sxu}%
  \BibitemOpen
  \bibfield  {author} {\bibinfo {author} {\bibfnamefont {K.}~\bibnamefont
  {Falls}},\ }\href {https://doi.org/10.1007/JHEP11(2025)126} {\bibfield
  {journal} {\bibinfo  {journal} {JHEP}\ }\textbf {\bibinfo {volume} {11}},\
  \bibinfo {pages} {126}},\ \Eprint {https://arxiv.org/abs/2504.17851}
  {arXiv:2504.17851 [hep-th]} \BibitemShut {NoStop}%
\bibitem [{\citenamefont {Fu}\ \emph {et~al.}(2025)\citenamefont {Fu},
  \citenamefont {Huang}, \citenamefont {Pawlowski}, \citenamefont {Tan},\ and\
  \citenamefont {Zhou}}]{Fu:2025hcm}%
  \BibitemOpen
  \bibfield  {author} {\bibinfo {author} {\bibfnamefont {W.-j.}\ \bibnamefont
  {Fu}}, \bibinfo {author} {\bibfnamefont {C.}~\bibnamefont {Huang}}, \bibinfo
  {author} {\bibfnamefont {J.~M.}\ \bibnamefont {Pawlowski}}, \bibinfo {author}
  {\bibfnamefont {Y.-y.}\ \bibnamefont {Tan}},\ and\ \bibinfo {author}
  {\bibfnamefont {L.-j.}\ \bibnamefont {Zhou}},\ }\href
  {https://doi.org/10.1103/4sh5-w4yc} {\bibfield  {journal} {\bibinfo
  {journal} {Phys. Rev. D}\ }\textbf {\bibinfo {volume} {112}},\ \bibinfo
  {pages} {054047} (\bibinfo {year} {2025})},\ \Eprint
  {https://arxiv.org/abs/2502.14388} {arXiv:2502.14388 [hep-ph]} \BibitemShut
  {NoStop}%
\bibitem [{\citenamefont {Helmboldt}\ \emph {et~al.}(2015)\citenamefont
  {Helmboldt}, \citenamefont {Pawlowski},\ and\ \citenamefont
  {Strodthoff}}]{Helmboldt:2014iya}%
  \BibitemOpen
  \bibfield  {author} {\bibinfo {author} {\bibfnamefont {A.~J.}\ \bibnamefont
  {Helmboldt}}, \bibinfo {author} {\bibfnamefont {J.~M.}\ \bibnamefont
  {Pawlowski}},\ and\ \bibinfo {author} {\bibfnamefont {N.}~\bibnamefont
  {Strodthoff}},\ }\href {https://doi.org/10.1103/PhysRevD.91.054010}
  {\bibfield  {journal} {\bibinfo  {journal} {Phys. Rev. D}\ }\textbf {\bibinfo
  {volume} {91}},\ \bibinfo {pages} {054010} (\bibinfo {year} {2015})},\
  \Eprint {https://arxiv.org/abs/1409.8414} {arXiv:1409.8414 [hep-ph]}
  \BibitemShut {NoStop}%
\bibitem [{\citenamefont {Tripolt}\ \emph {et~al.}(2014)\citenamefont
  {Tripolt}, \citenamefont {Strodthoff}, \citenamefont {von Smekal},\ and\
  \citenamefont {Wambach}}]{Tripolt:2014fpa}%
  \BibitemOpen
  \bibfield  {author} {\bibinfo {author} {\bibfnamefont {R.-A.}\ \bibnamefont
  {Tripolt}}, \bibinfo {author} {\bibfnamefont {N.}~\bibnamefont {Strodthoff}},
  \bibinfo {author} {\bibfnamefont {L.}~\bibnamefont {von Smekal}},\ and\
  \bibinfo {author} {\bibfnamefont {J.}~\bibnamefont {Wambach}},\ }\href
  {https://doi.org/10.1016/j.nuclphysa.2014.09.061} {\bibfield  {journal}
  {\bibinfo  {journal} {Nucl. Phys. A}\ }\textbf {\bibinfo {volume} {931}},\
  \bibinfo {pages} {790} (\bibinfo {year} {2014})},\ \Eprint
  {https://arxiv.org/abs/1407.8387} {arXiv:1407.8387 [hep-ph]} \BibitemShut
  {NoStop}%
\bibitem [{\citenamefont {Pawlowski}\ and\ \citenamefont
  {Strodthoff}(2015)}]{Pawlowski:2015mia}%
  \BibitemOpen
  \bibfield  {author} {\bibinfo {author} {\bibfnamefont {J.~M.}\ \bibnamefont
  {Pawlowski}}\ and\ \bibinfo {author} {\bibfnamefont {N.}~\bibnamefont
  {Strodthoff}},\ }\href {https://doi.org/10.1103/PhysRevD.92.094009}
  {\bibfield  {journal} {\bibinfo  {journal} {Phys. Rev. D}\ }\textbf {\bibinfo
  {volume} {92}},\ \bibinfo {pages} {094009} (\bibinfo {year} {2015})},\
  \Eprint {https://arxiv.org/abs/1508.01160} {arXiv:1508.01160 [hep-ph]}
  \BibitemShut {NoStop}%
\bibitem [{\citenamefont {Ihssen}\ \emph
  {et~al.}(2025{\natexlab{b}})\citenamefont {Ihssen}, \citenamefont {Kapust},\
  and\ \citenamefont {Pawlowski}}]{Ihssen:2025ybn}%
  \BibitemOpen
  \bibfield  {author} {\bibinfo {author} {\bibfnamefont {F.}~\bibnamefont
  {Ihssen}}, \bibinfo {author} {\bibfnamefont {R.}~\bibnamefont {Kapust}},\
  and\ \bibinfo {author} {\bibfnamefont {J.~M.}\ \bibnamefont {Pawlowski}},\
  }\href@noop {} {\  (\bibinfo {year} {2025}{\natexlab{b}})},\ \Eprint
  {https://arxiv.org/abs/2510.26678} {arXiv:2510.26678 [hep-lat]} \BibitemShut
  {NoStop}%
\bibitem [{\citenamefont {Falls}(2025{\natexlab{b}})}]{Falls:2025tid}%
  \BibitemOpen
  \bibfield  {author} {\bibinfo {author} {\bibfnamefont {K.}~\bibnamefont
  {Falls}},\ }\href@noop {} {\  (\bibinfo {year} {2025}{\natexlab{b}})},\
  \Eprint {https://arxiv.org/abs/2503.05869} {arXiv:2503.05869 [hep-th]}
  \BibitemShut {NoStop}%
\bibitem [{\citenamefont {Fukushima}\ \emph {et~al.}(2022)\citenamefont
  {Fukushima}, \citenamefont {Pawlowski},\ and\ \citenamefont
  {Strodthoff}}]{Fukushima:2021ctq}%
  \BibitemOpen
  \bibfield  {author} {\bibinfo {author} {\bibfnamefont {K.}~\bibnamefont
  {Fukushima}}, \bibinfo {author} {\bibfnamefont {J.~M.}\ \bibnamefont
  {Pawlowski}},\ and\ \bibinfo {author} {\bibfnamefont {N.}~\bibnamefont
  {Strodthoff}},\ }\href {https://doi.org/10.1016/j.aop.2022.169106} {\bibfield
   {journal} {\bibinfo  {journal} {Annals Phys.}\ }\textbf {\bibinfo {volume}
  {446}},\ \bibinfo {pages} {169106} (\bibinfo {year} {2022})},\ \Eprint
  {https://arxiv.org/abs/2103.01129} {arXiv:2103.01129 [hep-ph]} \BibitemShut
  {NoStop}%
\bibitem [{\citenamefont {Litim}\ and\ \citenamefont
  {Pawlowski}(2002{\natexlab{a}})}]{Litim:2002xm}%
  \BibitemOpen
  \bibfield  {author} {\bibinfo {author} {\bibfnamefont {D.~F.}\ \bibnamefont
  {Litim}}\ and\ \bibinfo {author} {\bibfnamefont {J.~M.}\ \bibnamefont
  {Pawlowski}},\ }\href {https://doi.org/10.1103/PhysRevD.66.025030} {\bibfield
   {journal} {\bibinfo  {journal} {Phys. Rev. D}\ }\textbf {\bibinfo {volume}
  {66}},\ \bibinfo {pages} {025030} (\bibinfo {year} {2002}{\natexlab{a}})},\
  \Eprint {https://arxiv.org/abs/hep-th/0202188} {arXiv:hep-th/0202188}
  \BibitemShut {NoStop}%
\bibitem [{\citenamefont {Litim}\ and\ \citenamefont
  {Pawlowski}(2002{\natexlab{b}})}]{Litim:2001ky}%
  \BibitemOpen
  \bibfield  {author} {\bibinfo {author} {\bibfnamefont {D.~F.}\ \bibnamefont
  {Litim}}\ and\ \bibinfo {author} {\bibfnamefont {J.~M.}\ \bibnamefont
  {Pawlowski}},\ }\href {https://doi.org/10.1103/PhysRevD.65.081701} {\bibfield
   {journal} {\bibinfo  {journal} {Phys. Rev. D}\ }\textbf {\bibinfo {volume}
  {65}},\ \bibinfo {pages} {081701} (\bibinfo {year} {2002}{\natexlab{b}})},\
  \Eprint {https://arxiv.org/abs/hep-th/0111191} {arXiv:hep-th/0111191}
  \BibitemShut {NoStop}%
\bibitem [{\citenamefont {Litim}\ and\ \citenamefont
  {Pawlowski}(2002{\natexlab{c}})}]{Litim:2002hj}%
  \BibitemOpen
  \bibfield  {author} {\bibinfo {author} {\bibfnamefont {D.~F.}\ \bibnamefont
  {Litim}}\ and\ \bibinfo {author} {\bibfnamefont {J.~M.}\ \bibnamefont
  {Pawlowski}},\ }\href {https://doi.org/10.1016/S0370-2693(02)02693-X}
  {\bibfield  {journal} {\bibinfo  {journal} {Phys. Lett. B}\ }\textbf
  {\bibinfo {volume} {546}},\ \bibinfo {pages} {279} (\bibinfo {year}
  {2002}{\natexlab{c}})},\ \Eprint {https://arxiv.org/abs/hep-th/0208216}
  {arXiv:hep-th/0208216} \BibitemShut {NoStop}%
\bibitem [{\citenamefont {Dupuis}\ \emph {et~al.}(2021)\citenamefont {Dupuis},
  \citenamefont {Canet}, \citenamefont {Eichhorn}, \citenamefont {Metzner},
  \citenamefont {Pawlowski}, \citenamefont {Tissier},\ and\ \citenamefont
  {Wschebor}}]{Dupuis:2020fhh}%
  \BibitemOpen
  \bibfield  {author} {\bibinfo {author} {\bibfnamefont {N.}~\bibnamefont
  {Dupuis}}, \bibinfo {author} {\bibfnamefont {L.}~\bibnamefont {Canet}},
  \bibinfo {author} {\bibfnamefont {A.}~\bibnamefont {Eichhorn}}, \bibinfo
  {author} {\bibfnamefont {W.}~\bibnamefont {Metzner}}, \bibinfo {author}
  {\bibfnamefont {J.~M.}\ \bibnamefont {Pawlowski}}, \bibinfo {author}
  {\bibfnamefont {M.}~\bibnamefont {Tissier}},\ and\ \bibinfo {author}
  {\bibfnamefont {N.}~\bibnamefont {Wschebor}},\ }\href
  {https://doi.org/10.1016/j.physrep.2021.01.001} {\bibfield  {journal}
  {\bibinfo  {journal} {Phys. Rept.}\ }\textbf {\bibinfo {volume} {910}},\
  \bibinfo {pages} {1} (\bibinfo {year} {2021})},\ \Eprint
  {https://arxiv.org/abs/2006.04853} {arXiv:2006.04853 [cond-mat.stat-mech]}
  \BibitemShut {NoStop}%
\bibitem [{\citenamefont {Reuter}\ and\ \citenamefont
  {Wetterich}(1994)}]{Reuter:1993kw}%
  \BibitemOpen
  \bibfield  {author} {\bibinfo {author} {\bibfnamefont {M.}~\bibnamefont
  {Reuter}}\ and\ \bibinfo {author} {\bibfnamefont {C.}~\bibnamefont
  {Wetterich}},\ }\href {https://doi.org/10.1016/0550-3213(94)90543-6}
  {\bibfield  {journal} {\bibinfo  {journal} {Nucl. Phys. B}\ }\textbf
  {\bibinfo {volume} {417}},\ \bibinfo {pages} {181} (\bibinfo {year}
  {1994})}\BibitemShut {NoStop}%
\bibitem [{\citenamefont {Reuter}(1998)}]{Reuter:1996cp}%
  \BibitemOpen
  \bibfield  {author} {\bibinfo {author} {\bibfnamefont {M.}~\bibnamefont
  {Reuter}},\ }\href {https://doi.org/10.1103/PhysRevD.57.971} {\bibfield
  {journal} {\bibinfo  {journal} {Phys. Rev. D}\ }\textbf {\bibinfo {volume}
  {57}},\ \bibinfo {pages} {971} (\bibinfo {year} {1998})},\ \Eprint
  {https://arxiv.org/abs/hep-th/9605030} {arXiv:hep-th/9605030} \BibitemShut
  {NoStop}%
\bibitem [{\citenamefont {Schaefer}\ and\ \citenamefont
  {Pirner}(1999)}]{Schaefer:1999em}%
  \BibitemOpen
  \bibfield  {author} {\bibinfo {author} {\bibfnamefont {B.-J.}\ \bibnamefont
  {Schaefer}}\ and\ \bibinfo {author} {\bibfnamefont {H.-J.}\ \bibnamefont
  {Pirner}},\ }\href {https://doi.org/10.1016/S0375-9474(99)00409-1} {\bibfield
   {journal} {\bibinfo  {journal} {Nucl. Phys. A}\ }\textbf {\bibinfo {volume}
  {660}},\ \bibinfo {pages} {439} (\bibinfo {year} {1999})},\ \Eprint
  {https://arxiv.org/abs/nucl-th/9903003} {arXiv:nucl-th/9903003} \BibitemShut
  {NoStop}%
\bibitem [{\citenamefont {Zappala}(2002)}]{Zappala:2002nx}%
  \BibitemOpen
  \bibfield  {author} {\bibinfo {author} {\bibfnamefont {D.}~\bibnamefont
  {Zappala}},\ }\href {https://doi.org/10.1103/PhysRevD.66.105020} {\bibfield
  {journal} {\bibinfo  {journal} {Phys. Rev. D}\ }\textbf {\bibinfo {volume}
  {66}},\ \bibinfo {pages} {105020} (\bibinfo {year} {2002})},\ \Eprint
  {https://arxiv.org/abs/hep-th/0202167} {arXiv:hep-th/0202167} \BibitemShut
  {NoStop}%
\bibitem [{\citenamefont {Schaefer}\ and\ \citenamefont
  {Wambach}(2007)}]{Schaefer:2006ds}%
  \BibitemOpen
  \bibfield  {author} {\bibinfo {author} {\bibfnamefont {B.-J.}\ \bibnamefont
  {Schaefer}}\ and\ \bibinfo {author} {\bibfnamefont {J.}~\bibnamefont
  {Wambach}},\ }\href {https://doi.org/10.1103/PhysRevD.75.085015} {\bibfield
  {journal} {\bibinfo  {journal} {Phys. Rev. D}\ }\textbf {\bibinfo {volume}
  {75}},\ \bibinfo {pages} {085015} (\bibinfo {year} {2007})},\ \Eprint
  {https://arxiv.org/abs/hep-ph/0603256} {arXiv:hep-ph/0603256} \BibitemShut
  {NoStop}%
\bibitem [{\citenamefont {Dietz}\ and\ \citenamefont
  {Morris}(2015)}]{Dietz:2015owa}%
  \BibitemOpen
  \bibfield  {author} {\bibinfo {author} {\bibfnamefont {J.~A.}\ \bibnamefont
  {Dietz}}\ and\ \bibinfo {author} {\bibfnamefont {T.~R.}\ \bibnamefont
  {Morris}},\ }\href {https://doi.org/10.1007/JHEP04(2015)118} {\bibfield
  {journal} {\bibinfo  {journal} {JHEP}\ }\textbf {\bibinfo {volume} {04}},\
  \bibinfo {pages} {118}},\ \Eprint {https://arxiv.org/abs/1502.07396}
  {arXiv:1502.07396 [hep-th]} \BibitemShut {NoStop}%
\bibitem [{\citenamefont {Labus}\ \emph {et~al.}(2016)\citenamefont {Labus},
  \citenamefont {Morris},\ and\ \citenamefont {Slade}}]{Labus:2016lkh}%
  \BibitemOpen
  \bibfield  {author} {\bibinfo {author} {\bibfnamefont {P.}~\bibnamefont
  {Labus}}, \bibinfo {author} {\bibfnamefont {T.~R.}\ \bibnamefont {Morris}},\
  and\ \bibinfo {author} {\bibfnamefont {Z.~H.}\ \bibnamefont {Slade}},\ }\href
  {https://doi.org/10.1103/PhysRevD.94.024007} {\bibfield  {journal} {\bibinfo
  {journal} {Phys. Rev. D}\ }\textbf {\bibinfo {volume} {94}},\ \bibinfo
  {pages} {024007} (\bibinfo {year} {2016})},\ \Eprint
  {https://arxiv.org/abs/1603.04772} {arXiv:1603.04772 [hep-th]} \BibitemShut
  {NoStop}%
\bibitem [{\citenamefont {Wetterich}(2018)}]{Wetterich:2017aoy}%
  \BibitemOpen
  \bibfield  {author} {\bibinfo {author} {\bibfnamefont {C.}~\bibnamefont
  {Wetterich}},\ }\href {https://doi.org/10.1016/j.nuclphysb.2018.07.002}
  {\bibfield  {journal} {\bibinfo  {journal} {Nucl. Phys. B}\ }\textbf
  {\bibinfo {volume} {934}},\ \bibinfo {pages} {265} (\bibinfo {year}
  {2018})},\ \Eprint {https://arxiv.org/abs/1710.02494} {arXiv:1710.02494
  [hep-th]} \BibitemShut {NoStop}%
\bibitem [{\citenamefont {Lippoldt}(2018)}]{Lippoldt:2018wvi}%
  \BibitemOpen
  \bibfield  {author} {\bibinfo {author} {\bibfnamefont {S.}~\bibnamefont
  {Lippoldt}},\ }\href {https://doi.org/10.1016/j.physletb.2018.05.037}
  {\bibfield  {journal} {\bibinfo  {journal} {Phys. Lett. B}\ }\textbf
  {\bibinfo {volume} {782}},\ \bibinfo {pages} {275} (\bibinfo {year}
  {2018})},\ \Eprint {https://arxiv.org/abs/1804.04409} {arXiv:1804.04409
  [hep-th]} \BibitemShut {NoStop}%
\bibitem [{\citenamefont {Bonanno}\ \emph {et~al.}(2020)\citenamefont
  {Bonanno}, \citenamefont {Lippoldt}, \citenamefont {Percacci},\ and\
  \citenamefont {Vacca}}]{Bonanno:2019ukb}%
  \BibitemOpen
  \bibfield  {author} {\bibinfo {author} {\bibfnamefont {A.}~\bibnamefont
  {Bonanno}}, \bibinfo {author} {\bibfnamefont {S.}~\bibnamefont {Lippoldt}},
  \bibinfo {author} {\bibfnamefont {R.}~\bibnamefont {Percacci}},\ and\
  \bibinfo {author} {\bibfnamefont {G.~P.}\ \bibnamefont {Vacca}},\ }\href
  {https://doi.org/10.1140/epjc/s10052-020-7798-9} {\bibfield  {journal}
  {\bibinfo  {journal} {Eur. Phys. J. C}\ }\textbf {\bibinfo {volume} {80}},\
  \bibinfo {pages} {249} (\bibinfo {year} {2020})},\ \Eprint
  {https://arxiv.org/abs/1912.08135} {arXiv:1912.08135 [hep-th]} \BibitemShut
  {NoStop}%
\bibitem [{\citenamefont {Falls}(2021)}]{Falls:2020tmj}%
  \BibitemOpen
  \bibfield  {author} {\bibinfo {author} {\bibfnamefont {K.}~\bibnamefont
  {Falls}},\ }\href {https://doi.org/10.1140/epjc/s10052-020-08803-0}
  {\bibfield  {journal} {\bibinfo  {journal} {Eur. Phys. J. C}\ }\textbf
  {\bibinfo {volume} {81}},\ \bibinfo {pages} {121} (\bibinfo {year} {2021})},\
  \Eprint {https://arxiv.org/abs/2004.11409} {arXiv:2004.11409 [hep-th]}
  \BibitemShut {NoStop}%
\bibitem [{\citenamefont {Mandric}\ and\ \citenamefont
  {Morris}(2023)}]{Mandric:2022dte}%
  \BibitemOpen
  \bibfield  {author} {\bibinfo {author} {\bibfnamefont {V.-M.}\ \bibnamefont
  {Mandric}}\ and\ \bibinfo {author} {\bibfnamefont {T.~R.}\ \bibnamefont
  {Morris}},\ }\href {https://doi.org/10.1103/PhysRevD.107.065012} {\bibfield
  {journal} {\bibinfo  {journal} {Phys. Rev. D}\ }\textbf {\bibinfo {volume}
  {107}},\ \bibinfo {pages} {065012} (\bibinfo {year} {2023})},\ \Eprint
  {https://arxiv.org/abs/2210.00492} {arXiv:2210.00492 [hep-th]} \BibitemShut
  {NoStop}%
\bibitem [{\citenamefont {Wetterich}(2025)}]{Wetterich:2024ivi}%
  \BibitemOpen
  \bibfield  {author} {\bibinfo {author} {\bibfnamefont {C.}~\bibnamefont
  {Wetterich}},\ }\href {https://doi.org/10.1016/j.physletb.2025.139435}
  {\bibfield  {journal} {\bibinfo  {journal} {Phys. Lett. B}\ }\textbf
  {\bibinfo {volume} {864}},\ \bibinfo {pages} {139435} (\bibinfo {year}
  {2025})},\ \Eprint {https://arxiv.org/abs/2403.17523} {arXiv:2403.17523
  [hep-th]} \BibitemShut {NoStop}%
\bibitem [{\citenamefont {Pagani}\ and\ \citenamefont
  {Sonoda}(2025)}]{Pagani:2024lcn}%
  \BibitemOpen
  \bibfield  {author} {\bibinfo {author} {\bibfnamefont {C.}~\bibnamefont
  {Pagani}}\ and\ \bibinfo {author} {\bibfnamefont {H.}~\bibnamefont
  {Sonoda}},\ }\href {https://doi.org/10.1103/PhysRevD.111.105006} {\bibfield
  {journal} {\bibinfo  {journal} {Phys. Rev. D}\ }\textbf {\bibinfo {volume}
  {111}},\ \bibinfo {pages} {105006} (\bibinfo {year} {2025})},\ \Eprint
  {https://arxiv.org/abs/2408.03625} {arXiv:2408.03625 [hep-th]} \BibitemShut
  {NoStop}%
\bibitem [{\citenamefont {Grossi}\ and\ \citenamefont
  {Wink}(2023)}]{Grossi:2019urj}%
  \BibitemOpen
  \bibfield  {author} {\bibinfo {author} {\bibfnamefont {E.}~\bibnamefont
  {Grossi}}\ and\ \bibinfo {author} {\bibfnamefont {N.}~\bibnamefont {Wink}},\
  }\href {https://doi.org/10.21468/SciPostPhysCore.6.4.071} {\bibfield
  {journal} {\bibinfo  {journal} {SciPost Phys. Core}\ }\textbf {\bibinfo
  {volume} {6}},\ \bibinfo {pages} {071} (\bibinfo {year} {2023})},\ \Eprint
  {https://arxiv.org/abs/1903.09503} {arXiv:1903.09503 [hep-th]} \BibitemShut
  {NoStop}%
\bibitem [{\citenamefont {Grossi}\ \emph {et~al.}(2021)\citenamefont {Grossi},
  \citenamefont {Ihssen}, \citenamefont {Pawlowski},\ and\ \citenamefont
  {Wink}}]{Grossi:2021ksl}%
  \BibitemOpen
  \bibfield  {author} {\bibinfo {author} {\bibfnamefont {E.}~\bibnamefont
  {Grossi}}, \bibinfo {author} {\bibfnamefont {F.~J.}\ \bibnamefont {Ihssen}},
  \bibinfo {author} {\bibfnamefont {J.~M.}\ \bibnamefont {Pawlowski}},\ and\
  \bibinfo {author} {\bibfnamefont {N.}~\bibnamefont {Wink}},\ }\href
  {https://doi.org/10.1103/PhysRevD.104.016028} {\bibfield  {journal} {\bibinfo
   {journal} {Phys. Rev. D}\ }\textbf {\bibinfo {volume} {104}},\ \bibinfo
  {pages} {016028} (\bibinfo {year} {2021})},\ \Eprint
  {https://arxiv.org/abs/2102.01602} {arXiv:2102.01602 [hep-ph]} \BibitemShut
  {NoStop}%
\bibitem [{\citenamefont {Koenigstein}\ \emph
  {et~al.}(2022{\natexlab{a}})\citenamefont {Koenigstein}, \citenamefont
  {Steil}, \citenamefont {Wink}, \citenamefont {Grossi},\ and\ \citenamefont
  {Braun}}]{Koenigstein:2021rxj}%
  \BibitemOpen
  \bibfield  {author} {\bibinfo {author} {\bibfnamefont {A.}~\bibnamefont
  {Koenigstein}}, \bibinfo {author} {\bibfnamefont {M.~J.}\ \bibnamefont
  {Steil}}, \bibinfo {author} {\bibfnamefont {N.}~\bibnamefont {Wink}},
  \bibinfo {author} {\bibfnamefont {E.}~\bibnamefont {Grossi}},\ and\ \bibinfo
  {author} {\bibfnamefont {J.}~\bibnamefont {Braun}},\ }\href
  {https://doi.org/10.1103/PhysRevD.106.065013} {\bibfield  {journal} {\bibinfo
   {journal} {Phys. Rev. D}\ }\textbf {\bibinfo {volume} {106}},\ \bibinfo
  {pages} {065013} (\bibinfo {year} {2022}{\natexlab{a}})},\ \Eprint
  {https://arxiv.org/abs/2108.10085} {arXiv:2108.10085 [cond-mat.stat-mech]}
  \BibitemShut {NoStop}%
\bibitem [{\citenamefont {Steil}\ and\ \citenamefont
  {Koenigstein}(2022)}]{Steil:2021cbu}%
  \BibitemOpen
  \bibfield  {author} {\bibinfo {author} {\bibfnamefont {M.~J.}\ \bibnamefont
  {Steil}}\ and\ \bibinfo {author} {\bibfnamefont {A.}~\bibnamefont
  {Koenigstein}},\ }\href {https://doi.org/10.1103/PhysRevD.106.065014}
  {\bibfield  {journal} {\bibinfo  {journal} {Phys. Rev. D}\ }\textbf {\bibinfo
  {volume} {106}},\ \bibinfo {pages} {065014} (\bibinfo {year} {2022})},\
  \Eprint {https://arxiv.org/abs/2108.04037} {arXiv:2108.04037
  [cond-mat.stat-mech]} \BibitemShut {NoStop}%
\bibitem [{\citenamefont {Koenigstein}\ \emph
  {et~al.}(2022{\natexlab{b}})\citenamefont {Koenigstein}, \citenamefont
  {Steil}, \citenamefont {Wink}, \citenamefont {Grossi}, \citenamefont {Braun},
  \citenamefont {Buballa},\ and\ \citenamefont
  {Rischke}}]{Koenigstein:2021syz}%
  \BibitemOpen
  \bibfield  {author} {\bibinfo {author} {\bibfnamefont {A.}~\bibnamefont
  {Koenigstein}}, \bibinfo {author} {\bibfnamefont {M.~J.}\ \bibnamefont
  {Steil}}, \bibinfo {author} {\bibfnamefont {N.}~\bibnamefont {Wink}},
  \bibinfo {author} {\bibfnamefont {E.}~\bibnamefont {Grossi}}, \bibinfo
  {author} {\bibfnamefont {J.}~\bibnamefont {Braun}}, \bibinfo {author}
  {\bibfnamefont {M.}~\bibnamefont {Buballa}},\ and\ \bibinfo {author}
  {\bibfnamefont {D.~H.}\ \bibnamefont {Rischke}},\ }\href
  {https://doi.org/10.1103/PhysRevD.106.065012} {\bibfield  {journal} {\bibinfo
   {journal} {Phys. Rev. D}\ }\textbf {\bibinfo {volume} {106}},\ \bibinfo
  {pages} {065012} (\bibinfo {year} {2022}{\natexlab{b}})},\ \Eprint
  {https://arxiv.org/abs/2108.02504} {arXiv:2108.02504 [cond-mat.stat-mech]}
  \BibitemShut {NoStop}%
\bibitem [{\citenamefont {Ihssen}\ \emph
  {et~al.}(2024{\natexlab{b}})\citenamefont {Ihssen}, \citenamefont
  {Pawlowski}, \citenamefont {Sattler},\ and\ \citenamefont
  {Wink}}]{Ihssen:2022xkr}%
  \BibitemOpen
  \bibfield  {author} {\bibinfo {author} {\bibfnamefont {F.}~\bibnamefont
  {Ihssen}}, \bibinfo {author} {\bibfnamefont {J.~M.}\ \bibnamefont
  {Pawlowski}}, \bibinfo {author} {\bibfnamefont {F.~R.}\ \bibnamefont
  {Sattler}},\ and\ \bibinfo {author} {\bibfnamefont {N.}~\bibnamefont
  {Wink}},\ }\href {https://doi.org/10.1016/j.cpc.2024.109182} {\bibfield
  {journal} {\bibinfo  {journal} {Comput. Phys. Commun.}\ }\textbf {\bibinfo
  {volume} {300}},\ \bibinfo {pages} {109182} (\bibinfo {year}
  {2024}{\natexlab{b}})},\ \Eprint {https://arxiv.org/abs/2207.12266}
  {arXiv:2207.12266 [hep-th]} \BibitemShut {NoStop}%
\bibitem [{\citenamefont {Ihssen}\ \emph {et~al.}(2023)\citenamefont {Ihssen},
  \citenamefont {Sattler},\ and\ \citenamefont {Wink}}]{Ihssen:2023qaq}%
  \BibitemOpen
  \bibfield  {author} {\bibinfo {author} {\bibfnamefont {F.}~\bibnamefont
  {Ihssen}}, \bibinfo {author} {\bibfnamefont {F.~R.}\ \bibnamefont
  {Sattler}},\ and\ \bibinfo {author} {\bibfnamefont {N.}~\bibnamefont
  {Wink}},\ }\href {https://doi.org/10.1103/PhysRevD.107.114009} {\bibfield
  {journal} {\bibinfo  {journal} {Phys. Rev. D}\ }\textbf {\bibinfo {volume}
  {107}},\ \bibinfo {pages} {114009} (\bibinfo {year} {2023})},\ \Eprint
  {https://arxiv.org/abs/2302.04736} {arXiv:2302.04736 [hep-th]} \BibitemShut
  {NoStop}%
\bibitem [{\citenamefont {Sattler}\ and\ \citenamefont
  {Pawlowski}(2024)}]{Sattler:2024ozv}%
  \BibitemOpen
  \bibfield  {author} {\bibinfo {author} {\bibfnamefont {F.~R.}\ \bibnamefont
  {Sattler}}\ and\ \bibinfo {author} {\bibfnamefont {J.~M.}\ \bibnamefont
  {Pawlowski}},\ }\href@noop {} {\  (\bibinfo {year} {2024})},\ \Eprint
  {https://arxiv.org/abs/2412.13043} {arXiv:2412.13043 [hep-ph]} \BibitemShut
  {NoStop}%
\bibitem [{\citenamefont {Zorbach}\ \emph {et~al.}(2026)\citenamefont
  {Zorbach}, \citenamefont {Koenigstein},\ and\ \citenamefont
  {Braun}}]{Zorbach:2024rre}%
  \BibitemOpen
  \bibfield  {author} {\bibinfo {author} {\bibfnamefont {N.}~\bibnamefont
  {Zorbach}}, \bibinfo {author} {\bibfnamefont {A.}~\bibnamefont
  {Koenigstein}},\ and\ \bibinfo {author} {\bibfnamefont {J.}~\bibnamefont
  {Braun}},\ }\href {https://doi.org/10.1103/z954-46md} {\bibfield  {journal}
  {\bibinfo  {journal} {Phys. Rev. D}\ }\textbf {\bibinfo {volume} {113}},\
  \bibinfo {pages} {036011} (\bibinfo {year} {2026})},\ \Eprint
  {https://arxiv.org/abs/2412.16053} {arXiv:2412.16053 [cond-mat.stat-mech]}
  \BibitemShut {NoStop}%
\bibitem [{\citenamefont {Inc.}()}]{Mathematica}%
  \BibitemOpen
  \bibfield  {author} {\bibinfo {author} {\bibfnamefont {W.~R.}\ \bibnamefont
  {Inc.}},\ }\href {https://www.wolfram.com/mathematica} {\bibinfo {title}
  {Mathematica, {V}ersion 14.2}},\ \bibinfo {note} {champaign, IL,
  2024}\BibitemShut {NoStop}%
\end{thebibliography}%
	
\end{document}